\documentclass[prd,twocolumn,nofootinbib,showpacs,superscriptaddress]{revtex4}
\pdfoutput=1
\usepackage{amsfonts}
\usepackage{amsmath}
\usepackage{amssymb}
\usepackage{bm}
\usepackage{dcolumn}
\usepackage{epsfig}
\usepackage{graphicx}
\usepackage{graphics}
\usepackage[latin1]{inputenc}
\usepackage{latexsym}
\usepackage{rotating}
\usepackage{hyperref}
\usepackage[all]{hypcap}
\usepackage{xspace} 
\usepackage[usenames]{color}

\usepackage{ulem}
\definecolor {darkgreen}{rgb}{0.2,0.7,0.2}

\newcommand\be{\begin{equation}}
\newcommand\ba{\begin{eqnarray}}
\newcommand\ee{\end{equation}}
\newcommand\ea{\end{eqnarray}}

\newcommand\bw{\begin{widetext}}
\newcommand\ew{\end{widetext}}

\newcommand{\mb}[1]{\mbox{\boldmath $#1$}}

\newcommand{\nn}{\nonumber}

\newcommand{\hGR}{h}
\newcommand{\hDef}{\mathfrak{h}}
\newcommand{\ppE}{{\mbox{\tiny ppE}}}

\newcommand{\GW}{{\mbox{\tiny GW}}}

\newcommand{\GR}{{\mbox{\tiny GR}}}

\newcommand{\IZ}{{\mbox{\tiny IZ}}}
\newcommand{\NZ}{{\mbox{\tiny NZ}}}
\newcommand{\FZ}{{\mbox{\tiny FZ}}}
\newcommand{\TT}{{\mbox{\tiny TT}}}
\newcommand{\MAT}{{\mbox{\tiny mat}}}

\newcommand{\CS}{{\mbox{\tiny CS}}}

\newcommand{\pont}{{\,^\ast\!}R\,R}
\newcommand{\met}{\mbox{g}}

\newcommand{\pd}{\partial}
\newcommand{\cd}{\nabla}
\newcommand{\LevTen}{\varepsilon}

\newcommand{\JJ}[4]{J^{(#1)}_{#2,#3,#4}}
\newcommand{\Jdef}[4]{\lim_{3\to #1}\pd^{(1)}_{#2}\pd^{(2)}_{#3}\pd^{(3)}_{#4}\mathcal{G}(ABC)}
\newcommand{\plusonetotwo}{+(1\leftrightarrow 2)}
\newcommand{\plusitoj}{+(i\leftrightarrow j)}

\newcommand{\self}{{\mbox{\tiny self}}}
\newcommand{\cross}{{\mbox{\tiny cross}}}

\newcommand{\Rbn}{\mathcal{R}_{\mathrm{IZ}}}
\newcommand{\Rnf}{\mathcal{R}_{\mathrm{NZ}}}

\begin{document}
\title{Post-Newtonian, Quasi-Circular Binary Inspirals in Quadratic Modified Gravity}

\author{Kent Yagi}
\affiliation{Department of Physics, Kyoto University, Kyoto, 606-8502, Japan.}

\author{Leo C. Stein}
\affiliation{Department of Physics and MIT Kavli Institute, Cambridge, MA 02139, USA.}

\author{Nicol\'as Yunes}
\affiliation{Department of Physics and MIT Kavli Institute, Cambridge, MA 02139, USA.}
\affiliation{Department of Physics, Montana State University, Bozeman, MT 59717, USA.}

\author{Takahiro Tanaka}
\affiliation{Yukawa Institute for Theoretical Physics, Kyoto University, Kyoto, 606-8502, Japan.}

\date{\today}

\begin{abstract} 

 We consider a general class of quantum gravity-inspired, modified
 gravity theories, where the Einstein-Hilbert action is extended
 through the addition of all terms quadratic in the curvature tensor
 coupled to scalar fields with standard kinetic energy. This class of
 theories includes Einstein-Dilaton-Gauss-Bonnet and Chern-Simons
 modified gravity as special cases. We analytically derive and solve the
 coupled field equations in the post-Newtonian approximation, assuming a
 comparable-mass, spinning black hole binary source in a quasi-circular,
 weak-field/slow-motion orbit. We find that a naive subtraction
 of divergent piece associated with the point-particle approximation is 
 ill-suited to represent compact objects in these theories. Instead, we model
 them by appropriate effective sources built so that known strong-field solutions 
 are reproduced in the far-field limit. In doing so, we prove that black holes in 
 Einstein-Dilaton-Gauss-Bonnet and Chern-Simons theory can have hair, while
 neutron stars have no scalar monopole charge, in diametrical opposition to results in scalar-tensor theories.
 We then employ techniques similar to the direct integration
 of the relaxed Einstein equations to obtain analytic expressions for
 the scalar field, metric perturbation, and the associated gravitational
 wave luminosity measured at infinity. We find that scalar field
 emission mainly dominates the energy flux budget, sourcing electric-type
 (even-parity) dipole scalar radiation and magnetic-type (odd-parity)
 quadrupole scalar radiation, correcting the General Relativistic
 prediction at relative $-1$PN and $2$PN orders. Such modifications lead
 to corrections in the emitted gravitational waves that can be mapped to
 the parameterized post-Einsteinian framework. Such modifications could be
 strongly constrained with gravitational wave observations. 

\end{abstract}

\pacs{04.30.-w,04.50.Kd,04.25.-g,04.25.Nx}


\maketitle

\section{Introduction}

The validity of Einstein's theory in the strong-gravity regime will soon be put to the most stringent tests yet, through the observation of gravitational waves (GWs) from compact object binary inspirals~\cite{lrr-2006-3,Schutz:2009tz,2010GWN.....4....3S}. Such waves carry detailed information about their source and the underlying gravitational theory in play. This information is primarily encoded in the evolution of the GW frequency, which in turn depends directly on the rate of energy transport away from the binary~\cite{lrr-2005-3}. In general relativity (GR), this transport is performed exclusively by GWs. In modified gravity theories, however, additional (scalar, vectorial or tensorial) degrees of freedom  can also carry energy and angular momentum away as they propagate.

Calculating how gravitational waves are corrected in modified gravity theories can be a gargantuan task
as the modification can increase the number of propagating degrees of freedom and the non-linearity of
the equations that control their propagation. For example, the amount of energy-momentum transported 
away from a binary system must be computed both from the GWs excited by the corresponding 
sources, as well as any additional waves associated with extra degrees of freedom~\cite{Misner:1973cw}. 
The sources that drive such waves can depend both on derivatives of the metric perturbation 
and the extra degrees of freedom, which, in turn, are specified by the solution to their
own equations of motion. The situation worsens if these are non-linearly coupled, 
e.g.~a scalar field equation of motion that depends on the metric tensor, whose evolution 
in turn depends on derivatives of the scalar field.

Such calculations, however, are feasible if one treats any GR deviations as
{\emph{small deformations}}~\cite{Yunes:2009hc}, which can be formalized
through the {\emph{small-coupling approximation}}, a common technique in perturbation theory 
to isolate physically relevant solutions in higher-derivative theories~\cite{1994PhRvD..49.5188C,Woodard:2006nt,2010PhRvD..82f4033C}. This is a reasonable approximation given that GR 
has passed a large number of tests, albeit in the weak-gravity regime. Even in the GW regime, signals will slowly transition from sampling weak fields to moderately strong fields during a full binary inspiral. The strongest GW events will not be able to sample anywhere close to the Plank regime, where one would expect completely new physics. The largest gravitational fields experienced by binaries occur when these merge, and even then, the metric curvature cannot exceed $m^{-2}$, where $m$ is the total mass of the binary. Earth-based detectors, such as LIGO~\cite{ligo}, VIRGO~\cite{virgo} and LCGT~\cite{lcgt}, and future space-borne detectors, such as LISA~\cite{lisa}, will only be able to sample gravitational fields up to this strength. 

Of the plethora of modified gravity theories, we choose to focus on a general class
that is characterized by the addition of quadratic curvature invariants to the
action, coupled to scalar fields with standard kinetic terms (see e.g.~Eq.~(\ref{exactaction}). 
Such theories are motivated from loop quantum
gravity~\cite{Ashtekar:2004eh,Rovelli:2004tv} and heterotic string
theory~\cite{1998stth.book.....P}, arising generically upon
four-dimensional compactification in the low-energy limit. Disjoint
sub-classes of quadratic theories reduce to
Einstein-Dilaton-Gauss-Bonnet (EDGB)
theory~\cite{Moura:2006pz,Pani:2009wy} and Dynamical Chern-Simons (CS)
modified gravity~\cite{jackiw:2003:cmo,Alexander:2008wi}. 

From a phenomenological standpoint, such quadratic gravity theories are
also interesting as straw-men to study small deviations from GR. This is
because the new quadratic terms are always small relative to the
Einstein-Hilbert term when considering merging binaries. In such
systems, the minimum radius of curvature is always larger than the
new scale introduced by the scalar fields. 
If this were not the case, astrophysical observations
would already have constrained quadratic gravity deviations. 

Quadratic gravity introduces an equation of motion for the scalar field and modifies the metric field equations. The former is a driven wave equation, whose sources are quadratic curvature invariants. The latter contains new terms that depend on the product of the scalar field and its derivatives with the Riemann tensor, Ricci tensor, Ricci scalar and their derivatives. As such, one might worry that higher derivative terms in the field equations
could render the system unstable. One must remember, however, that the action is 
a truncation (at quadratic order in the present case) of an {\emph{effective theory}} derived 
by integrating out heavy degrees of freedom contained in a more complete theory.  
Since we truncate the effective action, its validity is limited only to leading-order 
in the coupling parameters. Accounting for higher-order terms in the coupling would require 
the inclusion of higher-order terms (cubic, quartic, etc.) in the action~\cite{Woodard:2006nt}.
Therefore, the modified field equations should not be considered as an exact system, 
but rather as an effective one.

Given the above and using the {\emph{small-coupling}} approximation, the field equations become driven differential equations for the metric deformation and the scalar field. The source of the latter depends only on derivatives of the GR metric perturbation, while the source of the former depends both on the GR metric perturbation and the scalar field. We solve these equations in the post-Newtonian (PN) limit, where in particular we consider comparable-mass, spinning black hole (BH) binaries (electromagnetically uncharged), spiraling in a quasi-circular orbit. This forces the driven differential equations into driven wave equations, which can be studied with PN techniques~\cite{Damour:1983,Tagoshi:1994sm,Tanaka:1997dj,Blanchet:1992br,Blanchet:1995fr,Blanchet:1995fg,Blanchet:2002av} and then solved via retarded Green function methods. 

A complication arises when attempting to solve these equations,
as one must choose a prescription to describe BHs and neutron stars
(NSs). In standard PN theory and up to a certain high PN order, one can
choose a point-particle prescription, essentially because the exterior
gravitational field of a compact object is the same as that induced by a
point-particle. In modified quadratic gravity, however, both
non-spinning~\cite{Yunes:2011we} and spinning~\cite{Yunes:2009hc}, strong-field
BH solutions differ from that generated by simple point particles 
with a mass-monopole and a current-dipole moment; BHs in these theories 
have additional scalar multipole moments. (See Refs.~\cite{AliHaimoud:2011bk,AliHaimoud:2011fw} for similar discussions on NSs in CS gravity.) One can take these effects into account 
by constructing an \emph{effective} point-particle source that reproduces known, 
strong-field solutions to leading order in the weak-field region, sufficiently far away 
from the compact objects. With this effective point-particle prescription, we can then evaluate
the source of the driven wave equations and analytically solve them to
find the radiative part of the scalar field and metric perturbation. 

\subsection*{Executive Summary of Results}

Given the length of this paper, let us summarize the main results. We have devised
a framework in the small-coupling approximation to solve for compact binary inspirals
in modified quadratic gravity theories. One of the key ingredients in this framework is
the calculation of effective source terms that allow us to use the point-particle
approximation even for theories where such approximation is not valid. We applied this to 
modified quadratic gravity to find that both NSs and BHs have scalar hair, 
which leads to dipolar emission. EDGB and CS gravity are exceptions, where
although BHs retain scalar monopole and dipole charge, respectively, NSs shed 
the scalar monopole charge.
Therefore, BHs in EDGB 
generically contains dipolar GW emission, while CS gravity leads to modified quadrupolar emission. 

\begin{figure}
 \includegraphics[width=8.5cm,clip=true]{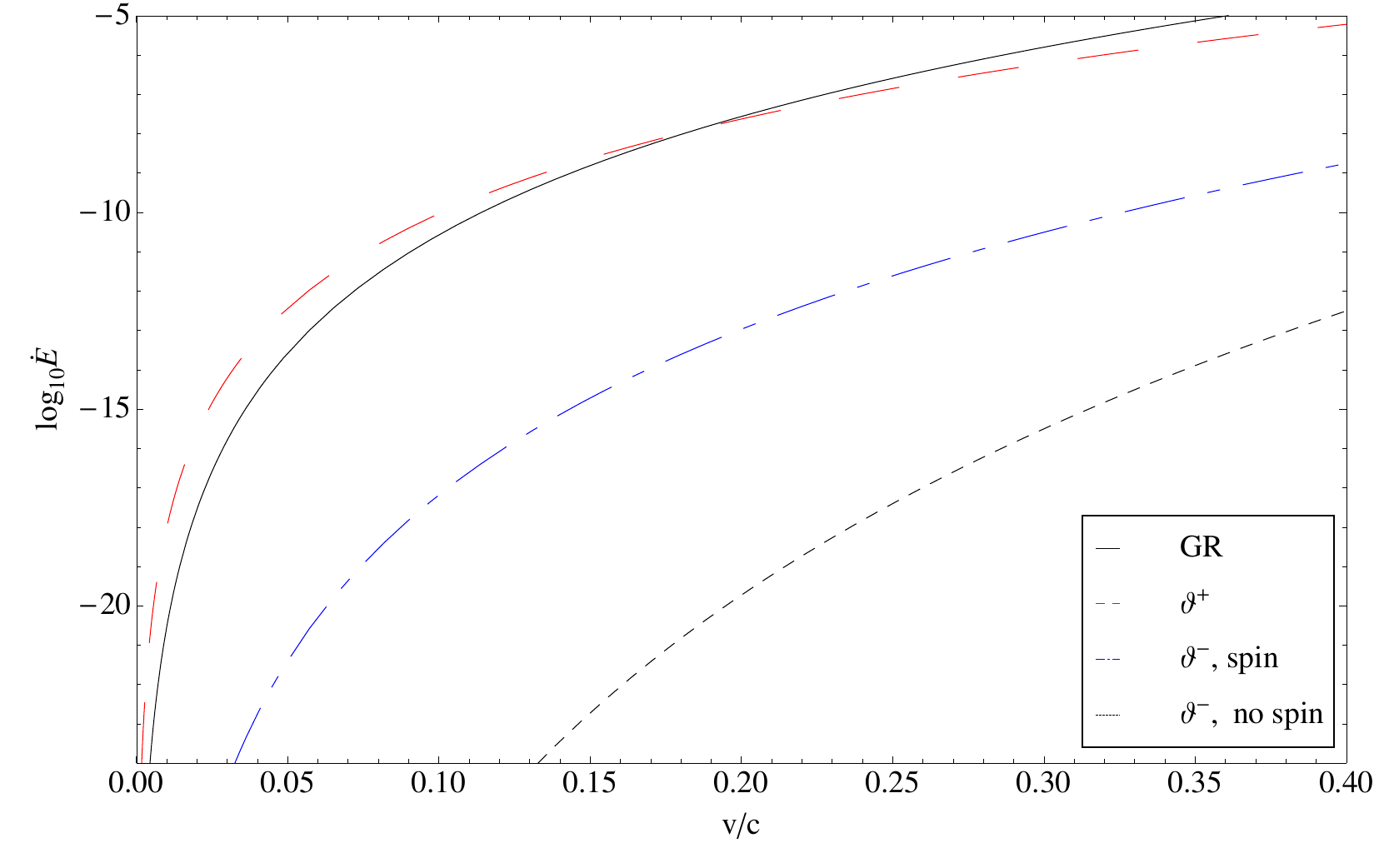}
 \caption{\label{fig:Edotall} Comparison of the energy flux carried by scalar fields of even-parity (dashed red), odd-parity and sourced by spinning BHs (blue dot-dashed) and odd-parity and sourced by non-spinning BHs (short dashed) relative to the GR prediction (solid black) as a function of orbital velocity. We here consider a quasi-circular, BH inspiral with $(m_{1},m_{2})=(8,20) M_{\odot}$, normalized spins $\hat{S}_{1}^{i} \equiv |S_{1}^{i}|/m_{1}^{2} = - \hat{S}_{2}^{i} \equiv - |S_{2}^{i}|/m_{2}^{2}$ perpendicular to the orbital plane, $|S_{A}^{i}| = m_{A}^{2}$ and coupling constants $\zeta_{3} = 6.25 \times 10^{-3} = \zeta_{4}$.}
\end{figure}

The presence of scalar monopole and dipole hair, and in particular the flux of energy-momentum carried by this hair, leads to a modification in the rate of change of the binary's binding energy. The even-parity sector of the theory leads to scalar hair, which modifies the energy flux at $-1$PN order relative to the GR quadrupole flux.   Of course, such a modification is proportional to the coupling parameter of the theory, which is assumed small. The odd-parity sector leads to dipole hair for spinning BH binaries, which modifies the energy flux at $2$PN relative order. If the BH binary components are non-spinning, they have no dipole hair but the binary orbital interaction generates a modification in the energy flux that enters at relative $7$PN order.  Figure~\ref{fig:Edotall} shows the energy flux carried by the even-parity scalar field (long dashed line), odd-parity scalar field (dot-dashed for spinning binaries and short dashed line for non-spinning binaries), and the GR quadrupole flux (solid line) as a function of orbital velocity. Observe that when one assumes that BHs are non-spinning, the scalar emission is greatly suppressed.

These energy flux corrections translate into changes to the
waveform observables. We explicitly calculate these and map them to
the parametrized post-Einsteinian (ppE)
framework~\cite{Yunes:2009ke,Cornish:2011ys}. Using the results of
Cornish~et~al.~\cite{Cornish:2011ys} we estimate that GW observations
could constrain the new length scale introduced in quadratic gravity
(related to the coupling constants of the theory) to roughly the BH
horizon scale. With a typical Ad.~LIGO stellar-mass BH inspiral
observation, one should be able to constrain the even-parity sector to
roughly $\mathcal{O}(10)$ km. With a typical LISA extreme-mass ratio
inspiral (EMRI) observation, one should be able to constrain the
odd-parity sector to roughly $\mathcal{O}(100)$ km. Such projected constraints 
are much stronger than current Solar System 
bounds~\cite{Alexander:2007vt,Smith:2007jm,Amendola:2007ni,AliHaimoud:2011fw}.

This paper is organized as follows:
Section~\ref{sec:QG} describes the action that will be considered in this paper and reviews the associated modified field equations and the scalar field equation of motion.  
Section~\ref{sec:Expansions} expands the field equations in the small-deformation approximation. 
Sections~\ref{sec:SF-Evolution} and~\ref{sec:Metric-Evolution} study the scalar field and metric deformation evolution, analytically solving the modified field equations.
Section~\ref{sec:E-flux} computes the energy flux carried by the scalar field and the metric deformation. 
Section~\ref{sec:Impact} considers the impact that such fluxes would have on gravitational waveform phase.
Section~\ref{sec:Discussions} concludes and points to future research. 

We have deferred many details of the computational techniques to the appendices. 
Appendix~\ref{sec:balding} shows the NSs in EDGB theory have no scalar monopole charge.
Appendix~\ref{int-tech} discusses specific integration techniques. 
Appendix~\ref{non-spin-reg-cont} estimates the order of the metric correction from the 
regularized contribution for non-spinning BHs in the odd-parity sector of the modified theory. 
Appendix~\ref{sec:Jtensors} discusses particular integrals that appear when solving the field equations.

Henceforth, we follow mostly the conventions of Misner, Thorne and Wheeler~\cite{Misner:1973cw}: Greek letters stand for spacetime indices; Latin letters in the middle of the alphabet $i,j,\ldots$, stand for spatial indices only. Parenthesis, square brackets and angled brackets in index lists denote symmetrization, antisymmetrization and the symmetric and trace free (STF) operator, respectively. Capital Latin letters usually refer to a multi-index, such as $x^{Q} = x^{ijk\ldots}$, where $x^{ijk\ldots} = x^{i} x^{j} x^{k} \ldots$. Partial derivatives are denoted with $\partial_{i} A = A_{,i} = \partial A/\partial x^{i}$, while covariant derivatives are denoted with the nabla $\nabla_{i} A$, for any quantity $A$. Deformations are labeled with the order-counting parameter $\varsigma$. Finally, we use geometric units, where $G=c=1$, except when denoting the order of certain terms in the PN approximation. Throughout, we performed analytic calculations with the \textsc{xTensor} package for \textsc{Mathematica}~\cite{2008CoPhC.179..597M,2009GReGr..41.2415B}.

\section{Modified gravity theories}
\label{sec:QG}

In this Section, we introduce the class of modified gravity theories that we study, by writing down its action and equations of motion. We then proceed to define the small deformation approximation more precisely.

\subsection{ABC of quadratic gravity}

Consider the following $4$-dimensional effective action:  
\ba 
S &\!\!
\equiv&\!\! \int d^4x \sqrt{-g} \left\{ \kappa R + \alpha_{1}
f_{1}(\vartheta) R^{2} + \alpha_{2} f_{2}(\vartheta) R_{\mu \nu} R^{\mu \nu}
\right. \nonumber \\ &&\!\! + \left. \alpha_{3} f_{3}(\vartheta) R_{\mu \nu \delta \sigma}
R^{\mu \nu \delta \sigma} + \alpha_{4} f_{4}(\vartheta) 
R_{\mu \nu \delta \sigma} \!{}^{*}R^{\mu \nu \delta \sigma}
\right. \nonumber \\ &&\!\! - \left. \frac{\beta}{2} \left[\nabla_{\mu}
\vartheta \nabla^{\mu} \vartheta + 2 V(\vartheta) \right] +
\mathcal{L}_{\rm mat} \right\}\,.
\label{exactaction} 
\ea
Here, $g$ stands for the determinant of the metric $g_{\mu \nu}$.
$R$, $R_{\mu \nu}$, $R_{\mu \nu \delta \sigma}$ and  ${}^{*}R_{\mu \nu \delta \sigma}$ 
are the Ricci scalar and tensor, the Riemann
tensor and its dual~\cite{Alexander:2009tp}, respectively, with the latter defined 
as\footnote{This definition is correct, in agreement with~\cite{Alexander:2009tp}, 
and fixing an inconsequential typo in~\cite{Sopuerta:2009iy}.}
$^{*}R^{\mu}{}_{\nu\delta\sigma} = (1/2) \LevTen_{\delta \sigma}{}^{\alpha \beta} R^{\mu}{}_{\nu \alpha \beta}$ and with $\LevTen^{\mu \nu \delta \sigma}$ the
Levi-Civita tensor. 
The quantity $\mathcal{L}_{\rm mat}$ is the external matter Lagrangian,
$\vartheta$ is a field, $(\alpha_{i},\beta)$ are coupling
constants and $\kappa = (16 \pi)^{-1}$.
This action contains all possible quadratic, algebraic curvature scalars
with running (i.e.~non-constant) couplings, where we assumed that all
quadratic terms are coupled to the {\emph{same}} field. All other
quadratic curvature terms are linearly dependent, such as the Weyl
tensor squared.

The theory defined by the action above is different from $f(R)$ theories
on several counts. First, $f(R)$ theories depend only on the Ricci
scalar, while the action above depends on the Ricci tensor, the Riemann
tensor and a dynamical field $\vartheta$. Second, $f(R)$ theories are
usually treated as exact, while the action presented above is an
{\emph{effective theory}}, truncated to quadratic order in the
Riemann tensor. The consequence of this is insisting on the use of
order-reduction in the field equations, where we treat 
all quantities that depend on $\alpha_{i}$ perturbatively. Such order reduction then
leads to the absence of additional polarization
modes~\cite{Sopuerta:2009iy,Stein:2010pn}, such as the longitudinal
scalar mode that arises in $f(R)$ theories. 

The field equations of dynamical quadratic gravity can be obtained by
varying the action with respect to all fields. For simplicity, we
restrict attention to coupling functions $f_{i}(\vartheta)$ that admit
the Taylor expansion $f_{i}(\vartheta) = f_{i}(0)+ f_{i}^\prime(0)
\vartheta + \mathcal{O}(\vartheta^{2})$ about small $\vartheta$, where
$f_{i}(0)$ and $f_{i}^\prime(0)$ are constants, and we assume 
that the asymptotic value of $\theta$ at spatial infinity vanishes. Let us further reabsorb $f_{i}(0)$ into the coupling constants $\alpha_{i}^{(0)} \equiv \alpha_{i} f_{i}(0)$ and $f_{i}^{\prime}(0)$ into the constants $\alpha_{i}^{(1)} \equiv \alpha_{i} f_{i}^{\prime}(0)$. Equation~\eqref{exactaction} then becomes
$S = S_{\GR} + S_{0} + S_{1}$:
\begin{align}
S_{\GR} \equiv & \int d^4x \sqrt{-g} \left\{ \kappa R + \mathcal{L}_{\rm mat} \right\}\,,\\
S_{0} \equiv & \int d^4x \sqrt{-g} \left\{ 
\alpha_{1}^{(0)} R^{2} 
+ \alpha_{2}^{(0)} R_{\mu \nu} R^{\mu \nu}
\right. \nonumber \\*
&+ \left. 
\alpha_{3}^{(0)} R_{\mu \nu \delta \sigma} R^{\mu \nu \delta \sigma} 
\right\}\,,\\
S_{1} \equiv & \int d^4x \sqrt{-g} \left\{ 
\alpha_{1}^{(1)} \vartheta R^{2} + \alpha_{2}^{(1)} \vartheta R_{\mu \nu} R^{\mu \nu}
\right. \nonumber \\*
&+ \left. 
\alpha_{3}^{(1)} \vartheta R_{\mu \nu \delta \sigma} R^{\mu \nu \delta \sigma} 
+ \alpha_{4}^{(1)} \vartheta  R_{\mu \nu \delta \sigma} \!{}^{*}R^{\mu \nu \delta \sigma}
\right. \nonumber \\*
&- \left. \frac{\beta}{2} \left[\nabla_{\mu}
\vartheta \nabla^{\mu} \vartheta + 2 V(\vartheta) \right] \right\}\,,
\label{action}
\end{align}
where clearly $S_{\GR}$ is the Einstein-Hilbert plus matter
 action. Notice that $S_{0}$ defines a GR correction that is 
decoupled from $\theta$.
The term proportional to $\alpha_4^{(0)}$ can not affect the classical
field equations since it is topological, i.e. the second Chern form,
so we have omitted it.
Similarly, if $\alpha_{i}^{(0)}$ are chosen to
 reconstruct the Gauss-Bonnet invariant,
$(\alpha_1^{(0)},\alpha_2^{(0)},\alpha_3^{(0)})=(1,-4,1)\alpha_{\mathrm{GB}}$,
 then these will not modify the field equations. On
 the other hand, $S_{1}$ defines a modification to GR with a direct
 (non-minimal) scalar field coupling, such that as the field goes to
 zero, the modified theory reduces to GR. 
We here restrict attention to the case $\alpha_{i}^{(0)} =
 0$. From this point forward, we will drop the superscript 
from $\alpha_i^{(1)}$.

The action above defines a class of modified gravity theories that
contains well-known GR extensions. For example, when $\alpha_{4} =
-\frac{1}{4} \alpha_{\CS}$ and all other $\alpha_{i} = 0$, quadratic
gravity reduces to dynamical CS gravity, where $\alpha_{\CS}$ is the CS
coupling parameter (see e.g.~\cite{Alexander:2009tp}). 
Alternatively,
when $\alpha_{4} = 0$, while 
$(\alpha_1,\alpha_2,\alpha_3)=(1,-4,1)\alpha_{\mathrm{EDGB}}$, 
quadratic gravity reduces to
Einstein-Dilaton-Gauss-Bonnet theory (see e.g.~\cite{Pani:2009wy}). Both
of these theories are motivated from fundamental physics; they
unavoidably arise as low-energy expansions of heterotic string
theory~\cite{Green:1987sp,Green:1987mn,Alexander:2004xd,lrr-2004-5}.
Dynamical CS gravity also arises in loop quantum gravity when the
Barbero-Immirzi parameter is promoted to a field in the presence of
fermions~\cite{Taveras:2008yf,Mercuri:2009zt,Gates:2009pt}. 

Variation of the action with respect to the metric yields the modified field equations: 
\ba
G_{\mu \nu} &\!\! + &\!\! \frac{\alpha_1 \vartheta}{\kappa} \mathcal{H}_{\mu \nu}^{(0)} +
\frac{\alpha_2 \vartheta }{\kappa} \mathcal{I}_{\mu \nu}^{(0)} 
+ \frac{\alpha_3 \vartheta}{\kappa} \mathcal{J}_{\mu \nu}^{(0)} 
\nonumber \\ \nonumber
&\!\!  &\!\! +
\frac{\alpha_1}{\kappa}
\mathcal{H}_{\mu \nu}^{(1)} + 
\frac{\alpha_2}{\kappa} \mathcal{I}_{\mu \nu}^{(1)} 
+ \frac{\alpha_3}{\kappa} \mathcal{J}_{\mu \nu}^{(1)} +
\frac{\alpha_4}{\kappa} \mathcal{K}_{\mu \nu}^{(1)} 
\nonumber \\
&=& \frac{1}{2\kappa} \left( T_{\mu \nu}^{\MAT} + T_{\mu \nu}^{(\vartheta)} \right)\,,
\label{FEs} 
\ea
where we have defined the short-hands\footnote{This corrects an error
  in Eq.~(5b) of~\cite{Yunes:2011we}.}
\begin{subequations}
\allowdisplaybreaks[1]
\begin{align}
\mathcal{H}_{\mu \nu}^{(0)} \equiv & 2 R R_{\mu \nu}  -
\frac{1}{2} g_{\mu \nu} R^{2} - 2 \nabla_{\mu \nu} R + 2 g_{\mu \nu} \square R\,,\\
\mathcal{I}_{\mu \nu}^{(0)} \equiv & \square R_{\mu \nu} + 2
R_{\mu \delta \nu \sigma} R^{\delta \sigma} - \frac{1}{2} g_{\mu \nu} 
R^{\delta \sigma} R_{\delta \sigma}
\nonumber \\*
&+ \frac{1}{2} g_{\mu \nu} \square R - \nabla_{\mu \nu} R\,,\\
\mathcal{J}_{\mu \nu}^{(0)}
\equiv & 8 R^{\delta \sigma} R_{\mu \delta \nu \sigma} - 
2 g_{\mu \nu} R^{\delta \sigma} R_{\delta \sigma} + 4 \square R_{\mu \nu}
\nonumber \\*
&- 2 R \, R_{\mu \nu} + \frac{1}{2} g_{\mu \nu} R^{2} - 2 \nabla_{\mu \nu} R\,,\\
\mathcal{H}_{\mu \nu}^{(1)} \equiv & -4
(\nabla_{(\mu} \vartheta) \nabla_{\nu)} R - 2 R \nabla_{\mu\nu} \vartheta 
\nonumber \\*
&+ g_{\mu \nu} \left[2 R \square \vartheta + 4 (\nabla^{\delta}\vartheta) \nabla_{\delta}R \right]\,,\\
\mathcal{I}_{\mu \nu}^{(1)} \equiv & -(\nabla_{(\mu}\vartheta) \cd_{\nu)} R - 2\nabla^\delta\vartheta
  \cd_{(\mu} R_{\nu)\delta} \nonumber\\*
&+ 2 \nabla^\delta\vartheta \cd_{\delta} R_{\mu \nu}
+R_{\mu\nu}\square \vartheta
- 2 R_{\delta(\mu}\cd^{\delta} \nabla_{\nu)}\vartheta \nonumber\\*
&+ \met_{\mu \nu} \left( \nabla^\delta \vartheta
  \cd_{\delta} R + R^{\delta \sigma} \cd_{\delta\sigma} \vartheta \right)\,,\\
\mathcal{J}_{\mu \nu}^{(1)}
\equiv & - 8 \left(\cd^\delta \vartheta \right) \left( \nabla_{(\mu} R_{\nu)\delta} - \nabla_{\delta} R_{\mu \nu}\right) + 4
R_{\mu \delta \nu \sigma} \nabla^{\delta\sigma} \vartheta\,, \\
\mathcal{K}_{\mu \nu}^{(1)}
\equiv & -4 \left( \cd^\delta\vartheta \right) \LevTen_{\delta \sigma \chi(\mu} \nabla^{\chi} R_{\nu)}{}^{\sigma} + 4
(\nabla_{\delta\sigma} \vartheta) {}^{*}\!R_{(\mu}{}^{\delta}{}_{\nu)}{}^{\sigma}\,,
\end{align}
\allowdisplaybreaks[0]%
\end{subequations}
where $\nabla_{\mu}$ is the covariant derivative, $\cd_{\mu\nu}\equiv\cd_\mu\cd_\nu$,
and $\square = \nabla_{\mu} \nabla^{\mu}$ is the d'Alembertian operator.  The $\vartheta$
field's stress-energy tensor is
\be
T_{\mu \nu}^{(\vartheta)} = \beta \left[(\nabla_{\mu}\vartheta) (\nabla_{\nu}\vartheta)
 - \frac{1}{2}g_{\mu \nu} \left(\nabla_{\delta}\vartheta \nabla^{\delta}\vartheta - 2
V(\vartheta) \right) \right]\,.
\label{theta-Tab}
\ee

Variation of the action with respect to $\vartheta$ yields the $\vartheta$ equation of motion:
\begin{align}
\beta \square \vartheta - \beta \frac{dV}{d\vartheta}
=&\, -\alpha_1 R^{2} - \alpha_2 R_{\mu \nu} R^{\mu \nu} \nonumber \\
&-\alpha_3 R_{\mu \nu \delta \sigma} R^{\mu \nu \delta \sigma} 
- \alpha_4 R_{\mu \nu \delta \sigma} \!{}^{*}R^{\mu \nu \delta \sigma}\,.
\label{EOM} 
\end{align}
Notice that when the spacetime is curved by some mass distribution, the
right-hand side will be proportional to density squared.

The parity of the field $\vartheta$ can be inferred from its equation
of motion. Since terms of the form $R^{2}$ are 
even-parity, while
terms of the form $R_{\mu \nu \delta \sigma} \!{}^{*}R^{\mu \nu \delta
  \sigma}$ are odd-parity, the field $\vartheta$ is of mixed
parity. Note however that the even and odd-parity couplings tend to have 
different origins from an underlying theory. In this paper we will
consider the even and odd-parity cases separately.

The inclusion of dynamics for the $\vartheta$ field
in the action guarantees that the field equations
are covariantly conserved without having to include any additional
constraints, 
i.e.~the covariant divergence of Eq.~\eqref{FEs} identically vanishes,
upon imposition of Eq.~\eqref{EOM}. This is a consequence
of the action being diffeomorphism invariant. 
Such invariance is in contrast to the preferred-frame effects present in
a non-dynamical theory~\cite{jackiw:2003:cmo}, i.e.~in the theory defined by the action in
Eq.~\eqref{action} but with $\beta = 0$.  In the latter, the field $\vartheta$ must be prescribed {\emph{a
priori}}. Moreover, the theory requires the existence of an
additional constraint (the right-hand side of \eqref{EOM} to vanish), 
which is an unphysical consequence of treating $\vartheta$ as prior
structure~\cite{Yunes:2007ss,Grumiller:2007rv}.

Before proceeding, let us further discuss the scalar field potential
$V(\vartheta)$.  This potential allows us to introduce additional
couplings, such as a mass term, to drive the evolution in
Eq.~\eqref{EOM}. However, there are reasons one might restrict such a
potential. If the mass is much larger than the inverse length scale of 
the system that we concern, the effect of such a field on the dynamics
of binaries is strongly suppressed. To the contrary, if the mass is
much smaller, the presence of mass does not give any significant
effects. Therefore we cannot expect to 
observe the effects of a finite mass without fine tuning.
No mass term may appear in a theory with a shift symmetry, which
is invariance under $\vartheta \to \vartheta +
{\rm{const}}$. Such theories are common in 4D, low-energy, effective string
theories~\cite{Boulware:1985wk,Green:1987mn,Green:1987sp,1992PhLB..285..199C,lrr-2004-5},
such as dynamical CS and EDGB.
For these reasons, and because the
assumption makes the resulting equations analytically tractable, we
will henceforth assume $V(\vartheta) = 0$.

\subsection{Small deformations}
\label{sec:small-deformation}

The ``unreasonable'' accuracy of GR to explain all experimental data to
date suggests that it is an excellent approximation to nature in
situations where the gravitational field is very weak and velocities are
very small relative to the speed of light. GW detectors will be
sensitive to events in situations where the field is stronger than ever
previously sampled. This, however, does not imply that GWs will ever
sample the Planck/string regime, where one
could expect large deviations from GR. 

We will here be interested in binary compact object coalescences up
until the binary reaches the innermost stable circular orbit
(ISCO). Even during merger, the largest curvature that GWs will
sample will be limited to the scale determined by the
horizon sizes, proportional to $m^{-2}$. Such scales are far removed from
high-energy ones, like the electroweak one, as GW detectors will not 
be sensitive to mergers of compact objects with masses below a solar 
mass. Even then, however, GWs can and will probe the {\emph{strong field}}, 
which has not been tested before. One is then justified in modeling GWs 
that may contain deviations from GR as {\emph{small deformations}}.

The small deformation scheme is also appealing for theoretical
reasons. As mentioned earlier, the theories we consider are
effective, valid only up to the truncation order. There are higher-order
terms that we have here neglected in the action, such as cubic and
quartic curvature combinations. Thus, one should not treat these
theories as exact nor   insist on solving the equations of motion
to higher orders in $\alpha_{i}$. If this is desired, then higher-order
curvature terms should also be included in the action.

One might be worried that such effective theories are unstable, since
they lead to field equations with derivatives higher than second
order. Such derivatives could lead to instabilities or ghost modes if
the Hamiltonian is not bounded from below. Linearization in the
coupling parameter, however, has the effect of recasting the field
equations in Einstein form with an effective stress-energy tensor that
depends on the GR solution, thus stabilizing the differential
equations~\cite{1994PhRvD..49.5188C}. Linearization removes modes
besides the two that arise in GR~\cite{Sopuerta:2009iy,Stein:2010pn}.

Small deformations can be treated similarly to how one models BH perturbations. 
That is, we expand the metric as
\be
g_{\mu \nu} = g_{\mu \nu}^{\GR} + \varsigma \, \hDef_{\mu \nu}
+O(\varsigma^2)\,,
\label{met-exp}
\ee
where the GR superscript is to remind us that this quantity is a GR
solution, while $\hDef_{\mu \nu}$ is a metric deformation away from
GR. The order-counting parameter $\varsigma$ is kept around only for
book-keeping purposes and is to be set to unity in the end. 

Applying such an expansion to Eq.~\eqref{EOM}, one finds
\be 
\beta {\square} {\vartheta}  = - \alpha_i \, \mathcal{S}({R}^{2}_{\GR}) + \mathcal{O}(\varsigma)\,,
\ee
where $\mathcal{S}(R^{2}_{\GR})$ stands for all source terms evaluated on the GR background $g_{\mu \nu}^{\GR}$. The solution to this equation will obviously scale as $\vartheta \propto \alpha_i/\beta$. Applying the decomposition and expansion of Eq.~\eqref{met-exp} to Eq.~\eqref{FEs} in vacuum, one finds
\be
G_{\mu \nu}[{\hDef}_{\mu \nu}] = - \frac{\alpha_i}{\kappa} 
C_{\mu \nu}[{\vartheta},{g}_{\mu \nu}^{\GR}] 
+ \frac{1}{2 \kappa}  T_{\mu \nu}^{(\vartheta)}[\vartheta]\,,
\ee
where the $\mathcal{O}(\varsigma^{0})$ terms automatically vanish, as
$g_{\mu \nu}^{\GR}$ satisfies the Einstein equations, and we have
grouped modifications into the tensor $C_{\mu \nu}$. This tensor
and $T_{\mu \nu}^{(\vartheta)}$ are to be evaluated on the GR metric and
act as sources for the metric deformation. Notice that, as a
differential operator acting on $\hDef_{\mu\nu}$, the principal part of
these differential equations continues to be strongly hyperbolic, as it
is still given by the $G_{\mu \nu}$ differential operator, with the
higher derivatives in $C_{\mu \nu}$ and the $T_{\mu \nu}^{(\vartheta)}$
acting as sources. Given this, the metric deformation is proportional to
$\xi_{i} \equiv \alpha_i^{2}/(\beta \kappa)$,
which is our actual perturbation parameter.  

Proper perturbation or deformation parameters should be dimensionless,
but  
the $\xi_i$ are dimensional. The dimensions of $\alpha$ and $\beta$, of course,
depend on the choice of dimensions for the scalar field. We here take the
viewpoint that $\vartheta$ is dimensionless, which then forces 
$\beta$ to be dimensionless as well as $\kappa$, 
and $\alpha$ to have dimensions of
length squared. Then, the deformation parameter $\xi$ has units of
length to the fourth power, which is why we define the dimensionless
\be
\zeta_{i} \equiv \xi_{i}/m^{4} = {\cal{O}}(\varsigma)\,,
\ee
as our proper deformation
parameter. One could choose different units
for the scalar field, but in all cases one arrives at the conclusion
that $\zeta_{i}$ is the proper deformation parameter~\cite{Yunes:2009hc}.

\section{Expansion of the field equations}
\label{sec:Expansions}

Let us decompose the GR metric tensor into a flat background plus a metric perturbation:
\be
\label{metric-exp}
g_{\mu \nu}^{\GR} = \eta_{\mu \nu} + \hGR_{\mu \nu}\,.
\ee
We emphasize here that throughout this paper, $\hGR_{\mu \nu}$ denotes the metric perturbation in GR while $\hDef_{\mu\nu}$ is the metric deformation away from GR. 

In expanding the modified field equations, we will also find it useful to define the standard trace-reversed metric perturbation in GR as
\be
\bar{\hGR}^{\mu\nu} \equiv \eta^{\mu\nu}-\sqrt{-g_{\GR}}g^{\mu\nu}_{\GR}\,.
\ee
In particular, notice that when the background is flat ${\bar{\hGR}}_{\mu\nu}={\hGR}_{\mu\nu}-\frac{1}{2}{\hGR} \eta_{\mu\nu}$ and ${\hGR}_{\mu\nu}={\bar{\hGR}}_{\mu\nu}-\frac{1}{2}{\bar{\hGR}} \eta_{\mu\nu}$ to linear order in GR. We also define the deformed trace reversed metric perturbation as
\be
\bar{\hDef}^{\mu\nu} \equiv \left( \eta^{\mu\nu}-\sqrt{-g}g^{\mu\nu}\right) - \bar{\hGR}^{\mu\nu}\,.
\ee
The harmonic gauge condition reduces to $\bar{\hGR}^{\mu\nu}{}_{,\nu}=0$
and $\bar{\hDef}^{\mu\nu}{}_{,\nu}=0$.  Throughout this paper, we only
study the GR deformation up to $
\mathcal{O}(\alpha_i/\beta)$ for $\vartheta$ 
and $\mathcal{O}(\zeta_i)$ for $\hDef_{\mu\nu}$.

\subsection{Scalar field}

The evolution equation for the scalar field at leading order in the metric perturbation becomes
\ba
\square_{\eta} \vartheta & = & 
- \frac{\alpha_1}{\beta} \left( \frac{1}{2\kappa} \right)^{2} T_{\mathrm{mat}}^2
- \frac{\alpha_2}{\beta} \left( \frac{1}{2\kappa} \right)^{2} T_{\mathrm{mat}}^{\mu\nu} T^{\mathrm{mat}}_{\mu\nu}
\notag \\
& & -  \frac{2\alpha_3}{\beta} (\hGR_{\alpha \beta ,\mu \nu} \hGR^{\alpha [\beta ,\mu] \nu} + \hGR_{\alpha \beta ,\mu \nu} \hGR^{\mu [\nu ,\alpha] \beta} ) \notag \\
 & &-  \frac{2 \alpha_4}{\beta} \epsilon^{\alpha \beta \mu \nu} \hGR_{\alpha \delta,\gamma \beta} \hGR_{\nu}{}^{[\gamma,\delta]}{}_{\mu}\,
\label{2nd-order-flat-pre}
\ea
with relative remainders of ${\mathcal{O}}(h)$. Here,
$\epsilon^{\mu \nu \delta \sigma}$ is the Levi-Civita symbol with
convention $\epsilon^{0123} = +1$ in an orthonormal, positively oriented frame, and we have used the harmonic gauge
condition.

\subsection{Metric perturbation}

Let us now perturb the metric field equations [Eq.~\eqref{FEs}] about
$\varsigma=0$. 
The deformed metric wave equation at linear order in $\hDef_{\mu\nu}$ becomes
\begin{align}
\frac{\kappa}{2} \square_{\eta} \hDef_{\mu\nu} =&\, 
\alpha_1 \vartheta \tilde{\mathcal{H}}_{\mu \nu}^{(0)} +
\alpha_2 \vartheta \tilde{\mathcal{I}}_{\mu \nu}^{(0)} 
+ \alpha_3 \vartheta \tilde{\mathcal{J}}_{\mu \nu}^{(0)}  
\nonumber \\
&+ \alpha_1
\tilde{\mathcal{H}}_{\mu \nu}^{(1)} + 
\alpha_2 \tilde{\mathcal{I}}_{\mu \nu}^{(1)} 
+ \alpha_3  \tilde{\mathcal{J}}_{\mu \nu}^{(1)} +
\alpha_4 \tilde{\mathcal{K}}_{\mu \nu}^{(1)} \nonumber \\
& -\frac{1}{2} \delta T_{\mu\nu}^{\MAT} -\frac{1}{2} T_{\mu\nu}^{(\vartheta)} \,,
\label{wave-eq-h-pert}
\end{align}
where the tensors on the right-hand side are given by
\allowdisplaybreaks[1]
\ba
\tilde{\mathcal{H}}_{\mu \nu}^{(0)} &=& -4\left( \hGR{}_{\rho} {}^{[ \sigma ,}{}^{\rho ]}{}_{\sigma \mu\nu} 
- \eta_{\mu\nu} \square_{\eta} \hGR{}_{\rho}{}^{[\sigma,} {}^{\rho]}{}_{\sigma}\right)\,,  
\\
\tilde{\mathcal{I}}_{\mu \nu}^{(0)}  &=& \square_{\eta}\hGR{}_{\nu [\rho,}^{} {}_{\mu]}{}^{\rho} - \square_{\eta}\hGR {}^{\rho}{}_{[\rho,}{}_{\mu]\nu} - 2 \hGR{}_{\rho} {}^{[ \sigma ,}{}^{\rho ]}{}_{\sigma \mu\nu} \nonumber \\ 
& & + \eta_{\mu\nu} \square_{\eta}\hGR{}_{\rho}{}^{[\sigma,} {}^{\rho]}{}_{\sigma}\,,  
\\
\tilde{\mathcal{J}}_{\mu \nu}^{(0)} & =& 4 \left(- \square_{\eta}\hGR{}_{\nu [\mu}^{} {}_{,\rho]}{}^{\rho} - \square_{\eta}\hGR {}^{\rho}{}_{[\rho,}{}_{\mu]\nu} -  \hGR{}_{\rho} {}^{[ \sigma ,}{}^{\rho ]}{}_{\sigma \mu\nu}  \right)\,, 
\\
\tilde{\mathcal{H}}_{\mu \nu}^{(1)} &=& -8 \hGR{}_{\rho} {}^{[ \sigma ,}{}^{\rho ]}{}_{\sigma (\mu} \vartheta_{,\nu) } -4 \hGR{}_{\rho} {}^{[ \sigma ,}{}^{\rho ]}{}_{\sigma } \vartheta_{,\mu\nu} \nn \\
& & + 4\eta_{\mu\nu} \left( 2 \hGR{}_{\rho} {}^{[ \sigma ,}{}^{\rho ]}{}_{\sigma \delta} \vartheta {}^{,\delta } + \hGR{}_{\rho} {}^{[ \sigma ,}{}^{\rho ]}{}_{\sigma } \square_{\eta}\vartheta \right)\,, 
\\
\tilde{\mathcal{I}}_{\mu \nu}^{(1)} &=& -2 \hGR{}_{\rho} {}^{[ \sigma ,}{}^{\rho ]}{}_{\sigma (\mu} \vartheta _{,\nu) } -2 \left( \hGR{}^{\delta}{}_{[\rho,} {}_{(\nu]\mu)}{}^{\rho} - \hGR{}^{\rho}{}_{[\rho,} {}_{(\nu]\mu)}{}^{\delta}   \right) \vartheta _{,\delta} \nn \\
& & - 2\left(\hGR{}_{(\nu [\mu),}^{} {}_{\rho]\delta}{}^{\rho} + \hGR {}^{\rho}{}_{[\rho,}{}_{(\mu]\nu)\delta} \right) \vartheta {}^{,\delta} \nonumber \\*
&&-2 \left( \hGR{}^{\delta}{}_{[\rho,} {}_{(\mu]}{}^{\rho} \vartheta_{,\nu) \delta} - \hGR{}^{\rho}{}_{[\rho,} {}_{(\mu]}{}^{\delta} \vartheta_{,\nu) \delta} \right) \nn \\
& &  + \eta_{\mu\nu} \left\{ 2 \hGR{}_{\rho} {}^{[ \sigma ,}{}^{\rho ]}{}_{\sigma \delta} \vartheta {}^{,\delta}+\left( \hGR{}^{\sigma} {}^{[ \rho ,}{}^{\delta ]}{}_{\rho} - \hGR{}_{\rho} {}^{[ \rho ,}{}^{\delta ]\sigma}  \right)  \vartheta _{,\sigma\delta}   \right\} \nonumber \\*
&&+ \square_{\eta}\vartheta \left( \hGR{}_{(\mu}{}^{\delta}{}_{,\nu)\delta} - \frac{1}{2} \square_{\eta}\hGR_{\mu \nu} - \frac{1}{2} \hGR_{,\mu \nu} \right)\,, 
\label{tilde-tilde-I} 
\\
\tilde{\mathcal{J}}_{\mu\nu}^{(1)} &=& -8 \left( \hGR{}^{\delta}{}_{[\rho,} {}_{(\nu]\mu)}{}^{\rho} + \hGR{}^{\rho}{}_{[\rho,} {}_{(\nu]\mu)}{}^{\delta} -\frac{1}{2} \hGR_{,\mu \nu}{}^{\delta} \right. \nn \\
& & \left. + \frac{1}{2} \square_{\eta}\hGR{}_{\mu \nu}{}^{,\delta}  \right) \vartheta _{,\delta} + 4 \left(\hGR_{\sigma [\mu,\delta] \nu} - \hGR_{\nu [\mu,\delta] \sigma}\right) \vartheta{}^{,\sigma\delta}\,,  \nn \\
\label{tilde-tilde-J} 
\\
\tilde{\mathcal{K}}_{\mu\nu}^{(1)} &=& \vartheta^{,\delta}{}_{,\sigma} \eta_{\nu\alpha}\bar{\varepsilon}^{\alpha\sigma\beta\gamma}
\left(\hGR{}_{\mu [\gamma,\beta] \delta}+\hGR{}_{\delta [\beta,\gamma] \mu}\right) \nn \\
& & - 2 \vartheta{}^{,\delta} \epsilon_{\delta\sigma\chi\mu} \hGR{}^{\sigma}{}_{[\alpha}{}^{,\alpha\chi}{}_{\nu]} +(\mu\leftrightarrow \nu)\,,
\label{tilde-tilde-K}
\ea
\allowdisplaybreaks[0]
%
where $\square_{\eta}$ is the d'Alembertian of flat spacetime, $\hGR = \hGR_{\mu}{}^{\mu}$, and $T_{\mu\nu}^{(\vartheta)}$ is given as
\be
T_{\mu\nu}^{(\vartheta)} = \beta\left( \vartheta_{,\mu} \vartheta_{,\nu} - \frac{1}{2} \eta_{\mu\nu} \vartheta_{,\delta} \vartheta{}^{,\delta} \right)\,.
\ee
The quantity $\delta T_{\mu\nu}^{\MAT}$ stands for the perturbation to
the energy-momentum tensor for matter. Even when dealing with BHs,
$\delta T_{\mu \nu}^{\MAT} \neq 0$ because we treat BHs as
distributional point particles and their trajectories are generically
modified at ${\mathcal{O}}(\varsigma)$. However, in this paper we
concentrate on the dissipative sector of the theory only, and not on
modifications to the shape of the orbits (conservative dynamics). The
latter does modify the GW phase evolution~\cite{Yunes:2011we,Yunes:2009hc}, 
as we discuss in Sec.~\ref{sec:Discussions}.

The evolution equation for the metric perturbation takes on the same
form (a sourced wave equation) as that for the scalar field. The source
terms in both of these equations depend on the GR metric perturbation,
which we here assume to be that of a compact binary quasi-circular inspiral 
in the PN approximation, i.e.~moving at small velocities relative
to the speed of light and producing weak gravitational fields. We
provide explicit expressions for the GR metric perturbation in the
subsequent subsection. 

\subsection{Post-Newtonian metric and trajectories}
\label{PN-metric-and-Trajectories}

In this subsection, we provide explicit expressions for the linear
metric perturbation in GR
that we use to evaluate all source terms. We are here interested in a binary system, composed of two compact objects with masses $m_{1}$ and $m_{2}$ and initially separated by a distance $r_{12} \equiv b$. The objects' trajectories can be parameterized via
\ba
\mathbf{x}_{1} \equiv x_{1}^{i} &\!\! =&\!\!
 +\frac{m_{2}}{m} b \left[ \cos{\omega t}, \sin{\omega t}, 0 \right]\,,
\\
\mathbf{x}_{2} \equiv x_{2}^{i} &\!\! =&\!\!
 -\frac{m_{1}}{m} b \left[ \cos{\omega t}, \sin{\omega t}, 0 \right]\,,
\ea
where $m \equiv m_{1} + m_{2}$ is the total mass and where we have
assumed they are located on the $x$--$y$ plane. Throughout this paper,
vectors are sometimes denoted with a boldface. We also
define 
\begin{align}
\mathbf{x}_{12} &\equiv x_{12}^{i} = x_{1}^{i} - x_{2}^{i},\\ 
\mathbf{n}_{12} &\equiv n_{12}^i = (x_{1}^{i} - x_{2}^{i})/b, \\
\mathbf{n}_{A} &\equiv n_{A}^{i} = (x^{i} -x_{A}^{i})/r_{A},
\end{align}
where we follow the conventions of~\cite{Blanchet:2002av},
with 
\be
r_{A} \equiv |x^{i} - x_{A}^{i}|.
\ee
We further
assume these objects are on a quasi-circular orbit with leading-order
angular velocity $\omega = (1/b) (m/b)^{1/2}$ and orbital velocity $v =
(m/b)^{1/2}$. The orbital separation $b$ is assumed constant, as its
time-evolution is driven by GW emission at high-order in $v/c$.

The GR spacetime metric for such a binary is expanded as in Eq.~\eqref{metric-exp}. In the near zone, the metric perturbation is given by
\ba
\label{NZmetric11}
\hGR_{00} &\!\! 
=&\!\!   2 U_1 \plusonetotwo + {\mathcal{O}}(v^{4}) \,,
\\
\hGR_{0i} &\!\! =&\!\!  -4 V_{1i} \plusonetotwo + {\mathcal{O}}(v^{5}) \,,
\\
\hGR_{ij} &\!\! =&\!\!  2 U_1  \delta_{ij} \plusonetotwo + {\mathcal{O}}(v^4) \,,
\label{NZmetric31}
\ea
where ${\mathcal{O}}(v^{A})$ stands for an $(A/2)$PN remainder, i.e.~a term of 
${\cal{O}}((v/c)^{A})$, and the notation $\plusonetotwo$ means that one should add the same terms 
with the labels $1$ and $2$
interchanged. The potentials $U_A$ and $V_{Ai}$ with $A=(1,2)$ are defined as
\be
U_A  =\!\!  \int \frac{\rho'_A}{|\bm{x}-\bm{x}'|} d^3x'\,,  \qquad
V_{Ai}  =\!\!  \int \frac{\rho'_A v'_{Ai}}{|\bm{x}-\bm{x}'|} d^3x'\,,
\ee
where $\rho_A$ and $v_{A}^{i} \equiv  \dot{x}_{A}^{i}$ are the 
density and the center of mass velocities of the respective objects, 
with the overhead dot standing for time differentiation. 
Field variables associated with a prime, e.g.~$\rho'_A$, 
are to be evaluated at $\bm{x}'$.
In the point-particle limit, the metric becomes
\ba
\label{NZmetric1}
\hGR_{00} &\!\! = &\!\!   \frac{2 m_{1}}{r_{1}} \plusonetotwo\,,
\\
\hGR_{0i} &\!\! = &\!\!  - \frac{4 m_{1}}{r_{1}} v_{1}^{i} \plusonetotwo\,,
\\
\hGR_{ij} &\!\! = &\!\!  \frac{2 m_{1}}{r_{1}}  \delta_{ij} \plusonetotwo\,,
\label{NZmetric3}
\ea
with remainders of relative $\mathcal{O}(v^{2})$.
We have kept the PN leading terms in the
metric that are proportional to $m_{A}$ only, but higher-order terms
can be found in~\cite{Blanchet:1998vx}, while terms proportional to the
spin of each BH can be found in~\cite{Tagoshi:2000zg}. 

\section{Scalar field evolution}
\label{sec:SF-Evolution}

In this section, we solve the evolution equation for the scalar field
both for field points in the far and near-zones, as defined in Sec.~\ref{subsec:Zones}. 
The former will allow us to evaluate the energy flux carried by the scalar field at 
infinity, while the latter will be essential to find effective source terms that reproduce the known strong field solutions and to solve the evolution equations for the metric deformation.

\subsection{Zones}
\label{subsec:Zones}

\begin{figure}
 \centerline{\includegraphics[width=6cm,clip=true]{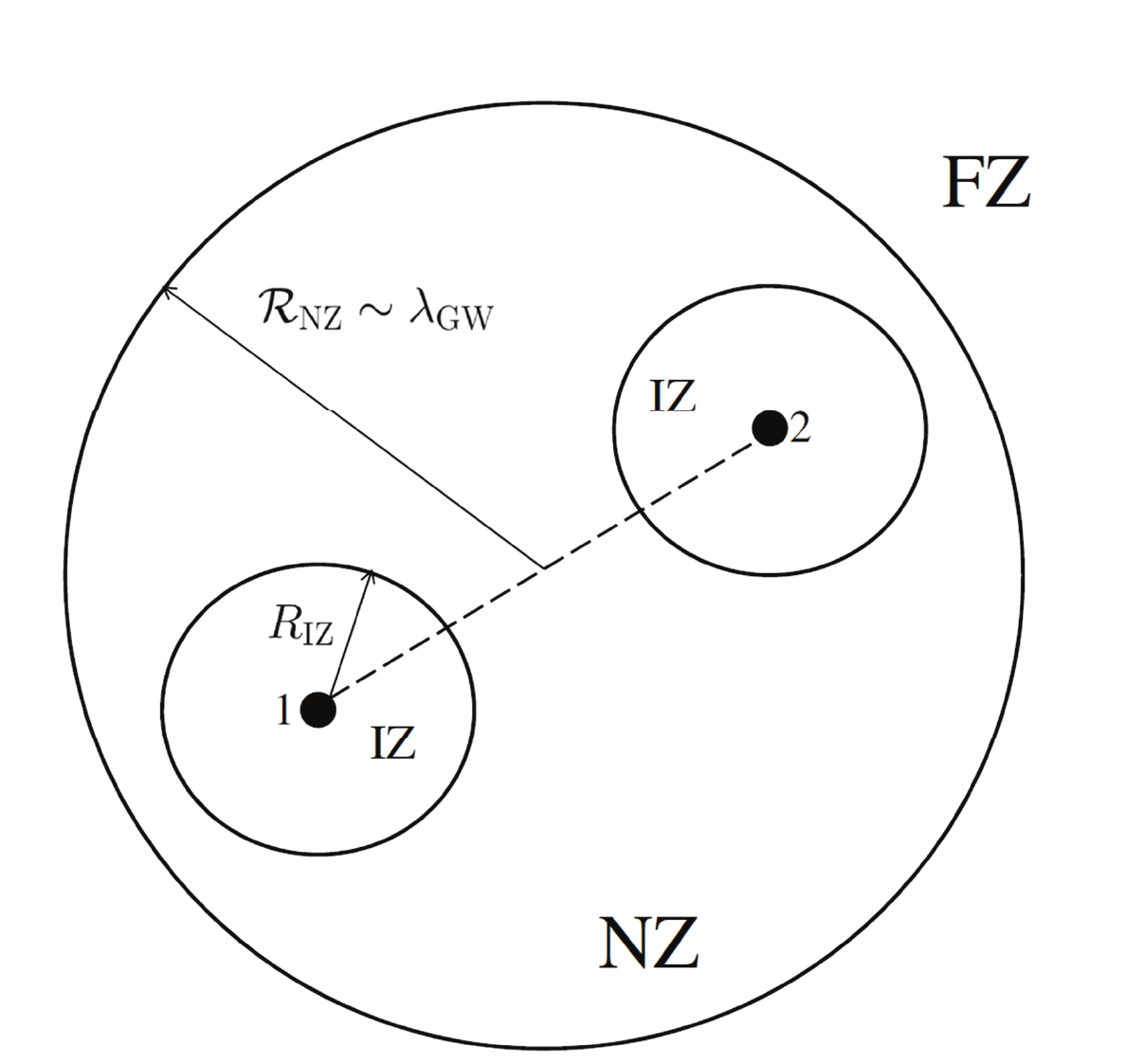}}
 \caption{\label{fig:zone} We consider three zones, inner zone (IZ), near zone (NZ) and far zone (FZ). The IZs are centered at each object and their radii $\Rbn$ satisfy $\Rbn \ll b$. The NZ is centered at the center of mass of the two bodies and the radius $\Rnf$ satisfies $\Rnf \sim \lambda_{\GW}$, where $\lambda_{\GW}$ is the GW wavelength.} 
\end{figure}

As shown in Fig.~\ref{fig:zone}, let us decompose the geometry into three
regions: an inner zone (IZ), a near zone (NZ) and a far zone
(FZ); see e.g.~\cite{Alvi:1999cw, Yunes:2005nn,JohnsonMcDaniel:2009dq} 
for further details. The IZs are centered at each object
with radii $\Rbn$. These radii are defined as the boundary inside which
either $T_{\mu\nu}^{\MAT}\ne 0$ or the usual PN approximation breaks
down due to strong-gravity effects. We here take them to be sufficiently larger than $m_A$ and much less than $b$.   The NZ is centered at the binary's center of mass with radius $\Rnf$ and excluding the IZs. This radius is defined as the boundary outside which time-derivatives cannot be assumed to be small compared with spatial derivatives due to the wave-like nature of the metric perturbation. We here take this boundary to be roughly equal to $\lambda_{\mathrm{GW}}$, where $\lambda_{\mathrm{GW}}$ denotes the GW wavelength. The FZ is also centered at the binary's center of mass, but it extends outside $\Rnf$.

One can only apply the PN formalism when the gravitational field is weak
and velocities are small. When we deal with strong field sources like
BHs and NSs, therefore, one can use the PN scheme in the NZ and FZ
only. In the IZs, one may not be able to use PN theory, since the
gravitational field may be too strong. In this case, we have to
asymptotically match our PN solution in the NZ with the strong field
solutions valid in the IZs, inside some buffer regions that overlap both NZ
and each IZ (see Refs.~\cite{D'Eath:1975qs,
Thorne:1984mz, Damour:1983} for a description of how to carry this out
in GR). The strong field solution for BHs was 
found in Refs.~\cite{Yunes:2011we} and~\cite{Yunes:2009hc} in the class
of theories considered here.

%

\subsection{Near zone solutions}
\label{sec:cheating}

Since the NZ is in the weak field regime, we can apply the PN formalism to compact binary systems. Let us consider the even and odd-parity sectors separately. 

\subsubsection{Even-parity sector}

The evolution equation for the even-parity sector is 
\begin{align}
\square_{\eta} \vartheta  = &
- 64\pi^2 \frac{\alpha_1}{\beta} \rho^2 - 64\pi^2 \frac{\alpha_2}{\beta} \rho^2 
\notag \nonumber\\
& -  \frac{2\alpha_3}{\beta} \left(\hGR_{\alpha \beta ,\mu \nu} h^{\alpha [\beta ,\mu] \nu} + \hGR_{\alpha \beta ,\mu \nu} h^{\mu [\nu ,\alpha] \beta} \right)\,,
\label{even-scalar-eq}
\end{align}
with $\rho\equiv \rho_1 + \rho_2$ and remainders of ${\mathcal{O}}(h^{3})$. 

First, let us consider weakly-gravitating objects, i.e.~not BHs or NSs,  
in which case the PN expansion is valid also in the IZ. 
By substituting the GR PN metric of 
Eqs.~\eqref{NZmetric11}-\eqref{NZmetric31}, 
the NZ solution to the above wave equation at leading
PN order becomes 
\ba
\vartheta &\! = &\!  16\pi \frac{\alpha_1}{\beta} \int_{\mathcal{M}} \rho'{}^2 \frac{d^3x'}{|\bm{x}-\bm{x}'|} + 16\pi \frac{\alpha_2}{\beta} \int_{\mathcal{M}} \rho'{}^2 \frac{d^3x'}{|\bm{x}-\bm{x}'|} \nonumber \\
&&\!\! + \frac{1}{\pi}\frac{\alpha_3}{\beta} \int_{\mathcal{M}} \left( 2 U'_{,ij} U'_{,ij} + \square_{\eta} U' \square_{\eta} U' \right) \frac{d^3x'}{|\bm{x}-\bm{x}'|}\,, 
\label{theta-sol-gen}
\ea
again with remainders of ${\mathcal{O}}(h^{3})$, with $U \equiv U_1 + U_2$
and $\mathcal{M}$ denoting the constant-time, NZ+IZ hypersurface. 
We can safely neglect the contribution from the FZ, 
since the fall-off of the source term is sufficiently fast. 

The solution in Eq.~\eqref{theta-sol-gen} can be simplified by integrating by parts several times and using that $\square U = -4\pi \rho$ and $ \square {|\bm{x}-\bm{x}'|}^{-1} = -4\pi \delta^{(3)} (\bm{x}-\bm{x}')$ to obtain 
\ba
\vartheta &\! = &\!  16\pi \frac{\alpha_1}{\beta} \int_{\mathcal{M}}
\rho'{}^2 \frac{d^3x'}{|\bm{x}-\bm{x}'|} + 16\pi \frac{\alpha_2}{\beta}
\int_{\mathcal{M}} \rho'{}^2 \frac{d^3x'}{|\bm{x}-\bm{x}'|} \nonumber \\
&&\!\! + 
48 \pi \frac{\alpha_3}{\beta} \int_{\mathcal{M}} \rho'^2  \frac{d^3x'}{|\bm{x}-\bm{x}'|}  
\nonumber \\ &&\!\!
- 8 \frac{\alpha_3}{\beta} \int_{\mathcal{M}} \rho' U'_{,i} 
\left( \frac{1}{|\bm{x}-\bm{x}'|} \right)_{\!\! ,i}   d^3x' 
\nonumber \\&&\!\!
-  4 \frac{\alpha_3}{\beta} \int_{\mathcal{M}}  U'_{,i} U'_{,i} 
\delta^{(3)} (\bm{x}-\bm{x}')  d^3x'\,.
\label{theta-sol-gen-2}
\ea
Expanding this solution in terms of particles $1$ and $2$, we 
arrive at
\be
\vartheta = \vartheta_{\self} + \vartheta_{\cross}\,,
\ee
with 
\ba
\vartheta_{\self} &\!\! = &\!\! 
{16\pi\over \beta} \left(\alpha_1 + \alpha_2 + 3 \alpha_3 \right) 
\int_{\mathcal{M}} \rho'{}_{1}^2 \frac{d^3x'}{|\bm{x}-\bm{x}'|} 
\nonumber \\
&&\!\! - 8 \frac{\alpha_3}{\beta} 
\int_{\mathcal{M}} 
\rho'_{1} U'_{1,i}    \left( \frac{1}{|\bm{x}-\bm{x}'|} \right)_{\!\! ,i}
d^3x' 
\nonumber \\
&&\!\! - 4 \frac{\alpha_3}{\beta} U_{1,i} U_{1,i} 
\plusonetotwo \,, \qquad
\label{theta-self}
\ea
and
\ba
\vartheta_{\cross} &= &
-8 \frac{\alpha_3}{\beta} \Biggl[
\int_{\mathcal{M}} 
\left(
\rho'_{1} U'_{2,i}+ \rho'_{2} U'_{1,i}
\right)  \left( \frac{1}{|\bm{x}-\bm{x}'|} \right)_{\!\! ,i}   d^3x' 
\nonumber \\
&&\qquad + U_{1,i} U_{2,i} \Biggr] \,.
\label{theta-cross}
\ea
$\vartheta_{\self}$ is the part of $\vartheta$ that  
can be evaluated by considering a single object only, 
while $\vartheta_{\cross}$ is the part that depends on 
the fields of both bodies.  

The integrals that define both $\vartheta_{\self}$ and $\vartheta_{\cross}$ 
have  support in the IZs only, and thus, the NZ integral operator 
is homogeneous (source-free). 
When we discuss the NZ behavior of fields associated 
with compact objects, such as BHs or NSs, we cannot directly evaluate 
such IZ integrals. These are derived under the assumption 
that the PN expansion is valid everywhere, 
which fails for compact objects in the
IZs. Instead, we need to determine these 
homogeneous solutions through asymptotic matching. 
Before doing so, it is helpful to study the 
meaning of each term for weakly-gravitating objects. 

Neglecting the size of the weakly-gravitating objects, 
the first term in Eq.~(\ref{theta-self}) in the NZ is 
evaluated as 
\be
\int_{\mathcal{M}} \rho'_1\!{}^2 \frac{d^3x'}{|\bm{x}-\bm{x}'|}
\approx \frac{1}{r_1} \int_{\mathcal{M}} \rho'_1\!{}^2 d^3x'\,,
\label{int-exp-1}
\ee
with remainders of relative ${\mathcal{O}}(m/r)$,
while the second term becomes 
\ba
\int_{\mathcal{M}} &\!\! 
\rho'_1& \!\! 
U'_{1,i} \left( \frac{1}{|\bm{x}-\bm{x}'|} \right)_{\!\! ,i}   d^3x' 
\approx \frac{n_1^i}{r_1^2} \int_{\mathcal{M}} \rho'_1 U'_{1,i} d^3x' \nonumber \\
&=& \!\! 
- \frac{n_1^i}{r_1^2} \int_{\mathcal{M}} \rho_1 (\bm{x}') \left(
\int_{\mathcal{M}} \rho_1 (\bm{y}) \frac{x'{}^i-y^i}{|\bm{x}'-\bm{y}|^2}
d^3y \right) d^3x' \cr
&=& \!\! 0\,. \qquad
\label{int-exp-2}
\ea
The last equality can be shown 
by exchanging the integration variables\footnote{In
fact, this integral vanishes to all orders in $x$. This is because
$(\rho_{1} U_{1,i})_{,i}$ is spherically symmetric, and thus, when it
acts as a source to a wave equation, the solution should either scale as
$1/r$ or it should vanish identically. We have here shown that there is
no $1/r$ part.}. 
Thus, 
one can approximate 
$\vartheta_{\self}$ as 
\be
\vartheta_{\self} = {q_1\over r_1}- 
4 \frac{\alpha_3 m_1^2}{\beta r_1^4} \plusonetotwo\,,
\label{final-theta-self}
\ee
with the scalar monopole charge defined by 
\be
q_A\equiv 
{16\pi\over \beta} \left(\alpha_1 + \alpha_2 + 3 \alpha_3 \right) 
\int_{\IZ} \rho'_{A}\!\!{}^2 d^3x'\,,
\ee
with $A=(1,2)$. Here we put ``IZ'' to the integral to emphasize that  
the integration can be restricted to both IZs because the integrand is localized.

The first term in Eq.~(\ref{final-theta-self}) 
represents the monopole field around object 1. 
These monopole fields give the leading PN contribution in the NZ 
unless both monopole charges $q_1$ and $q_2$ vanish. 
This is indeed the case in EDGB theory,  where
$(\alpha_1,\alpha_2,\alpha_3)=(1,-4,1)\alpha_{\mathrm{EDGB}}$. 
We will later show that this cancellation does really survive 
even if we consider NSs. 
If this cancellation occurs, the higher order 
terms of ${\mathcal{O}}(m^{2}/r^{2})$ in
the expansion of Eq.~\eqref{int-exp-1} become the dominant
contribution to $\vartheta$. 
The second term in Eq.~(\ref{final-theta-self}) is 
much higher PN order compared with the first term and 
hence sub-dominant in the NZ. 

Let us now consider $\vartheta_{\self}$ for compact objects, 
where the IZ integrals must be treated carefully.
Since the PN expansion is no longer valid in 
the IZ, one cannot use the simple extrapolation of the above result. 
In Sec.~\ref{sec:matching}, we match the NZ solution to the
one obtained for isolated BHs in the 
strong-field~\cite{Yunes:2011we,Yunes:2009hc}. 
We will not discuss the matching for NSs in this 
paper, but the order of magnitude estimate 
\be
q_A= \sum_{i=1}^{3} q_{i,A} = \sum_{i=1}^{3} {\alpha_i \over \beta}{\mathcal{O}}\left({m_A^2\over R_A^3}\right)
\ee
should still be valid, where $R_A$ is the radius of the $A$th NS. 
When $\alpha_1 + \alpha_2 + 3 \alpha_3 =0$, 
the cancellation observed in the weakly gravitating objects 
may still persist even for NSs. However, the cancellation 
will not in general be exact, except for the EDGB subcase.
In EDGB theory, the NS scalar monopole charge vanishes 
independently of the equation of state. Mathematically speaking, this is
because the monopole charge is given by the integral of 
the Gauss-Bonnet invariant $R_{\mathrm{GB}} \equiv R^2 - 4 R_{\mu\nu}
R^{\mu\nu} + R_{\mu\nu\rho\sigma}R^{\mu\nu\rho\sigma}$, which vanishes 
for any simply-connected, asymptotically flat geometry. 
A more explicit proof is given in Appendix~\ref{sec:balding}.

Let us now return to the $\vartheta_\cross$ contribution and consider first 
weakly-gravitating objects. To evaluate Eq.~(\ref{theta-cross}),
one can use point-particle expansions of the potentials and the density,
i.e.~$\rho_{A} = m_A \delta^{(3)} (\bm{x}-\bm{x}_A) $ and $U_A =
m_A/r_A$. Simple substitution leads to 
\ba
\vartheta_{\cross} 
&\!\! \approx &\!\! 
 8 \frac{\alpha_3 m_1m_2}{\beta m^4}\left[m^4 
  \left(
 \frac{n_{1}^j}{r_1^2} \frac{n_{2}^j}{r_2^2} +
\frac{n_{12}^j}{b^2} \frac{n_{2}^j}{r_2^2}
\right. \right. 
\nn \\
&& - \left. \left. 
\frac{n_{12}^j}{b^2}  \frac{n_{1}^j}{r_1^2} \right) +
{\mathcal{O}}\left(\frac{m^{5}}{r^{5}} \right)\right]\,. 
\label{theta-cross-final}
\ea
The first term in parentheses comes from the term 
$U_{1,i} U_{2,i}$ in Eq.~\eqref{theta-cross}. The remaining two terms 
come from the integral in Eq.~(\ref{theta-cross}). 
The second and third terms in parentheses look
like scalar dipole moments for bodies 2 and 1 respectively.
However, a Taylor expansion about the center of mass of each body, 
shows that the $1/r_{A}^{2}$ piece of $\vartheta_{\cross}$ cancels, 
which implies that there is no scalar dipole. 

Let us now consider $\vartheta_{\cross}$ for compact objects. 
As discussed in the previous paragraph, one might expect a scalar dipole charge 
induced by the acceleration of object 1 due to  the gravitational field 
of object 2 ($\propto U_{2,i}(\mb{x}_1)$). In GR, however, acceleration 
is understood as geodesic motion in a perturbed geometry.
The deviation of the local geometry from the unperturbed 
isolated geometry originates due to tides, and this is a relative 
$4$PN effect. This is much smaller than the scalar monopole charge contribution
from $\vartheta_{\self}$. 

To summarize, the dominant contribution to $\vartheta$ comes from the 
monopole charge associated with each object, which depends on its
internal structure.

\subsubsection{Odd-parity sector}

In the odd-parity case, the scalar
field evolution equation is
\be
\square_{\eta} \vartheta = - \frac{2 \alpha_4}{\beta} \epsilon^{\alpha \beta \mu \nu} \hGR_{\alpha \delta,\gamma \beta} \hGR_{\nu}{}^{[\gamma,\delta]}{}_{\mu}\,,
\label{scalar-odd-eq}
\ee
plus terms of ${\mathcal{O}}(h^{3})$.  
Again, we first consider weakly gravitating objects. 
At leading PN order, the above equation becomes
\ba
\square_{\eta} \vartheta &\!\!  = &\!\!  \frac{2 \alpha_4}{\beta} \epsilon_{ijk} (
\hGR_{00,mi} \hGR_{k0,jm} + \hGR_{0l,jm} \hGR_{kl,im})\, \nonumber \\
&=& -32 \frac{\alpha_4}{\beta} \epsilon_{ijk} U_{,im} V_{k,jm}\,,
\label{scalar-wave-eq-CS}
\ea
with remainders of relative ${\mathcal{O}}(v^2)$.  
As in the even-parity case, we write the solution to this wave equation as
\be
\vartheta = \vartheta_{\self} + \vartheta_{\cross}\,,
\ee
where
\be
\vartheta_{\self} = \frac{8}{\pi} \frac{\alpha_4}{\beta} \epsilon_{ijk} \int_{\mathcal{M}} U_{1,im}' V_{1k,jm}' \frac{d^3x'}{|\bm{x}-\bm{x}'|} \plusonetotwo\,,
\label{odd-self}
\ee
and
\be
\vartheta_{\cross} = \frac{8}{\pi} \frac{\alpha_4}{\beta} \epsilon_{ijk} \int_{\mathcal{M}} U_{1,im}' V_{2k,jm}' \frac{d^3x'}{|\bm{x}-\bm{x}'|} \plusonetotwo \,.
\label{odd-cross}
\ee

Let us first consider self-interaction terms $\vartheta_{\self}$. 
Integrating by parts several times, we find
\ba
\!\!\!\! \vartheta_{\self} &\!\! =&\!\! -16 \frac{\alpha_4}{\beta} \epsilon_{ijk} 
\bigg[ \int_{\mathcal{M}} \rho'_1 V_{1k,j}' \left(
\frac{1}{|\bm{x}-\bm{x}'|} \right)_{\!\! ,i} d^3x' 
\nonumber \\
&&\!\! +  
\int_{\mathcal{M}} U'_{1,i} \rho_1' v_{1k}' \left(
\frac{1}{|\bm{x}-\bm{x}'|} \right)_{\!\! ,j} d^3x' \nonumber \\
&&\!\! + \int_{\mathcal{M}} U'_{1,i} V_{1k,j}' \delta^{(3)} (\bm{x}-\bm{x}') d^3x' \plusonetotwo \bigg]\,,
\ea
where we have used the relations $\square U_1 = -4\pi \rho_1$, $\square
V_1^{k} = -4\pi \rho_1 v_1^{k}$ and $\square {|\bm{x}-\bm{x}'|}^{-1} =
-4\pi \delta^{(3)} (\bm{x}-\bm{x}')$. 
The third term vanishes when we take the point-particle 
limit\footnote{\label{footnotesym}
The vanishing of this term is a general consequence of the symmetry 
of the system. The source term contains an $\epsilon^{ijk}$ symbol, 
which must be contracted with other vectors to produce a scalar. 
We here have only two possible vectors to contract with, 
i.e. the velocity $v_1^{i}$ and the unit vector $n_1^{i}$ from object 1.  
Hence, any contraction with the Levi-Civita symbol should vanish.}, 
i.e.~$\rho_A=m_A \delta^{(3)} (\bm{x}-\bm{x}_{A})$,
$U_A=m_A/r_A$ and $V_{Ai}=m_A v_{Ai}/r_A$.

Let us evaluate the first and the second terms in the NZ. 
Keeping only the leading PN term in the NZ, we find
\ba
\vartheta_{\self } \!\!  &=& \!\! 16 \frac{\alpha_4}{\beta}
\epsilon_{ijk} \frac{n_{1,i}}{r_{1}^2}  \int_{\mathcal{M}} \!\!\!
\rho_1' ( V_{1k,j}' - U'_{1,j} v_{1k}' ) d^3x' \plusonetotwo \nn \\
&=& \!\! 
\frac{  n_{1,i} }{r_{1}^2} 
\mu_{i}^{(1)} \plusonetotwo \,,
\label{final-NL-odd}
\ea
where we have defined
\be
\mu_{i}^{(A)} \equiv 32 \frac{\alpha_4}{\beta} \epsilon_{ijk}  
\int_{\IZ}  \rho'_A V_{Ak,j}'  d^3x'\,. 
\ee
This leading-order PN term in $\vartheta_{\self}$ 
represents a magnetic-type dipole.
 
As in the even-parity case, to extend this result to 
compact objects we have to determine 
the value of $\mu_{i}^{(A)}$ by matching the NZ solution in 
Eq.~\eqref{final-NL-odd} to a strong field solution. 
This will be carried out in Sec.~\ref{sec:matching} 
for the BH case. For NSs, we just present an order of magnitude 
estimate based on a simple extrapolation of weakly-gravitating results:
\be
 \mu^i_{(A)} = \frac{\alpha_4}{\beta} \mathcal{O}\left({
    m_A S_A^i\over R_A^3}\right)\,, 
\ee
where $S_A^i$ is the spin angular momentum of the object. 
Following the procedure in Appendix~\ref{sec:balding}, we can show that NSs cannot have scalar monopole charge in the dynamical CS case.

Next, we consider the cross term $\vartheta_{\cross}$ in the
weakly-gravitating case. Integrating by parts several times, we find
\ba
\vartheta_{\cross} &\!\! =&\!\! 
-16 \frac{\alpha_4}{\beta} \epsilon_{ijk}
 \bigg[ \int_{\mathcal{M}} \rho'_1 V_{2k,j}' \left(
 \frac{1}{|\bm{x}-\bm{x}'|} \right)_{\!\! ,i} d^3x' 
\nonumber \\
&&\!\! + \int_{\mathcal{M}} U'_{1,i} \rho_2' v_{2k}' \left(
\frac{1}{|\bm{x}-\bm{x}'|} \right)_{\!\! ,j} d^3x' \nonumber \\
&&\!\! + \int_{\mathcal{M}} U'_{1,i} V_{2k,j}' \delta ^{(3)}
(\bm{x}-\bm{x}') d^3x' \plusonetotwo \bigg]\,.
\cr&&
\ea
One can take the point-particle
limit of this expression without any trouble to obtain
\ba
\vartheta_{\cross} &\!\! =&\!\! -16 \frac{\alpha_{4}m_1 m_2}{\beta m^{4}} \;
\epsilon_{ijk} v_{12k}  
\left[m^{4}\left(
 \frac{n_{12}^i n_{1}^j  }{r_{1}^2 b^2} 
  +\frac{n_{12}^i n_{2}^j}{r_{2}^2 b^2} 
\right. \right.
\nonumber \\
&&\qquad + \left. \left.
\frac{n_{1}^i n_{2}^j}{r_{1}^2 r_2^2} 
  \right)
+ {\mathcal{O}}\left( \frac{m^{5}}{r^{5}} \right) \right] \,. 
\label{final-PP-odd}
\ea
Since $\mu_i^{(A)}$ in Eq.~\eqref{final-NL-odd} do not depend on $v$,
these terms are of relative $\mathcal{O}(v^5)$ compared to the 
leading-order term of $\vartheta_{\self}$. 

As for compact objects, the results found in the even-parity case also
apply here. Terms proportional to $1/r_A^2$ in the above expression suggest 
that each object has a dipole component induced by the companion.  
When we expand this expression around $r_A\ll b$, however, 
the terms proportional to $1/(r_A^2 b^2)$ 
cancel each other, as in the even-parity case, leading to 
no induced dipole moment. Even if this were not the case, however, the
corrections to the dipole moment would be higher order than the contributions
from $\vartheta_{\self}$.

To conclude, the dominant contribution to $\vartheta$ is clearly that of
$\vartheta_{\self}$ given in Eq.~\eqref{final-NL-odd}, which again
depends on the structure of the source and thus violates the effacement
principle. 

\subsection{Matching near zone and strong-field solutions and finding the effective source terms}
\label{sec:matching}

In alternative theories of gravity, the point-particle limit is not
always valid and the multipole moments of compact objects may 
depend on the internal structure of the source. In
the previous subsections, we found that the dominant contributions to
the scalar field come from self-interaction terms, which in turn depend
on certain structure constants. 
In this subsection, we determine these constants by
matching the $\vartheta$ solution to that of an isolated BH.

\subsubsection{Even-parity sector}
\label{sec:matching-even}

In the even-parity case, the monopole charges $q_1$ and $q_2$ 
in Eq.~(\ref{final-theta-self}) must be determined by matching 
to a BH solution. An isolated BH sources a scalar 
field~\cite{Yunes:2011we}, whose leading PN behavior is
\be
{\vartheta}_{\rm YS} = \frac{2 \alpha_{3}}{\beta m_A^{2}} \frac{m_{A}}{r_{A}}\,.
\ee
Matching this solution to the NZ solution of Eq.~(\ref{final-theta-self}) 
we obtain
\be
q_{A} = \frac{2 \alpha_{3}}{\beta m_{A}}\,.
\label{eq:BHscalarcharge}
\ee
Notice that this monopole charge does not depend on $(\alpha_{1},\alpha_{2})$, as
for pure BH spacetimes, 
these coupling constants appear in combination with the Ricci scalar and tensor,
which vanishes. This is to be contrasted with 
the NS case, in which $q_{A}$ depends on $\alpha_{1}$ and $\alpha_2$ as
well as $\alpha_3$ and vanishes in EDGB theory. 
Interestingly, BHs do not have scalar hair in more traditional (Brans-Dicke type)
scalar-tensor theories, while NSs do possess them. 
This situation is reversed in EDGB theory. 

The matching carried out above dealt with the
monopole part of $\vartheta$. That is, we have
ignored any tidal deformation of either BH induced 
by its binary companion. In BH perturbation theory, one can
calculate the deformation of the isolated BH metric to find that it
depends on the sum of electric and magnetic tidal tensors, leading to a
metric deformation that scales as $(r_{1}/b)^{2} (m_{2}/b)$ for $r_{1}
\ll b$~\cite{Alvi:1999cw,Yunes:2005nn,Martel:2005ir,Poisson:2005pi,Yunes:2006iw,JohnsonMcDaniel:2009dq}.
Thus, in the IZ of object 1, tidal deformations lead to corrections of
${\mathcal{O}}(m^{3}/b^{3})$, which are much smaller than the effects
considered here. Therefore, it suffices in this section to consider an
isolated BH when matching the scalar fields.  

With this at hand, we can now treat BHs in even-parity, quadratic
modified gravity as delta function sources of matter energy density, 
and with effective scalar charge density
\be
\rho_{\vartheta} = q_{A} \delta^{(3)}(\bm{x}-\bm{x}_{A}).
\ee
In the PN expansion such sources reproduce the BH
solution found by Yunes and Stein~\cite{Yunes:2011we} at leading order.  

Let us make a few observations about the effective source term approach.
First, notice that the scalar field diverges as $m_{A} \to 0$, 
which violates the small-coupling approximation. 
This is related to the fact that as one shrinks a BH, the
radius of curvature at the horizon also goes to zero, probing
increasingly shorter length scales. When the small-coupling
approximation is violated, one can no longer neglect the scalar field's
stress-energy tensor and the $({\mathcal{H}}_{\mu
\nu},{\mathcal{I}}_{\mu \nu},{\mathcal{J}}_{\mu \nu},{\mathcal{K}}_{\mu
\nu})$ tensors that would dominate over the Einstein tensor. Of course, one
cannot take this limit seriously, as we are considering here a
low-energy effective theory, which is missing higher-curvature terms
that would need to be included. Notice also that this is
different from the behavior of scalar fields in 
traditional scalar-tensor theories, where the scalar
field vanishes in the $m_{A} \to 0$ limit.

\subsubsection{Odd-parity sector}

In the odd-parity case, the dipole charges of the respective objects    
in Eq.~(\ref{final-NL-odd}) are to be determined by matching 
against the appropriate BH solutions. An isolated
non-spinning BH in the odd-parity case does not support a scalar
field. By contrast, a spinning BH does, and in the slow-rotation
limit, neglecting higher order PN corrections, it is given by 
~\cite{Yunes:2009hc} 
\be
{\vartheta}_{\rm YP} = -\frac{5}{2} \frac{\alpha_4}{\beta
r_A^2} n_{A}^i \chi_{A}^i\,,
\ee
where $\chi_A\equiv  S_A^i/m_A^2$ 
is the normalized spin angular momentum vector of the $A$th BH.
Matching this solution to the NZ $\vartheta_\self$ in 
Eq.~(\ref{final-NL-odd}), we obtain
\be
\mu^{i}_{A} = \frac{5}{2} \frac{\alpha_{4}}{\beta}
\chi_A^{i}\,.
\label{eq:BHmu}
\ee
With this at hand, we can now treat BHs in odd-parity, quadratic
modified gravity as distributional sources 
of matter energy density and 
effective scalar charge density
\be
\rho_{\vartheta} =- \mu_{A}^i
\delta^{(3)}(\bm{x}-\bm{x}_{A})_{,i}\,.
\ee
In the PN expansion, such sources reproduce the BH
solution found by Yunes and Pretorius~\cite{Yunes:2009hc} 
at leading order.  

Let us make a few observations about this solution. First, notice that
the pseudo-scalar dipole charge is well behaved in the limit $m_{A} \to
0$, because there is a maximum BH spin $|\chi_A^i|<1$. 
Second, notice that in the $|\chi_A^i| \to
0$ limit, this dipole charge vanishes, which is a consequence of Birkhoff's
theorem holding in CS 
gravity~\cite{Yunes:2007ss,Grumiller:2007rv,Alexander:2009tp}. 
Namely, non-spinning BHs in CS theory are the same as BHs in GR
(i.e. Schwarzschild BHs). Therefore, in this case the point-particle
limit is well-justified and the metric deformation or the scalar field
does not depend on the internal structures of non-spinning sources.

\subsection{Far-zone field point solutions}
\label{sec:NZ-Evolution}

Let us assume that we have the wave equation
\be
\square_{\eta} \vartheta = \tau(t,x)\,,
\ee
where $\tau$ denotes the source term.
The far-zone field point solution to this wave equation is given as~\cite{Will:1996zj,2000PhRvD..62l4015P}
\be
\vartheta^{\FZ} = - \frac{1}{4\pi} \sum_{m=0}^\infty
\frac{(-1)^m}{m!} \partial_M \left[ \frac{1}{r}\int_{\mathcal{M}}
  \tau (u,x'^i) x'^M \right]\,,
  \label{DIRE-FZ1}
\ee

with $u\equiv t-r$. 
By using $u_{,i}=-n_i $ and by keeping only terms proportional to $1/r$, the above solution reduces to
\be
\vartheta^{\FZ} = -\frac{1}{4\pi} \frac{1}{r} \sum_{m=0}^{\infty} \frac{1}{m!} \frac{\partial^{m}}{\partial t^{m}} \int_{\mathcal{M}} \tau (u,x'^{i}) \left(n_{j} x'{}^{j}\right)^{m} d^{3}x'\,.
\label{DIRE-FZ}
\ee
Here, the region $\mathcal{M}$ denotes the hypersurface of $t-r=\mathrm{const}$.
In the following, we apply these formulas to the even and odd parity cases separately.

\subsubsection{Even-parity sector}

Following the discussion in Sec.~\ref{sec:cheating}, the evolution equation for the scalar field is dominantly 
\be
\square_{\eta} \vartheta = - 4 \pi q_{1} \delta^{(3)}(\bm{x}- \bm{x}_{1}) \plusonetotwo \,
\label{NZ-FZ-EOM-theta}.
\ee
From Eq.~\eqref{DIRE-FZ}, this wave equation can be solved as
\begin{align}
\vartheta^{\FZ} =&\, \frac{1}{r} \sum_{m} \frac{1}{m!}
 \frac{\partial^m}{\partial t^m} \int_{\mathcal{M}} q_{1}
 \delta^{(3)}(\bm{x}'-\bm{x}_1)  ({n}_{j} \; {x}'^{j})^m d^3x'
 \nonumber \\
&\plusonetotwo \,.
\end{align}

The $m=0$ term gives 
\be
\vartheta^{\FZ} = \frac{q}{r}\,,
\ee
where we have defined the total scalar monopole charge $q \equiv q_{1} +
q_{2}$. Recall that this monopole charge $q$ refers to the scalar field,
and not to an electromagnetic one. For a BH binary or a NS binary in a
quasi circular orbit, $q$ only changes during merger, as mass is
carried away in radiation. Thus, monopole radiation is inefficient and suppressed.

For the $m=1$ case, we find
\be
\vartheta^{\FZ} = \frac{\dot{D}_{i} n^{i}}{r}\,,
\label{eq:thetaFZdipole}
\ee
where we have defined the total scalar dipole moment as
\be
D^{i} \equiv q_{1} x_{1}^{i} + q_{2} x_{2}^{i}\,.
\label{eq:dipoledef}
\ee
When we evaluate this for circular orbits, we find
\be
\vartheta^{\FZ} =  \frac{1}{r} \left(  q_{1} \frac{m_{2}}{m} - q_{2} \frac{m_{1}}{m} \right)v_{12i} n^{i}\,,
\label{theta-NZFZ-even}
\ee
where we have defined the relative velocity $v^{k}_{12} \equiv
v_1^k-v_2^k$. 

The $m=1$ term clearly leads to dipole radiation in the FZ, which is less relativistic than GR quadrupole radiation, becoming stronger at smaller velocities. Of course, this term is proportional to the coupling constants of the theory, which are assumed much smaller than one. . Reference~\cite{Yunes:2011we} failed to recognize such dipolar emission because they considered the motion of test particles that had no scalar charge. We cannot think of any mechanism that would suppress such dipolar radiation. 

\subsubsection{Odd-parity sector: spinning bodies}

As in the previous Section, the evolution equation for the scalar field is dominantly 
\be
\square_{\eta} \vartheta = 4 \pi  \mu_{1}^{i} \delta^{(3)}(\bm{x} - \bm{x}_{1})_{,i} \plusonetotwo \,.
\ee
By using Eq.~\eqref{DIRE-FZ1}, the far-zone field point solution is obtained as 
\begin{align}
\vartheta^{\FZ} = - \sum_{m=0}^\infty \frac{(-1)^m}{m!} \pd_M \biggl[
\frac{1}{r} &\int_{\mathcal{M}} \mu_1^i 
\delta^{(3)}(\bm{x}'-\bm{x}_1)_{,i} x'^M d^3x' \nn\\
& {}\plusonetotwo
\biggr].
\label{eq:FZintgeneral}
\end{align}

When $m=0$ there is obviously no contribution to the scalar field. When $m=1$, 
\be
\int_{\mathcal{M}} \delta^{(3)}(\bm{x}-\bm{x}_1)_{,i} x^j d^3x = - \delta_{ij}\,, 
\ee
and thus  
\be
\vartheta^{\FZ} = \frac{\mu_{i} n^i}{r^2}  + \frac{\dot{\mu}_{i} n^i}{r}  \,,
\ee
with $\mu_i \equiv \mu_{1i} + \mu_{2i} $.
Notice that we recover the solution of
Yunes and Pretorius~\cite{Yunes:2009hc} for the first term of the 
above equation with $\mu_A^i$ given as in Eq.~\eqref{eq:BHmu}.
These terms will not strongly radiate because $\dot{\mu_{i}}$ 
is non-vanishing only for spin-precessing systems. Even then, such radiation
would be suppressed by the ratio of the orbital timescale to the precession timescale. 

The $m=2$ contribution, by contrast, depends on the much shorter
orbital timescale. 
We look for terms of $\mathcal{O}(r^{-1})$ since they are the only ones that contribute to the energy flux at infinity.
Keeping in mind that the function being differentiated depends on retarded time, we can rewrite Eq.~\eqref{eq:FZintgeneral}
as
\begin{align}
\vartheta^{\FZ} =&\, -\frac{1}{r} \sum_{m}
\frac{1}{m!} \frac{\partial^m}{\partial t^m} \int_{\mathcal{M}}
\mu_1^i \delta^{(3)}(\bm{x}'-\bm{x}_1)_{,i}  ({n}_{k} \, {x}'^{k})^m d^3x' \nonumber \\
&\plusonetotwo \,.
\end{align}
When $m=2$, we have that
\be
\mu_1^i \int_{\mathcal{M}} \delta^{(3)} (\bm{x}-\bm{x}_1)_{,i} x^p x^q
d^3x \plusonetotwo = -2\mu^{pq}\,,
\ee
where the pseudo-tensor quadrupole
moment (not to be confused with $\mu^i \mu^j$)
is defined as
\be
\mu^{ij} \equiv x_1^{(i} \mu_1^{j)} + x_2^{(i} \mu_2^{j)} \,.
\label{eq:quadrupoledef}
\ee
The $m=2$ contribution becomes
\be
\vartheta^{\FZ} = \frac{1}{r} \ddot{\mu}_{ij}  n^{ij}
=-\frac{1}{r} \omega^2 \mu_{ij} n^{ij} \,,
\label{eq:thetaFZquadrupole}
\ee
where the final equality is evaluated on a circular orbit.
Notice that such a scalar field will strongly radiate because $\mu^{ij}$ 
depends on the orbital timescale.

\subsubsection{Odd-parity sector: non-spinning bodies}

When both objects  are non-spinning, the self-interaction terms produced by
the effective source identically vanish. One is then left with the
source term constructed from the product of the gravitational fields
of objects 1 and 2. These terms
will be proportional to $m_{1} m_{2}$.
As we will see, there are many contributions that turn out to vanish upon NZ
integration. For pedagogical reasons, we will show here explicitly how
this happens and eventually arrive at contributions that do not vanish.

The evolution equation for the scalar field to leading PN order is
\be
\square_{\eta} \vartheta_{\FZ} = - 32 \frac{\alpha_4}{\beta} \epsilon_{ijk} m_1 m_2 v_{12k} \left( \frac{1}{r_1} \right)_{\!\!,im} \left( \frac{1}{r_2} \right)_{\!\!,jm}\,,
\label{wave-eq-scalar-leading}
\ee
where we substituted the NZ metric components in the point-particle approximation.
The leading order term of the solution to this differential equation,
i.e.~the $m=0$ term in the sum of Eq.~\eqref{DIRE-FZ}, is evaluated as
\ba
\vartheta_{\FZ} &\!\! =&\!\! \frac{8}{\pi} \frac{\alpha_4}{\beta} m_{1} m_{2}
\epsilon_{ijk} \frac{v^{k}_{12}}{r}   \int_\mathcal{M} \left(
\frac{1}{r_1} \right)_{\!\! ,im} \left( \frac{1}{r_2} \right)_{\!\! ,jm} d^{3}x
\nonumber \\
&\!\! =&\!\!   -16 \frac{\alpha_4}{\beta} m_{1} m_{2} \epsilon_{ijk}
\frac{v^{k}_{12}}{r} \partial^{(1)}_{i} \partial^{(2)}_{j} \partial^{(1)}_{m}
\partial^{(2)}_{m} Y = 0\,.
\ea

Here we integrated over the NZ+IZ hypersurface ${\mathcal{M}}$ without taking any 
care of the strong gravity region in the IZs.
One can easily show that the contribution from the IZs is not large in the
present case.
In the second line, we replaced partial derivatives with respect to
$x^{i}$
acting on $1/r_A$ 
 with (minus the) particle derivatives with respect
 to $x_{A}^{i}$:
\be
\frac{\partial}{\partial x^{i}}
\to -\frac{\partial}{\partial x_A^i} \equiv -\partial^{(A)}_{i},
\ee
with $A=(1,2)$.
We commuted these particle derivatives with
 the integral, and finally obtained a typical NZ integral,
 discussed in Appendix~\ref{int-tech}. From Eq.~\eqref{Y0-Y1}, we know
 that $Y=b$, and by taking all particle derivatives, the last
 equality is established. 

We could have inferred  
that the $m=0$ term in the sum does not contribute for non-spinning BHs
without any explicit
calculations.
The argument here is similar to that in footnote~\ref{footnotesym}.
Possible vectors to contract with the Levi-Civita symbol include the
velocities $v_A^{i}$ and the unit vectors $n_A^{i}$, 
but not spin vectors $S_A^i$, as we here consider non-spinning BHs. 
In particular, for the $m=0$ case, there cannot be any FZ
vectors $n^i$ present.
Thus, all vectors
that can be contracted onto the Levi-Civita symbol 
must lie in the same
orbital plane and this obviously vanishes. 
This argument should be true at all PN orders\footnote{One may think
  that one can construct a vector that does not lie in the orbital
  plane by taking the cross product of two vectors that lie on this
  plane, e.g. $\bm{n}_{12} \times \bm{v}_{12}$. However, since GR is
  parity even, such a vector cannot be present in the PN metric.}.

Let us then consider the next-order term. This will arise from the leading-order source term [right-hand side of Eq.~\eqref{wave-eq-scalar-leading}] with $m=1$ in the NZ sum:
\ba
\vartheta_{\FZ} &\!\!  = &\!\!   \frac{8}{\pi} \frac{\alpha_4}{\beta} \frac{m_{1} m_{2}}{r} n^p \epsilon_{ijk} v^{k}_{12} \nonumber \\
&&\!\! \times \frac{\partial}{\partial t}\int_\mathcal{M} \left(
\frac{1}{r_1} \right)_{\!\!,im} \left( \frac{1}{r_2} \right)_{\!\! ,jm} x^p d^{3}x  \nn \\ 
&\!\! =&\!\!   - 16 \frac{\alpha_4}{\beta} \frac{m_{1} m_{2}}{r} n^p \epsilon_{ijk} v^{k}_{12} \frac{\partial}{\partial t} \partial^{(1)}_{i} \partial^{(2)}_{j} \partial^{(1)}_{m} \partial^{(2)}_{m} Y_p\,, \nn \\
\ea
where we have used Eq.~\eqref{Y}, which defines $Y_p$. By direct
evaluation, one can show that this term also identically vanishes. The
first non-vanishing contribution coming from an $m=1$ term must then be
${\mathcal{O}}(v^{3})$ smaller than the ordering of the $m=0$ term.

Finally, let us consider the (next)$^{2}$-order term. This can arise only from the leading-order source term with $m=2$ in the NZ sum:
\ba
\vartheta_{\FZ} &\!\! = &\!\!  \frac{4}{\pi} \frac{\alpha_4}{\beta} \frac{m_{1} m_{2}}{r} n^{pq} \epsilon_{ijk}  \frac{\partial ^2}{\partial t^2} v^{k}_{12} 
\nonumber \\
&&\!\! \times 
\int_\mathcal{M} \left( \frac{1}{r_1} \right)_{\!\! ,im} \left(
\frac{1}{r_2} \right)_{\!\! ,jm} x^p x^q d^{3}x \nonumber \\
&\!\! =&\!\!  - 8 \frac{\alpha_4}{\beta} \frac{m_{1} m_{2}}{r} n^{pq}
\epsilon_{ijk} \frac{\partial^2}{\partial t^2} v^{k}_{12}  \partial^{(1)}_{i}
\partial^{(2)}_{j} \partial^{(1)}_{m} \partial^{(2)}_{m} 
\nonumber \\
&&\!\! \times 
\left( Y_{\langle pq\rangle} + \frac{1}{3} \delta_{pq} S \right)\,,
\ea
which simplifies to
\be
\vartheta_{\FZ} = 16 \frac{\alpha_4}{\beta} \frac{1}{r}
\frac{\eta m
\delta m}{b}  \epsilon_{ijk} n_{ip} \omega^2  v_{12}^k n_{12}^{jp}\,,
\label{theta-sol-CS}
\ee
where we have defined the mass difference $\delta m\equiv m_1-m_2$ 
and the symmetric mass ratio $\eta \equiv m_{1} m_{2}/m^{2}$.  We have here used Kepler's law and expanded the STF tensors. This is the dominant FZ behavior of the scalar field, which as we see is much suppressed relative to the odd-parity solution we found for spinning BHs. 

\subsection{Summary of this section}
\label{sec:summary}

Let us summarize the results found so far for later use.
In the even-parity case, generically at least one of the binary
component objects will have a scalar monopole charge.
Since the scalar field excitation due to the induced monopole
is dominant, we neglect all the other less important contributions.
Weakly gravitating objects need not have a scalar monopole charge if
$\alpha_1+\alpha_2+3\alpha_3 = 0$, and BHs have no scalar monopole charge if
$\alpha_3=0$. In EDGB theory, NSs have no scalar monopole charge.
In the odd-parity case, the dominant contribution is the magnetic-type
scalar dipole moment induced by spins. Generically, astrophysical
objects will possess spin, but we will continue to include
non-spinning results to compare with previous work.

In the NZ, we can parametrize 
the leading PN terms of the scalar field as
\be
\vartheta_{\NZ} = \frac{A}{r_{1}^{a} b^{b}} + \frac{B}{r_{1}^{c}
r_{2}^{d}}  \plusonetotwo\,,
\label{summery:NZ-scalar}
\ee
where $(A,B,a,b,c,d)$ are given in 
Table~\ref{tab:theta-par} and for compactness of the Table we define
\be
\sigma_{\NZ}^{pq} \equiv -16 \frac{\alpha_4}{\beta} \eta m^{2} \epsilon_{pqs} v_{12}^s\,.
\label{NZ-q12}
\ee

In the FZ, we can parametrize the scalar field as
\be
\vartheta_{\FZ} = \frac{C}{r}\,,
\label{summery:FZ-scalar}
\ee
where $C$ is also given in Table~\ref{tab:theta-par} and we define
\be
\sigma_{\FZ}^{pq} \equiv 16 \frac{\alpha_{4}}{\beta} \eta\, 
m \, \delta m \frac{\omega^2}{b}\epsilon_{qjk}v_{12}^{k} n_{12}^{jp}\,,
\label{FZ-q12}
\ee

{\renewcommand{\arraystretch}{1.2}
\begin{table}
\capstart
\begin{centering}
\begin{tabular}{rcccccccc}
\hline
\hline
\noalign{\smallskip}
\multicolumn{1}{c}{$~$} && $A$ & \multicolumn{1}{c}{$B$}
& \multicolumn{1}{c}{$C$}
&  \multicolumn{1}{c}{$a$} &  \multicolumn{1}{c}{$b$}
&  \multicolumn{1}{c}{$c$} &  \multicolumn{1}{c}{$d$}
  \\
\hline
Even-P && $q_{1}$ & $0$ &$\dot{D}_{i} n^{i}$& $1$ & $0$ & $-$ & $-$\\
Odd-P, Spins && $\mu_{1}^{i} n_{1}^{i}$ & $0$ &$\ddot{\mu}_{i} n^{ij}$& $2$ & $0$ & $-$ & $-$ \\
Odd-P, No Spins && $\sigma_{\NZ}^{pq} n_{12}^{p} n_{1}^{q}$ & $\frac{1}{2}
 \sigma_{\NZ}^{pq} n_{1}^{p} n_{2}^{q}$ & $\sigma_{\FZ}^{pq} n^{pq}$&
 $2$ & $2$ & $2$ & $2$\\
\noalign{\smallskip}
\hline
\hline
\end{tabular}
\end{centering}
\caption{Scalar field parameters, as defined in Eqs.~\eqref{summery:NZ-scalar} and~\eqref{summery:FZ-scalar}. The quantities $q_{1}$ and $\mu_{1}^{i}$ are defined in Eqs.~\eqref{eq:BHscalarcharge} and~\eqref{eq:BHmu}, while $\sigma_{\NZ}^{pq}$ is defined in Eq.~\eqref{NZ-q12}. The quantities $D_{i}$ and $\mu_{i}$ are defined in Eqs.~\eqref{eq:dipoledef} and~\eqref{eq:FZintgeneral}, while $\sigma_{\FZ}^{pq}$ is given in Eq.~\eqref{FZ-q12}.}
\label{tab:theta-par}
\end{table}
}

\section{Metric evolution}
\label{sec:Metric-Evolution}

In this section, we solve the evolution equations for the metric
deformation in the FZ, so that we can calculate the gravitational
energy flux at infinity. 
Note that throughout, we use the Newtonian relationship $v^2 =
m/b$ (and similarly for the acceleration). This relationship must be
corrected at higher PN order or 
at $\mathcal{O}(\varsigma)$. 
As we mentioned earlier, here we do not take into account 
the corrections to the orbital motion due to the conservative force 
at $\mathcal{O}(\varsigma)$.  
These conservative effects do not interfere 
at $\mathcal{O}(\varsigma)$ with the radiative effects that 
we are concerned with in this paper. Therefore the corrections to the 
GW waveform become a simple summation of these two different types of 
effects.

For the FZ field points, the solution
to the metric deformation equation of motion [Eq.~\eqref{wave-eq-h-pert}] can be read from Eq.~\eqref{DIRE-FZ}:
\be
\hDef_{ij}  = -\frac{8}{r} \sum_{m=0}^{\infty} \frac{1}{m!}\frac{\partial^m}{\partial t^m} \int_{\mathcal{M}} {\tilde{\mathcal{C}}}_{ij} \;
(n^k x'{}^k)^m d^3x'  + O\left( r^{-2} \right)\,,
\label{NZ-FZ-metric-sol}
\ee
where we have defined the source term as
\begin{align}
\tilde{\mathcal{C}}_{ij} =&\, 
\alpha_1 \left( \vartheta \tilde{\mathcal{H}}^{(0)}_{ij} + \tilde{\mathcal{H}}^{(1)}_{ij} \right) + \alpha_2  \left( \vartheta \tilde{\mathcal{I}}^{(0)}_{ij} + \tilde{\mathcal{I}}^{(1)}_{ij} \right) 
\nn \\
&+ \alpha_3  \left( \vartheta \tilde{\mathcal{J}}^{(0)}_{ij} + \tilde{\mathcal{J}}^{(1)}_{ij} \right) + \alpha_{4} \tilde{{\mathcal{K}}}^{(1)}_{ij}-\frac{1}{2} T_{ij}^{(\vartheta)}\,.
\label{Cij-source-term-def}
\end{align}
Notice that this corresponds to an IZ+NZ integration for FZ field points, where we have neglected the FZ integration because it is subdominant.

The integrals presented above have to be carried out also in the IZ,
where the PN expansion is not valid anymore.  In GR, however, such
divergences can be ignored, using a regularization scheme. Since both
the true solution and an appropriately regularized solution satisfy
the field equations in the NZ, their difference due to the IZ
contribution is only through a homogeneous solution. Such homogeneous
solutions are regular in the NZ and FZ, but can be divergent in the
IZ.  They are characterized by the multipole moments of the
respective objects, which can be determined by studying tidal
perturbations around a strongly gravitating object.  One can then
perform matching of the metric solution, as for the scalar solution,
but the metric matching is beyond the scope of this paper. In what
follows, we only consider the regularized contribution,
following Hadamard \emph{partie finie} (FP)
regularization~\cite{Blanchet:2000nu}. 
We comment more on the divergent contribution at the end of this Section.

\subsection{Even-parity sector}
\label{sec:heven}

Let us focus on the metric perturbation in the even-parity
sector first. The leading order term both in the PN and $1/r$ expansion at infinity 
is formally given by
\begin{align}
\hDef_{ij} &= \hDef_{ij}^{T} + \hDef_{ij}^{{\cal{J}}}\,,
\\  
\hDef_{ij}^{T} &\equiv \frac{4}{r} \int_{{\cal{M}}} T_{ij}^{(\vartheta)}  d^{3}x\,,
\\
\hDef_{ij}^{{\cal{J}}} &\equiv -\frac{8\alpha_{3}}{r} \int_{{\cal{M}}} 
\tilde{\cal{J}}_{ij} 
d^{3}x\,,
\end{align}
where we have defined $\tilde{\cal{J}}_{ij} \equiv \vartheta \tilde{\mathcal{J}}^{(0)}_{ij} + \tilde{\mathcal{J}}^{(1)}_{ij}$.
The source terms $\tilde{\cal{H}}_{\mu\nu}$ and $\tilde{\cal{I}}_{\mu\nu}$ 
do not contribute to this expression 
since they identically vanish in the NZ where $R_{\mu\nu}=0$.

We can estimate the order of magnitude of both $\hDef_{ij}^{{\cal{J}}}$ and $\hDef_{ij}^{T}$ as follows:
\begin{align}
\label{orderT1-even}
\hDef_{ij}^{T} 
&\sim {\mathcal{O}}\left(\beta \frac{m}{r} v^{-2} \vartheta^{2}  \right) 
=  \zeta_{3} \frac{m}{r} v^{2}\times 
 {\mathcal{O}}\left(1 \right)\,,
\\
\label{order2-even}
\hDef_{ij}^{{\cal{J}}} 
&\sim {\mathcal{O}}\left(\frac{\alpha_{3}}{m^{2}} \frac{m}{r}
v^{4} \vartheta \right) 
= \zeta_{3} \frac{m}{r} v^{2}\times 
  {\mathcal{O}}\left(v^{4}\right)\,.
\end{align}
Here we factored out $v^2$ in the final expressions, 
 since the GR leading 
quadrupolar field is also proportional to $v^2$. 
Clearly, the dominant contribution comes from Eq.~\eqref{orderT1-even}. 

Let us now make this computation more precise. The stress-energy
tensor will contain self-interactions of the form $\vartheta_{A,i}
\vartheta_{A,j}$ and cross terms of the form $\vartheta_{1,i}
\vartheta_{2,j}$. The former case leads to divergent integrals, which
must be determined by strong-field matching, so we do not consider
them here. Let us concentrate on the latter, which take
the form
\begin{align}
T_{ij}^{(\vartheta)} 
&=  \beta \left( \vartheta_{,i} \vartheta_{,j} -  \frac{1}{2}
\delta_{ij} \vartheta_{,\mu} \vartheta^{,\mu}  \right)\,
\label{hdef-T}
\\ 
&\approx  \beta  q_1 q_2 \left[ 2 \left( \frac{1}{r_1}
\right)_{\!\! ,(i} \!\! \left( \frac{1}{r_2} \right)_{\!\! ,j)} \!\!\!\! -
\delta_{ij} \left( \frac{1}{r_1} \right)_{\!\! ,k} \!\! \left( \frac{1}{r_2}
\right)_{\!\! ,k} \right]\,,
\end{align}
which sources the metric perturbation 
\ba
\hDef_{ij} &\!\! =&\!\!  
\frac{4}{r} \int_{\mathcal{M}} T_{ij}^{(\vartheta)} d^3x \,, \nonumber \\
&\!\! =&\!\!  - \frac{4\pi}{r} \beta  q_1q_2 \left( 2 \partial^{(1)}_{i} \partial^{(2)}_{j} b - \delta_{ij} \partial^{(1)}_{k} \partial^{(2)}_{k} b \right) \plusonetotwo \,, \nonumber \\
&\!\! =&\!\!  -\frac{16\pi}{r} \beta \frac{q_1 q_2}{b} n_{12}^{ij} \,,
\label{h-even-original}
\ea
where we used an integration formula for the triangle potential 
given in Appendix~\ref{int-tech}.
We can see that this correction is
0PN relative to the radiative metric perturbation in GR, just as we
predicted in Eq.~\eqref{orderT1-even}. However, this correction 
turns out to be still smaller in the energy flux 
than the dipole scalar radiation, which gives a -1PN correction.

\subsection{Odd-parity sector}
\label{odd-section}

We now focus on the odd-parity sector, for which the solution is given by the term
proportional to $\alpha_{4}$ in Eq.~\eqref{NZ-FZ-metric-sol}, namely
\ba
\label{hDef-NZ}
\hDef_{ij} &\!\! =&\!\!  \hDef_{ij}^{T} + \hDef_{ij}^{{\cal{K}}}\,,
\\  
\hDef_{ij}^{{\cal{K}}} &\equiv& -\frac{8\alpha_{4}}{r} \int_{{\cal{M}}} \tilde{\cal{K}}^{(1)}_{ij} d^{3}x\,.
\ea
The stress-energy contribution $\hDef_{ij}^{T}$ is the same as in Eq.~\eqref{hdef-T}.

The $\mathcal{K}$ contribution to Eq.~\eqref{hDef-NZ} is more
involved. The leading-order behavior of the $\mathcal{K}$ tensor is
\ba
\tilde{\mathcal{K}}^{(1)}_{ij} &\!\! =&\!\!  \dot{\vartheta}^{}_{,k} \epsilon_{jkl} \hGR_{00,il} +  \dot{\vartheta}^{}_{,k} \epsilon_{jlm} (\hGR_{im,lk} + \hGR_{lk,im})
\nonumber \\
&&\!\! +
\vartheta^{}_{,kl} \epsilon_{jlm} (\hGR_{i0,mk}+\hGR_{mk,i0}-\hGR_{k0,im} -\hGR_{im,k0})
\nonumber \\
&&\!\! - 
\vartheta^{}_{,k} \epsilon_{ikl} (2 \hGR_{0[m,j]lm} - 2 \dot{\hGR}_{l[j,m]m} - \dot{\hGR}_{00,jl}  )
\nonumber \\
&&\!\! -2 \dot{\vartheta}^{} \epsilon_{ikl} \hGR_{k[j,m]lm}\plusitoj\,.
\label{cij-leading}
\ea
Other terms are of higher PN order. By applying the Lorenz or harmonic gauge condition $h^{\mu\nu}{}_{,\nu}=0$, substituting $\hGR_{ij}=\hGR_{00} \delta_{ij}$ into Eq.~\eqref{cij-leading}, and using $\epsilon_{jkl}\hGR^{\mu\nu,kl}=\epsilon_{jkl} \vartheta_{,kl}=0$, we get
\begin{align}
\tilde{\mathcal{K}}^{(1)}_{ij} =&\, 2  \dot{\vartheta}_{,k} \epsilon_{jkl} \hGR_{00,il} - 2 \vartheta_{,km} \epsilon_{jkl} \hGR_{0[m,i]l}  
\nonumber \\
&-2\vartheta_{,k} \epsilon_{jkl} \hGR_{0[m,i]lm}   + 2  \vartheta_{,k} \epsilon_{jkl} \dot{\hGR}_{00,il}  \plusitoj\,. 
\label{cij-simplified}
\end{align}
The $\tilde{\cal{K}}_{ij}$ term in Eq.~\eqref{hDef-NZ} is then a sum
of four terms, namely
\be
\hDef_{ij}^{{\cal{K}}} = \sum_{n=1}^{4} \hDef_{ij}^{(n)} \,,
\ee
where we have defined
\allowdisplaybreaks[1]
\begin{align}
\label{hij-cs-1}
\hDef_{ij}^{(1)} =&\, - \frac{16 \alpha_4}{r} \int_{\mathcal{M}} 
\dot{\vartheta}_{,k} \epsilon_{jkl} \hGR_{00,il} d^3x  \plusitoj\,,\\
\label{hij-cs-2} 
\hDef_{ij}^{(2)} =&\, + \frac{16 \alpha_4}{r} \int_{\mathcal{M}} 
\vartheta_{,km} \epsilon_{jkl} \hGR_{0[m,i]l} d^3x  \plusitoj\,, \\
\label{hij-cs-3} 
\hDef_{ij}^{(3)} =&\, + \frac{16 \alpha_4}{r} \int_{\mathcal{M}} 
\vartheta_{,k} \epsilon_{jkl} \hGR_{0[m,i]lm} d^3x  \plusitoj\,,\\
\label{hij-cs-4} 
\hDef_{ij}^{(4)} =&\, - \frac{16 \alpha_4}{r} \int_{\mathcal{M}} 
\vartheta_{,k} \epsilon_{jkl} \dot{\hGR}_{00,il} d^3x  \plusitoj\,.
\end{align}
\allowdisplaybreaks[0]

When we substitute the PN metric into the above terms, the
right-hand sides depend on the velocity vectors $v_A^i$ (which
depend on time only). The field $\vartheta$ is
given in Eq.~\eqref{summery:NZ-scalar} and its derivative can be
computed simply from that equation. Since this field is a NZ one, it 
depends on time through the positions of the objects, 
which implies that its time derivative can be converted into a 
spatial derivative via $\partial_{t} f(r_{1})= - v_{1}^{i} \partial_{i}f(r_{1})$. 

Let us begin by making a simple order of magnitude estimate of how large the regularized contribution is. For this, it suffices to look at Eqs.~\eqref{orderT1-even}  and~\eqref{hij-cs-1}: 
\ba
\hDef_{ij}^{T} &\!\! \sim &\!\! {\mathcal{O}}\left(\beta \frac{m}{r} v^{-2} \vartheta^{2} \right)\,,
\\
\hDef_{ij}^{{\cal{K}}} &\!\! \sim &\!\! {\mathcal{O}}\left(\frac{\alpha_{4}}{m^{2}} \frac{m}{r} v^{5} \vartheta \right)\,.
\ea
The $\vartheta$ field here is that of the NZ, and hence 
\ba
\label{orderT1}
\hDef_{ij}^{T} 
&\!\! \sim &\!\! 
\zeta_{4} \frac{m}{r} v^2 \times {\mathcal{O}}\left(
\chi^2 v^{4} 
+ \eta \chi v^{9} 
+ \eta^{2}v^{14} 
\right)\,, 
\\
\label{order2}
\hDef_{ij}^{{\cal{K}}} 
&\!\! \sim &\!\! 
\zeta_4 \frac{m}{r} v^2 \times
{\mathcal{O}}\left(
\chi v^{7} +
 \eta v^{12} \right)\,, 
\ea
where $\chi$ stands for the magnitude of $\chi_1^i$ and 
$\chi_2^i$.
From this analysis, $\hDef_{ij}^T$ 
is clearly larger for rapidly spinning objects, leading to a 2PN effect. 

For the non-spinning case, one might expect the ${\cal{K}}$
contribution to lead to a 6PN effect, but as we explain in
Appendix~\ref{non-spin-reg-cont}, these leading-order effects actually
vanish. This cancellation can also rather easily be seen by integrating
by parts in Eqs.~\eqref{hij-cs-1}-\eqref{hij-cs-4}. After discarding
boundary terms (taking into account the boundary term is equivalent to adding homogeneous solutions,
corresponding to deformed multipole moments of compact objects), we obtain
expressions of the form $\epsilon_{jkl}\, \vartheta\,
h_{00,kl\ldots}$, which obviously vanishes by the antisymmetry of the
Levi-Civita tensor.  We carry out a more careful analysis in
Appendix~\ref{non-spin-reg-cont}, where we explicitly show that the
leading and first sub-leading order terms vanish\footnote{In
  Appendix~\ref{non-spin-reg-cont}, we only show this for non-spinning
  BHs, but a similar calculation can be performed for spinning BHs to
  $\mathcal{O}(\chi)$.}. The first non-vanishing term is then of
${\mathcal{O}}(v^{2})$ smaller than the order of magnitude
estimates in Eqs.~\eqref{orderT1} and~\eqref{order2}, leading to 7PN
and $4.5$PN contributions at $\mathcal{O}(\chi^0)$ and $\mathcal{O}(\chi^1)$, respectively.

Since the largest contribution seems to arise for spinning BHs from
the $\hDef_{ij}^T$ term, let us consider this in more detail. Two
possible contributions are generated here: one that depends only on
self-interaction terms, and one that depends on the cross-interaction.
The former leads to divergent integrals, which need to be matched from
strong-field solutions, and we do not consider these here.
The latter leads to the metric deformation
\ba
\hDef_{ij}^{T} &\!\!  = &\!\!  - \frac{4\pi}{r} \beta  \mu_1^k \mu_2^l \left( 2 \partial^{(1)}_{ik} \partial^{(2)}_{jl} Y - \delta_{ij} \partial^{(1)}_{pk} \partial^{(2)}_{pl} Y \right)  \plusonetotwo \nonumber \\
&\!\!  = &\!\!   \frac{8 \pi \beta}{r b^{3}} \left\{2 \mu_{1}^{(i} \mu_{2}^{j)} -12 n_{12}^{(i} \mu_{1}^{j)} \left(n_{12}^{k} \mu_{2k}\right) 
\right. 
 \nonumber \\ 
&&\!\! + \left. 3 n_{12}^{ij} \left[5 \left(n_{12}^{k} \mu_{1k} \right) \left(n_{12}^{l} \mu_{2l} \right) - \mu_{1k} \mu_{2}^{k} \right] \right\} \plusonetotwo\,, \nn \\
\label{h-odd-a2}
\ea
which is clearly of the order predicted in Eq.~\eqref{orderT1}, i.e.~2PN order relative to GR. This is of the same order as the energy flux correction carried by the pseudo-scalar radiation. 

\subsection{Multipole moments}
\label{sec:multipolemoments}

In this Subsection, we discuss the additional contribution from the IZs, 
which enter as additional homogeneous solutions in the NZ and FZ, 
These contributions are homogeneous in the sense that they arise
from sources that have support only in the IZs, and thus they vanish
in the NZ and FZ (see e.g.~the discussion prior to Eq.~\eqref{int-exp-1}). 
The homogeneous solutions are characterized by the mass and 
current multipole moments of the strong-field bodies, which must be
determined by matching to strong gravity solutions in the IZ. 
When we solve the non-linear equations of motion iteratively, 
the source terms in general can be classified into two pieces: 
a self-interaction part and a cross-interaction part, as in the case of 
$\vartheta$ in Sec.~\ref{sec:SF-Evolution}. 
The cross-interaction part is sourced by the companion, 
while the self-interaction part is not. 

The self-interaction part is rather easy to handle because matching involves only
a single isolated object. As described in Sec.~\ref{sec:cheating}, these self-interaction
terms can be thought of as homogeneous solutions that have support only in the IZ. 
As such, in the small-coupling approximation, they satisfy homogenous field equations 
that take Einstein form. If the spin of the object is neglected, the only possible linear 
perturbation to such a homogeneous solution that is compatible with asymptotic flatness 
is a shift of the body's mass (in the $1/r$ piece of the $(t,t)$ and diagonal parts of the metric). 
In essence, this is a consequence of Birkhoff's theorem, which holds for homogeneous
solutions. Such a shift is consistent with the strong-field, non-spinning BH solution 
in EDGB theory found in~\cite{Yunes:2011we}. In that case, the mass shift is simply 
$m_{A} \to (49/80)\zeta_3 m_{A}$.

For spinning objects, one expects there to be higher multipole
moments in the strong-field solution. However, one should be able to 
absorb current dipole moment modifications 
by a redefinition of the spin parameter, while the mass dipole moment will be 
absorbed by the redefinition of the position of the center of mass. 
Therefore, the leading-order corrections that survive 
are the mass quadrupole moment, which produces a metric perturbation 
in the NZ proportional to $1/r^3$. 
As we will see, when we consider FZ solution, there is an additional 
factor of $v^2$ that enters. 
 
Therefore, contributions to the energy flux from the quadrupole or 
higher multipole moments are at least 3PN order relative to that from 
the GR quadrupole formula. 
We will later find that corrections to the energy flux
due to scalar radiation appear at -1PN and 2PN relative order 
for the even and odd-parity cases, respectively. 
Hence, the contributions from the multipole moments that we discussed here 
are definitely smaller than those introduced by scalar radiation 
in the even-parity case, and at most, the same order in the odd-parity case. 

Let us take a look at spinning BHs in the odd-parity sector in more detail. At ${\mathcal{O}}(\chi)$ there is freedom in adding a homogeneous solution proportional to $1/r^{2}$ in the $h_{0i}$ component. This corresponds to a freedom in shifting the Kerr parameter measured at infinity. Reference~\cite{Yunes:2009hc} set this homogeneous solution to zero so that there is no shift in the Kerr parameter. At $\mathcal{O}(\chi^2)$, there should be corrections proportional to $1/r^{3}$ in $h_{ij}$ which shifts the quadrupole moment. Since there is no parameter in the Kerr geometry that can absorb this correction in the quadrupole moment, this $1/r^{3}$ correction cannot be eliminated. 

The effective source term that reproduces this correction should look like
\ba
\square \hDef_{ij} &=& -4\pi  Q_1 u_i u_j (\delta_{kl}-3\hat{S}_{1,k}
\hat{S}_{1,l} ) 
{\delta^{(3)}}(\bm{x}-\bm{x}_1)_{,kl} \nn \\
& & \plusonetotwo \,,
\ea
where $Q_A=\mathcal{O}(\zeta_4 m_A a_A^2)$ and $\hat{S}_{A,k} \equiv S^{i}_{A}/m_{A}^{2}$ is a unit spin angular momentum vector.
The solution of this wave equation at $\mathcal{O}(1/r)$ is given by
\ba
\hDef_{ij} &\!\!  = &\!\!  \frac{1}{r} \sum_{m=0}^{\infty} \frac{1}{m!} \frac{\partial^m}{\partial t^m}  u_i u_j (\delta_{kl}-3\hat{S}_{1,k} \hat{S}_{1,l} ) Q_1  \nn \\ 
&&\!\!  \times \int_{\mathcal{M}} {\delta^{(3)}} (\bm{x}-\bm{x}_1)_{,kl} (\bm{n}\cdot \bm{x})^m d^3x  \plusonetotwo \,. \nn \\
\ea
The leading-order contributions at $m=0$ (2PN) and $m=1$ ($2.5$PN) vanish, leading
to the first non-zero contribution at $m=2$
\be
\hDef_{ij} = \mathcal{O}\left( \frac{1}{r} Q \omega^2 v^2 \right) =
\zeta_4 \frac{m}{r}v^2\times \mathcal{O}\left( 
{\chi^2} v^6 \right)
\,,
\label{quadrupole-rad-FZ}
\ee
which is 3PN relative to GR. 
Therefore, the self-interacting correction in the metric at $\mathcal{O}(\chi^2)$ is smaller compared to the corrections in the energy flux carried by the scalar field and the metric field with regularized modification. 

The cross-interaction part is more complicated. In this case, we have 
to consider the induced multipole moments due to the presence 
of the secondary object. Thus, even if we consider 
non-spinning objects, higher multipole moments might be 
induced. Another important difference is that neither  
the mass monopole nor the spin dipole can be simply absorbed 
by a redefinition of the mass and spin of each object. This is because 
the shifts of these multipole moments depend on the orbital 
parameters, such as separation $b$. 
Notice, however, that the effects of the secondary object 
propagate only through the scalar field or the gravitational 
tidal force.

The order of magnitude of the former scalar field effect 
is more complicated to estimate and it depends on the situation. 
In the even-parity case, $\vartheta$ sourced by the secondary body 
at the position of the primary body is proportional to $1/b$. 
In EDGB theory, since $\vartheta$ has shift symmetry 
within the context of the classical theory, 
the effects are suppressed by the gradient of the field, i.e.~they are 
proportional to $1/b^2$. In the odd-parity case, there is again shift symmetry 
and the monopole scalar charge is absent. Because of these two 
reasons, the suppression is proportional to $1/b^3$ in CS theory. 
These suppressions will be sufficient to conclude that the effects are 
relatively at least 1PN and 3PN in the even and odd-parity cases respectively, 
which is smaller than the effects induced by scalar radiation. 

In the odd-parity non-spinning case, the latter gravitational tidal force 
dominates over the scalar propagation effect.
To calculate this tidal force properly requires asymptotic matching between the IZ
solution and a strong-field, perturbed Schwarzschild solution in CS gravity. Perturbations
of the Schwarzschild spacetime can be decomposed as a sum over electric and magnetic
tidal tensors (see e.g.~\cite{Poisson:2005pi}). 
The former scale as $1/b^{3} (1 + v + v^{2} + \ldots)$, while the latter scales
as $v/b^{3} (1 + v + v^{2} + \ldots)$~\cite{JohnsonMcDaniel:2009dq}. 
Such tidal deformations will induce gravitational waves
that will scale as the second-time derivatives of the electric and magnetic quadrupole deformations, 
i.e.~they will scale as $\omega^{2}/b^{3} (1 + v + v^{2} + \ldots)$ and 
$\omega^{2} v/b^{3} (1 + v + v^{2} + \ldots)$. In GR, the leading order effect is induced by the
electric quadrupole moment and it scales as $\omega^{2}/b^{3}$, a 5PN order effect. In CS, 
we expect the magnetic quadrupole moment to provide the leading-order deformation, and
at the level of energy flux, this couples to the GR metric perturbation produced by either the mass octupole or current dipole moment, leading to 6PN correction.
This interpretation seems to be consistent with the results of Pani et al.~\cite{Panietal}.
 
 
\section{Energy flux}
\label{sec:E-flux}

The inspiral of a compact binary system is controlled by the system's change in binding energy and angular momentum. The binding energy changes according to the dissipation of energy carried by all dynamical fields, which here includes the metric perturbation and the scalar field. The stress-energy tensor (SET) associated with each field quantifies the density and flux of energy and momentum. The energy loss is calculated as the integral of the energy flux through a 2-sphere of radius $r$ in the limit $r\to\infty$ and in the
direction of the sphere's outward unit normal $n^i$. That is, for some field $\varphi$ (be it $\hGR_{ij}$, $\hDef_{ij}$, or $\vartheta$) with SET $T^{(\varphi)}_{\mu\nu}$,
\be
\label{eq:Edotdefinition}
\dot{E}^{(\varphi)} = \lim_{r\to\infty} \int_{S^2_r} 
\left<T^{(\varphi)}_{ti} n^i \right>_{\omega}  r^2 d\Omega\,,
\ee
where the angle brackets with subscript $\omega$ stand for orbit averaging. 

The total energy flux can be ordered in powers of $\varsigma$ as
\be
\dot{E} = \dot{E}_{\GR} + \varsigma \, \delta \dot{E} + {\cal{O}}(\varsigma^{2})\,.
\ee
The GR energy flux $\dot{E}^{\GR}$ is given by the GR metric perturbation only, without any contributions from the scalar field at ${\cal{O}}(\varsigma^{0})$, as there is no scalar field in GR. For circular orbits, this is
\be
\dot{E}_{\GR} = - \frac{32}{5} \eta^{2} v^{10}\,. 
\label{EdotGR}
\ee
The $\mathcal{O}(\varsigma)$ correction, $\delta \dot{E}$, can be decomposed into
\be
\delta \dot{E} = \delta \dot{E}^{(\vartheta)} + \delta \dot{E}^{(\hDef)}\,,
\ee
where the first term is the scalar field contribution and the second term is the contribution of the deformed metric perturbation. 

The scalar field contribution is calculated with the SET given by Eq.~\eqref{theta-Tab}:
\be
\label{eq:EdotdefinitionTheta}
\delta \dot{E}^{(\vartheta)} = \beta \lim_{r\to\infty} \int_{S^2_r} 
\left< \dot{\vartheta} \; n^i \; \partial_{i} \vartheta \right>_{\omega} r^2 d\Omega\,.
\ee
Since we are taking the $r \to \infty$ limit, $\vartheta$ must be that valid in the FZ.

The metric deformation contribution to the energy flux is slightly more subtle. 
This modification to the GR flux can have three distinct sources:
(i) the effective SET in terms of $\hGR_{ij}$ and $\hDef_{ij}$ may be functionally different, but as shown in~\cite{Stein:2010pn}, this is not so for the class of theories we consider here\footnote{Reference~\cite{Stein:2010pn} showed that the TT gauge exists in quadratic gravity as $r\to\infty$. Any non-TT propagating mode that is sourced in the NZ vanishes in the FZ at all orders. This is in contrast to scalar-tensor theories in the Jordan frame, where the scalar ``breathing'' mode is present in the metric. This difference comes from the way the metric deformation and the scalar field couple in the field equations. In the quadratic gravity case, $\vartheta$ does not multiply $G_{\mu\nu}$ in the field equations (the Einstein-Hilbert sector of the action is unmodified), while the opposite is true in scalar-tensor theories in the Jordan frame. Therefore, in the former $\hDef_{\mu\nu}$ and $\vartheta$ decouple in the $r\rightarrow \infty$ limit and there is no breathing mode. In contrast, in the latter the coupling between $\hDef_{\mu\nu}$ and $\vartheta$ remains in the limit $r \rightarrow \infty$, leading to a non-vanishing breathing mode and a modification to the effective SET.};
(ii) The orbital equations of motion, and the associated relations
$m/b=v^2$ and $\omega=v^3/m$, might be modified at
$\mathcal{O}(\varsigma)$, as was partially calculated in~\cite{Yunes:2011we};
(iii) The generation mechanism of the FZ metric perturbation is modified, i.e.~the radiative part of the metric perturbation is deformed. We consider here only the dissipative modifications introduced by (iii), as (ii) would require an analysis of the equations of motion, which is beyond the scope of this paper\footnote{  The distinction between (ii) and (iii) can be ambiguous at higher PN order, because how the orbital parameters are modified depends on the gauge choice. However, as long as we impose the harmonic gauge condition on both GR and the deformed metric perturbations, we do not have to worry about this gauge issue at least up to next-to-leading PN order.
}. 

Letting $H_{\alpha \beta} = h_{\alpha \beta} + \varsigma \hDef_{\alpha \beta} + {\cal{O}}(\varsigma^{2})$, the effective SET of GWs is given by~\cite{Stein:2010pn}
\be
T_{\mu \nu}^{{(H)}} = \frac{1}{32 \pi}  \left< H^{\TT}_{\alpha \beta,(\mu} H_{\TT}^{\alpha \beta}{}_{,\nu)}\right>_{\lambda} \,,
\ee
where the angle brackets with a subscript $\lambda$ stand   for a quasi-local average over several wavelengths and
TT stands for the transverse-traceless projection 
\be
H^{\TT}_{ij} = \Lambda_{ij,kl} H_{kl}\,, \qquad
\Lambda_{ij,kl}=P_{ik} P_{jl}-\frac{1}{2} P_{ij} P_{kl}\,,
\ee
with $P_{ij}=\delta_{ij}-n_{ij}$ the projector onto the plane perpendicular to the line from the source to a FZ field point. Expanding this SET in orders of $\varsigma$, the ${\cal{O}}(\varsigma^{0})$ part leads to $\dot{E}_{\GR}$, while the ${\cal{O}}(\varsigma)$ part is
\be
T_{\mu \nu}^{{(\hDef)}} = \frac{1}{16 \pi}  \left< \hGR^{\TT}_{\alpha \beta,(\mu} \hDef_{\TT}^{\alpha \beta}{}_{,\nu)}\right>_{\lambda} \,,
\ee
which leads to
\be
\label{eq:EdotdefinitionhDef}
\delta \dot{E}^{(\hDef)} = \frac{1}{16 \pi}  \lim_{r\to\infty} \int_{S^2_r} 
\left< \left< \hGR^{\TT}_{\alpha \beta,(t} \hDef_{\TT}^{\alpha \beta}{}_{,i)}\right>_{\lambda} 
 \; n^i \right>_{\omega} r^2 d\Omega\,.
\ee
As before, the $\hGR_{\alpha \beta}$ and $\hDef_{\alpha \beta}$ are those valid in the FZ.

\subsection{Scalar field correction to the energy flux}

\subsubsection{Even-parity sector}
\label{sec:Edottheta-even}

In the even-parity case, $\vartheta^{\FZ}$ is dominated by the
dipole component [Eq.~\eqref{eq:thetaFZdipole}], which we
repeat here for convenience: $\vartheta^{\FZ} = \dot{D}_i n^i/r$,
where $D_i$ is the NZ dipole given in
Eq.~\eqref{eq:dipoledef}. This is inserted into the energy
loss formula, Eq.~\eqref{eq:EdotdefinitionTheta}. Since 
the FZ scalar field depends on retarded time, both time
and spatial derivatives can be written as time derivatives
of the NZ moments. This gives
\begin{equation}
\delta \dot{E}^{(\vartheta)} = -\beta \int_{S^{2}_{\infty}} \left\langle \ddot{D}_i \ddot{D}_j n^{ij} \right\rangle_{\omega} d\Omega = -\frac{4\pi}{3}\beta\left<\ddot{D}^i\ddot{D}_i\right>_\omega\,,
\end{equation}
which for circular orbits gives
\begin{equation}
\delta \dot{E}^{(\vartheta)} = -\frac{4\pi}{3}\beta\omega^4|D|^2 
= -\frac{4\pi}{3}{\beta\over m^4}(m_2 q_1 - m_1 q_2)^2 v^8\,.
\label{eq:Edotdipolecirc}
\end{equation}
Note that here, as before, the $m\to 0$ limit diverges, 
because the effective theory breaks down on short length scales
and $\varsigma \ll 1$ is violated.

When the compact bodies are BHs, their scalar monopole charges
are given by Eq.~\eqref{eq:BHscalarcharge}, $q_A = 2\alpha_3/(\beta
m_A)$, which then leads to 
\be
\delta \dot{E}^{(\vartheta)} = - \frac{1}{3}\zeta_3 \frac{1}{\eta^2}
\frac{\delta m^2}{m^2} v^8\,. 
\ee
 
Comparing this with the GR energy
flux, we find
\be
\label{eq:ratioeven}
\frac{\delta  \dot{E}^{(\vartheta)}}{\dot{E}^\GR} = \frac{5}{96} \zeta_3
\frac{1}{\eta^4} \frac{\delta m^2}{m^2} v^{-2}\,,
\ee
a relative -1PN effect. That is, the energy lost to the scalar field due to
dipole radiation would enter as a lower-order in $v$ effect than the energy
loss in GR.
If one takes the limit $m_2\to\infty$ while keeping $(m_1,v)$ fixed,
then the above ratio scales as $m_1^{-4}$; i.e. the energy flux ratio
is sensitive to the smallest horizon scale of the system. The effect
is of a similar size for comparable stellar-mass binary and EMRI
system. A SMBH-SMBH binary experiences the smallest effect.

\subsubsection{Odd-parity sector: spinning bodies}
\label{sec:Edottheta-odd-spins}

The scalar field $\vartheta^{\FZ}$ is here dominated by the
quadrupole component [Eq.~\eqref{eq:thetaFZquadrupole}], 
which we repeat here for convenience 
$\vartheta^{\FZ} = \ddot{\mu}_{ij}  n^{ij}/r = -\omega^2 \mu_{ij} n^{ij}/r$,
where the quadrupole tensor $\mu_{ij}$ is defined in
Eq.~\eqref{eq:quadrupoledef}. Inserting this into the energy loss
formula [Eq.~\eqref{eq:EdotdefinitionTheta}] gives
\begin{align}
\delta  \dot{E}^{(\vartheta)} =&\, -\beta \int_{S^2_\infty}
\left\langle \dddot{\mu}_{ij}\dddot{\mu}_{kl} n^{ijkl} \right\rangle_{\omega} d\Omega\,, \nn\\
=&\, -\frac{4\pi}{15}\beta
\left\langle \left[ 2 \dddot{\mu}_{ij}\dddot{\mu}^{ij} + \left(\dddot{\mu}^i{}_i\right)^2 \right] \right\rangle_{\omega}\,.
\end{align}

Let us evaluate this for quasi-circular orbits with non-precessing spins. The third time derivative of the quadrupole tensor $\mu_{ij}$ becomes
\be
\dddot{\mu}^{ij} = b^{-3} 
  \left( m_1 v_{12}^{(i} \mu_2^{j)} - m_2
  v_{12}^{(i} \mu_1^{j)} \right)\,,
\ee
and the total energy flux is
\be
\delta  \dot{E}^{(\vartheta)} = - \frac{5}{48} \zeta_4 \left[ \bar{\Delta}^2+2
\left<  (\bar{\Delta}\cdot \hat{v}_{12})^2\right>_{\omega} \right] v^{14}\,,
\label{deltaEdotvartheta-odd-spins}
\ee
where $\hat{v}_{12}$ is the unit vector in the direction of the
relative velocity and the dimensionless quantity $\bar{\Delta}$ is defined as
\be
\bar{\Delta}^i \equiv \frac{m_2}{m} \chi_1 \hat{S}_1^i - \frac{m_1}{m} \chi_2 \hat{S}_2^i\,.
\label{delta}
\ee
Notice that $\delta \dot{E}^{(\vartheta)}$ in Eq.~\eqref{deltaEdotvartheta-odd-spins} is finite in the EMRI limit. Note also that when both spins are perpendicular to the orbital plane, $\bar{\Delta}$ is as well, and the second term of $\delta \dot{E}^{(\vartheta)}$ vanishes.  Comparing Eq.~\eqref{deltaEdotvartheta-odd-spins} with GR,
\be
\frac{\delta \dot{E}^{(\vartheta)}}{\dot{E}^\GR} =
\frac{25}{1536}\zeta_4 \frac{1}{\eta^2}\left[ \bar{\Delta}^2+2
\left<  (\bar{\Delta}\cdot \hat{v}_{12})^2 \right>_{\omega} \right] v^4\,,
\label{theta-Edot-spin}
\ee
hence scalar radiation in the odd-parity sector is clearly a relative
2PN effect.
This effect was not included in the work of 
Pani~et~al.~\cite{Panietal}, who found a $7$PN correction, 
since their simulations did not include spins.
If one takes the limit $m_2\to\infty$ while keeping $(m_1,v)$ fixed,
then the above ratio scales as $m_1^{-2} m_2^{-2}$; i.e. the energy
flux ratio is sensitive to the geometric mean of the two horizon
scales in the system. This implies that the effect is greatest for
comparable stellar-mass binaries.

\subsubsection{Odd-parity sector: non-spinning bodies}
\label{sec:Edottheta-odd-no-spin}

The odd-parity $\vartheta_{\FZ}$ in Eq.~\eqref{theta-sol-CS} 
can be used to evaluate the energy loss in Eq.~\eqref{eq:EdotdefinitionTheta}:%
\begin{align}
\delta \dot{E}^{(\vartheta)} 
=&\, -256 \kappa \zeta_4 \delta m^{2} \eta^{2}  \left(\frac{m}{b} \right)^{8} \! 
\int_{S_{\infty}^{2}} \! \! \! \! \! d\Omega \left[\partial_{t} \! \left(\epsilon^{ijk} n^{ip}  v_{12}^{k} n_{12}^{jp}\right) \right]^{2}
\nonumber \\
=&\,  -256  \kappa \zeta_4 \eta^{2} \frac{\delta m^{2}}{m^{2}} \left(\frac{m}{b} \right)^{10} \! \!
\int_{S_{\infty}^{2}} d\Omega \left(\epsilon^{ijk} n^{ip} v_{12}^{kp} n_{12}^{j} \right) ^{2}
\nonumber \\
=&\, -\frac{64}{15} \zeta_4 \eta^2 \frac{\delta m^{2}}{m^{2}} \left( \frac{m}{b} \right)^{12}\,.
\end{align}
Compared to the GW radiation in GR [Eq.~\eqref{EdotGR}], this scalar radiation becomes
\be
\frac{\delta \dot{E}^{(\vartheta)}}{\dot{E}^{\GR}} =  \frac{2}{3} \frac{\delta m^{2}}{m^{2}} \zeta_4 \; v^{14}\,,
\label{eq:Edottheta-odd-nospin-rel}
\ee
which shows that this is a relative 7PN effect. In contrast with the
cases of even-parity and odd-parity with spins, this effect is
dominantly controlled by the total mass, rather than the mass
ratio. The effect is greatest for a system of stellar-mass BHs.

\begin{figure}
 \includegraphics[width=8.5cm]{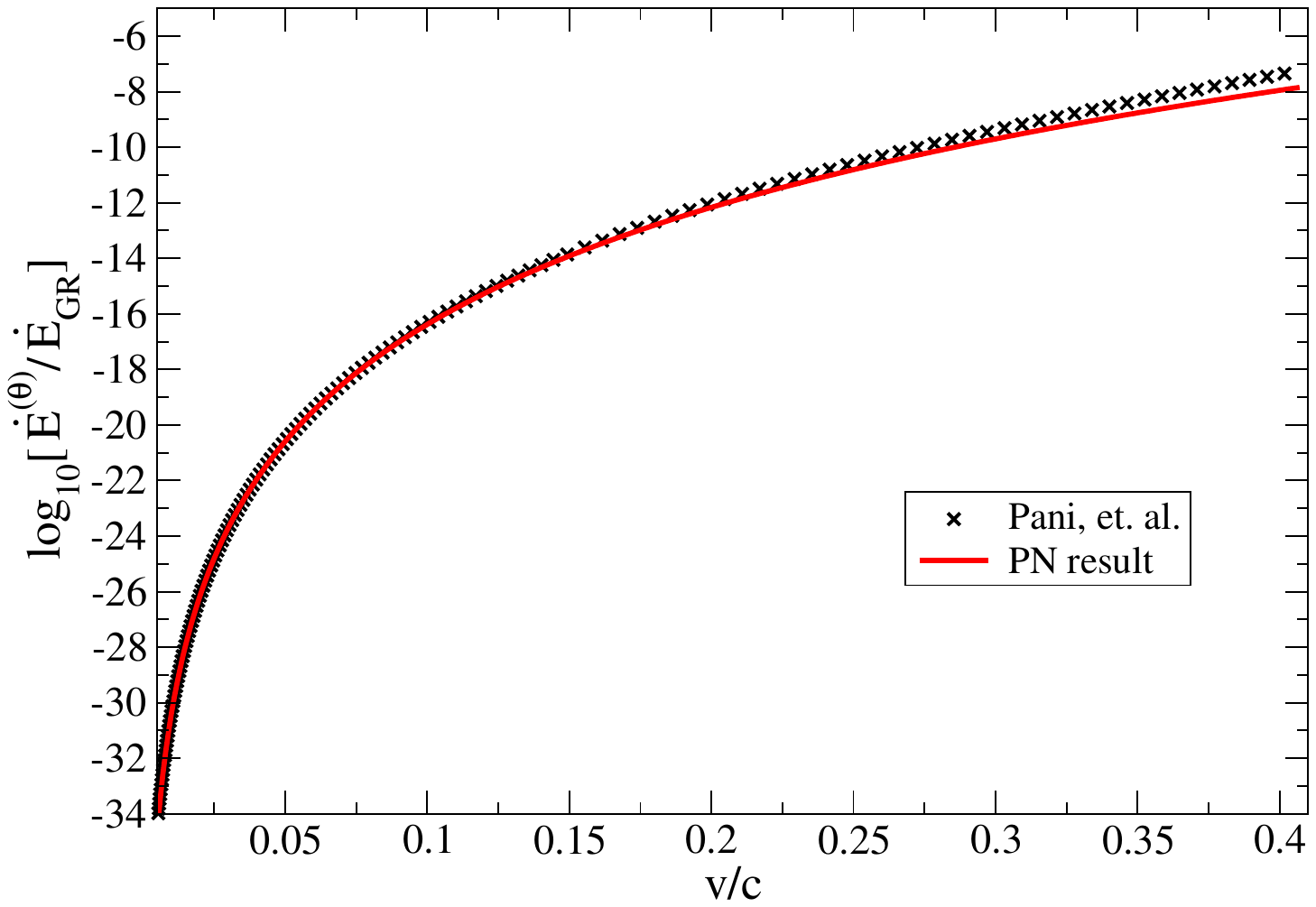}
 \caption{\label{fig:comparison} Comparison of
Eq.~\eqref{eq:Edottheta-odd-nospin-rel} to the numerical results of Pani~et~al.~\cite{Panietal}. The latter can be mapped to the generic
quadratic gravity action of Eq.~\eqref{action} by letting $\alpha_4 =
-\alpha_{\CS}/4$, which then implies that $\zeta_4 = -\zeta_{\CS}/16$. 
We here used $\zeta_4 = 6.25 \times 10^{-3}$, which is equivalent to their parameter $\zeta_{\CS} = 0.01$. Observe that at low velocities, in the regime where the PN approximation is valid, the two curves agree. }
\end{figure}

The above result can be compared to numerical calculations recently
performed by Pani~et~al.~\cite{Panietal}. They estimated the effect of
scalar radiation in dynamical CS gravity~\cite{Alexander:2009tp} 
for non-spinning, circular EMRIs. 
They numerically solved the master perturbation equations on 
a Schwarzschild background to obtain the time evolution of the 
scalar field and the metric perturbation, caused by a non-spinning 
point particle. Figure~\ref{fig:comparison} compares their results to
ours, found in Eq.~\eqref{eq:Edottheta-odd-nospin-rel}. Observe that the numerical results of Pani~et~al.\ are in excellent agreement with our post-Newtonian calculation, which extends it to comparable mass-ratios (notice the factor of $\delta m/m$).  
%

\subsection{Metric deformation correction to the energy flux}

For the even-parity case, the correction to the energy flux that arises
from the deformation to the gravitational metric perturbation is at least of 0PN
order relative to GR. This is higher PN order compared to the
scalar dipole radiation found in Sec.~\ref{sec:Edottheta-even}, and thus, we will not consider it further.

For the odd-parity case with spinning BHs, one of the leading contribution comes from the metric correction sourced by $T_{ij}^{(\vartheta)}$, which is given in Eq.~\eqref{h-odd-a2}. Inserting this metric perturbation into Eq.~\eqref{eq:EdotdefinitionhDef}, the energy flux correction relative to GR becomes
\begin{align}
\frac{\delta\dot{E}^{(h)}}{\dot{E}^{\GR}} &= \frac{75}{16} 
\frac{\zeta_{4}}{\eta} \chi_1\chi_2 
\left<\hat{S}_{1}^{i} \hat{S}_{2}^{j} \left( 2 \hat{v}^{12}_{ij} - 3 n^{12}_{<ij>} \right)
\right>_\omega v^{4}\,,
\end{align}
which is of relative 2PN order, just as the contribution due to scalar radiation
in Eq.~\eqref{theta-Edot-spin}. Notice that both the metric deformation and scalar field corrections to the energy flux are of $\mathcal{O}(\chi^2)$, but 
the latter is larger by a factor of ${\cal{O}}(\eta^{-1})$.
 
We expect $\mathcal{O}(\chi)$ corrections to the energy flux due to the metric deformation to be higher PN order. For very slowly spinning binaries, however, they may give larger corrections compared to
the $\mathcal{O}(\chi^2)$ 2PN ones presented here.

In the odd-parity sector with non-spinning objects, 
the regularized contributions to the metric deformation can only
provide energy flux corrections of at least 7PN order. However,
as explained in Sec.~\ref{sec:multipolemoments}, we expect that
matching strong-field solutions to the non-regular NZ ones may
generate 6PN corrections in the energy flux, similar to those found
by Pani et al.~\cite{Panietal}.

\section{Impact on gravitational wave phase}
\label{sec:Impact}

How do all these modifications to the energy flux affect the GW observable? To answer this question, we compute the Fourier transform of the phase of the GW response function in the stationary phase approximation (SPA), where we assume the GW phase changes much more rapidly than the GW amplitude~\cite{Droz:1999qx}.

We begin by parameterizing all the corrections to the energy flux that we have studied so far via the following power law: 
\be
\dot{E}=\dot{E}_{\mathrm{GR}}(1+Av^a)\,,
\label{eq:PPE}
\ee
where $(A,a)$ are summarized in Table~\ref{tab:PPEparams} for the
four different sectors considered.

With the generic energy flux parameterization, the orbital phase for a quasi-circular inspiral becomes
\ba
\phi(F) &\!\!  = &\!\!  \int^{F}\frac{dE}{d\omega} \left( \frac{dE}{dt} \right)^{-1} \omega d\omega \nonumber \\
&\!\!  = &\!\!  \phi_{\mathrm{GR}}(F) \left[ 1+\frac{5}{a-5}A (2\pi m F)^{a/3} \right]\,,
\label{phiofF}
\ea
where $F$ and $\omega=2 \pi F$ are the linear and angular orbital frequency, $\phi_{\mathrm{GR}}=-1/(32\eta)(2\pi m F)^{-5/3}$ is the GR orbital phase and $E(\omega) = -(\mu/2) (m\omega)^{2/3}$ is the binary's binding energy to Newtonian order. Recall here that $m=m_{1}+m_{2}$ is the total mass of the binary, while $\mu = m_{1} m_{2}/m$ is the reduced mass and $\eta = \mu/m$ is the symmetric mass ratio. Equation~\eqref{phiofF} is not valid when $a = 5$ (a 2.5PN correction), as then the integrand becomes proportional to $\omega^{-1}$, which leads to a log term. 

Before we compute the Fourier phase, we must first define $t_0$, the time at which the stationary phase condition is satisfied $F(t_0)=f/2$, where $f$ is the GW frequency. This condition can be solved to yield
\be
t_0=t_{0,\mathrm{GR}}\left(1-\frac{8}{8-a}A(\pi mf)^{a/3} \right)\,,
\ee
where $t_{0,\mathrm{GR}}$ is the GR $t_0$. Again, this expression is not valid at $a = 8$, because once more the correction to $t_{0}(f)$ would be a log term.

With this at hand, we can now compute the Fourier phase in the SPA:
\begin{align}
\Psi_{\mathrm{GW}} =&\, 2\phi(t_0)-2 \pi f t_0 \nonumber \\
=&\,  \Psi_{\mathrm{GR}}\left[ 1-\frac{40}{(a-5)(a-8)}A \eta^{-a/5} (\pi \mathcal{M} f)^{a/3}  \right]\,,
\end{align}
where $\Psi_{\mathrm{GR}} \equiv (3/128) (\pi \mathcal{M} f)^{-5/3}$, and
where $\mathcal{M} = \eta^{3/5} m$ is the chirp mass. Again, these expressions are not valid when $a = 5$ or $a = 8$, for the reasons described above. 

The corrections to the GW phase found here map directly to the parameterized post-Einsteinian (ppE) framework~\cite{Yunes:2009ke}. In that framework, one postulates that modified gravity theories affect the Fourier phase of the GW response function in the SPA via
\be
\label{ppE-phase}
\Psi^{\ppE}_{\GW} = \Psi_{\mathrm{GR}} + \beta_\ppE \left( \pi \mathcal{M} f\right)^{b_\ppE} \,,
\ee
where $(\beta_\ppE,b_\ppE)$ are ppE parameters. We see that this is identical to the corrections introduced by a change in the energy flux, with the mapping
\be
\beta_\ppE = - \frac{15}{16} \frac{A}{(a-5)(a-8)} \eta^{-a/5}\,,
\qquad
b_\ppE = \frac{a-5}{3}\,.
\label{eq:ppEmapping}
\ee
This is not surprising, as the ppE framework was in part motivated by studying power-law (in velocity) modifications to the energy flux and the binding energy~\cite{Yunes:2009ke}. 

We have then found that a large number of energy flux corrections associated with extra gravitational and scalar field emissions can be mapped to the ppE framework. In the even parity case, the leading-order frequency exponent $b_{\ppE} = -7/3$, while in the odd-parity case $b_{\ppE} = -1/3$, unless the binary is non-spinning in which case $b_{\ppE} = +3$. 

{
\renewcommand{\arraystretch}{1.2}
\begin{table}
\capstart
\begin{centering}
\begin{tabular}{rccl}
\hline
\hline
\noalign{\smallskip}
\multicolumn{1}{c}{Sector} && $A$ & \multicolumn{1}{c}{$a$}  \\
\cline{1-1}\cline{3-4}
\noalign{\smallskip}
Even-Parity && $\frac{5}{96} \zeta_3\frac{1}{\eta^4} \frac{\delta m^2}{m^2}$ & $-2$ \\
Odd-P, Spins ($\vartheta$),  && $\frac{25}{1536}\zeta_4 \frac{1}{\eta^2}\left[ \bar{\Delta}^2+2
  \left<(\bar{\Delta}\cdot \hat{v}_{12})^2\right>_\omega \right]$ & $+4$ \\
Odd-P, Spins ($h$),  &&
$\frac{75}{16} 
\zeta_{4}\frac{1}{\eta} 
\langle
{S}_{1}^{i}{S}_{2}^{j} \left( 2 \hat{v}^{12}_{ij} - 3 n^{12}_{<ij>} \right)
\rangle_\omega
$
& $+4$ \\
Odd-P, No Spin && $\frac{2}{3}  \zeta_4\frac{\delta m^{2}}{m^{2}} $ 
& $+14$ \\
\noalign{\smallskip}
\hline
\hline
\end{tabular}
\end{centering}
\caption{Coefficients of the relative energy flux.}
\label{tab:PPEparams}
\end{table}
}

The results found in this paper could help in the generalization of the ppE framework to more generic quasi-circular inspirals. The original framework considered only non-spinning, equal mass inspirals, while recently
Cornish~et~al.~\cite{Cornish:2011ys} generalized it
to non-spinning, unequal mass systems through $A \to A\eta^c$.
In this paper we have found that $A$ does not only 
depend on a simple power law of $\eta$, but also on the mass
difference $\delta m/m =\sqrt{1-4\eta}$ and on combinations of the spins. 
For single detections, however, such a generalization is not needed
as one only measures a single number, $\beta_\ppE$, and one 
cannot extract the dependencies on $\eta$,
$\delta m/m$, and the spins.

Although we currently lack any GW detections, we can still estimate the projected constraints that such detections would place on quadratic gravity. According to Table~\ref{tab:PPEparams}, the even-parity sector leads to the strongest deviations from GR, since $a$ is the most negative. Therefore, we consider EDGB theory, $(\alpha_1,\alpha_2,\alpha_3,\beta) = (1,-4,1,\alpha_{\rm{EDGB}}^{-1})\alpha_{\mathrm{EDGB}}$, as a simple sub-case of the even-parity sector. Let us first imagine that we have detected a GW with Ad.~LIGO and signal-to-noise ratio (SNR) of 20 that is consistent with GR and that originates from a non-spinning BH binary with masses $(m_1,m_2)=(6,12)M_{\odot}$. Given such a detection, Cornish \textit{et al.}~\cite{Cornish:2011ys} estimated the projected bound $|\beta_{\mathrm{ppE}}| \lesssim 5 \times 10^{-4}$ for $b_{\mathrm{ppE}} = -\frac{7}{3}$, which implies $|\alpha_{\mathrm{EDGB}}|^{1/2} \lesssim 4\times 10^5 \; {\mathrm{cm}}$. Let us now assume that we have detected a GW with LISA classic with and SNR of $879$   and still consistent with GR, but that originates from a non-spinning BH binary with masses  $(m_1,m_2)=(10^6,3 \times 10^6)M_{\odot}$ at $z=1$. Given such a detection, Cornish \textit{et al.}~\cite{Cornish:2011ys} estimated a bound on $|\beta_{\mathrm{ppE}}| \lesssim 10^{-6}$ for the same value of $b_{\mathrm{ppE}}$ as before, which leads to $\alpha_{\rm{EDGB}}^{1/2} \lesssim 10^{10} \; {{\rm cm}}$. In both cases, notice that these projected bounds are consistent with the small-coupling requirement $\zeta_i \ll 1$; i.e.~saturating the projected Ad.~LIGO and LISA constraints we have $\zeta_{\rm{Ad.~LIGO}} \sim 3 \times 10^{-2}$ and $\zeta_{\rm{LISA}} \sim 10^{-5}$ for those particular binary systems, which is clearly much less than unity.

Comparing these results with the current constraint obtained by the
Cassini satellite, $|\alpha_{\mathrm{EDGB}}|^{1/2} < 1.3 \times
10^{12}$cm~\cite{Amendola:2007ni}, we see that Ad.~LIGO and LISA could
constrain $\alpha_{\mathrm{EDGB}}$ much more strongly. Unfortunately,
it seems difficult to put constraints on EDGB with binary pulsar
observations, since NSs have no scalar monopole charge in this theory. We
emphasize again that this is opposite to the expectation from
scalar-tensor theories, in which NSs have scalar monopole charges while BHs do
not. Finally, one cannot estimate the bounds one could place on
dynamical CS gravity, since one would have to properly account for
modifications to the conservative equations of motion, which we have
not calculated here.

\section{Conclusions and discussions}
\label{sec:Discussions}

We have studied the binary inspiral problem in a wide class of
quadratic gravity theories in the slow-motion, weak-gravity regime.
The structure of a compact object in such theories affects the
exterior scalar field sourced by the object. Despite this, we 
can model a compact object by an effective scalar field source 
characterized by its scalar monopole and dipole moments.
The scalar monopole charge is enhanced inversely proportional to 
the mass of the object, while the dipole charge 
is independent of the mass for a fixed dimensionless spin parameter. 
With this effective source, we then derived and
solved the modified field equations for the scalar field and metric
deformation.

We find that the scalar field generically emits 
dipole radiation in the even-parity sector, 
and quadrupole radiation in the odd-parity
sector. Such radiation affects the rate of change of the binary energy
at relative $-1$PN order in the even-parity case and relative $2$PN order in
the odd-parity case. The quadrupole contribution depends
quadratically on the BH spins,   and thus it is suppressed for non-spinning
binaries. In that case, the odd-parity contribution becomes of relative 
$7$PN order, as found numerically in~\cite{Panietal}. We have found excellent
agreement between their numerical results and our analytical calculations.

We have also calculated the metric perturbation in the FZ and its associated energy flux. 
In the even-parity sector, the dominant metric contribution leads to a 0PN relative correction
in the energy flux, which is smaller than the -1PN correction induced by scalar dipolar radiation.
In the odd-parity sector and for spinning BHs, the metric perturbation leads to a 2PN modification 
to the energy flux, which is of the same order as that induced by quadrupolar scalar radiation.
In the odd-parity sector and for non-spinning BHs, we expect the energy flux correction due to the metric
deformation is suppressed to at least of 6PN order, as found by Pani \textit{et al.}~\cite{Panietal}.

Whether these corrections can be measured or constrained depends on whether they are degenerate with GR
terms in the physical observable, i.e.~the waveform. A $-1$PN effect cannot be degenerate, as
there are no such terms predicted in GR. A $2$PN effect, however, could be degenerate with a 
spin-spin interaction for quasi-circular inspirals with aligned or counter-aligned spin components. 
That is, a renormalization of the spin magnitudes of both bodies can eliminate this $2$PN
effect, assuming one truncates the waveform at that order. If higher-order PN waveforms are used,
or if the orbit is more generic (i.e.~if there is precession or eccentricity), then this degeneracy can be broken.

We also calculated the effects of such energy flux modifications on the
gravitational waveform. The waveform phase depends sensitively on the rate of change
of the orbital frequency, which in turn is governed by the rate of
change of energy. We calculated the corrections that would be induced in
the waveform and mapped them to the ppE framework. We then used a recent
ppE study~\cite{Cornish:2011ys} to estimate the constraints that
Ad.~LIGO and LISA could potentially place on quadratic gravity
theories. Given a GW detection, we found that the magnitude of the new 
length scale introduced by quadratic gravity theories (associated with a ratio of their coupling 
constants) could constrain at a level controlled by the smallest length-scale probed in the 
inspirals, i.e.~the size of the smallest compact object's event horizon or surface. The best
projected bounds achievable with Ad.~LIGO will thus come from stellar-mass 
BH or NS inspirals, while LISA will benefit the most from EMRIs. Since NSs have no scalar 
monopole charge in EDGB theory, this theory cannot be constrained from binary pulsar observations. 
This property is diametrically opposite to scalar-tensor theories where BHs have no hair. 

There are several possible avenues for future work. Since we here mainly considered corrections due to the dissipative sector of the theory, one possibility is to calculate the non-dissipative corrections that would modify the binding energy (here, we mean both gravitational and scalar binding energy) and the equations of motion. There are two effects that should be accounted for: new scalar-scalar forces and metric deformations. Let us consider the former first. In the even parity case, compact objects have an associated scalar monopole charge, and thus, there is an additional scalar force with a $1/r$ potential that should lead to a relative 0PN non-dissipative correction. Similarly, in the odd-parity case, a spinning compact body possesses a current dipole charge, and hence, dipole-dipole interactions should arise. Since the dipole potential is proportional to $1/r^{2}$, while the dipole charge couples to the first derivative of the potential, the binding energy and the equations of motion should be corrected at relative 2PN order.  

Another non-dissipative modification is induced by deformations of the background metric tensor.
In the even-parity sector, such corrections enter at relative 0PN order, as found by Yunes and Stein~\cite{Yunes:2011we}. In the odd-parity sector, there is no metric deformation for isolated non-spinning BHs, but for spinning ones there is a correction proportional to $r^{-4}$ to the $(t,i)$ components~\cite{Yunes:2009hc}, which then leads to a $4.5$PN correction in the equations of motion when we consider boosted BHs. This then implies the following: (i) in the even-parity case, the conservative corrections to the equations of motion do not affect the leading-order modification to the waveforms, since this is dominated by the $-1$PN scalar radiation effect; (ii) in the odd-parity case, the conservative corrections from the metric deformation can be neglected, but those due to the scalar-scalar force will contribute at the same order as the effect calculated here. A complete analysis of the waveform observable would thus require the calculation of such a scalar-scalar, conservative effect. 

Another possibility could be to study modified quadratic gravity in the context of BH perturbation theory. This would be a tremendous effort that would have to be split into separate parts. First, one would have to find an analytic, strong-field solution for arbitrarily-fast rotating BHs in quadratic gravity. This has only been found in the slow-rotation limit both in the even-parity~\cite{Yunes:2011we}   and odd-parity sectors~\cite{Yunes:2009hc}. Once this is accomplished, one would have to study the evolution of metric perturbations away from this solution. Such evolution equations would have to be decoupled in terms of some master function to derive Teukolsky-like master equations. Finally, with these equations at hand, one would have to solve them numerically, when the perturbations are sourced by a small object in a tight orbit. Such an analysis would be interesting because one would be able to derive not only the corrections to the energy flux carried out to infinity, but also that which is absorbed by the BH horizons and which we ignored in this paper. 

A final follow-up would be to study how NS solutions are modified in quadratic gravity~\cite{Pani:2011xm,AliHaimoud:2011fw}   and how the energy flux from NS binaries is modified. This could then lead to direct constraints on quadratic gravity theories from double binary pulsar observations. Such constraints could be stronger  , relative to current Solar System constraints, as they could potentially provide constraints of roughly the order of magnitude of the NS radius. Of course, in the case of EDGB theory or dynamical CS gravity, these constraints might not be stronger as NSs have no scalar monopole charge in such theories.

\acknowledgments We would like to thank Cliff Will, Eric Poisson and
Luc Blanchet for invaluable teachings regarding the PN
approximation. We also acknowledge Richard O'Shaughnessy, Yanbei Chen,
Yacine Ali-Ha\"{i}moud, and Misao Sasaki for useful comments on the
manuscript, and Paolo Pani, Vitor Cardoso and Leonardo Gualtieri for
making some of their data accessible to us. NY thanks the Yukawa
Institute for Theoretical Physics for their hospitality. KY is
supported by the Japan Society for the Promotion of Science (JSPS)
grant No.~$22 \cdot 900$. LCS acknowledges support from NSF Grant
PHY-0449884 and from MIT's Solomon Buchsbaum fund.  NY acknowledges
support from NSF grant PHY-1114374, as well as support provided by the
National Aeronautics and Space Administration through Einstein
Postdoctoral Fellowship Award Number PF0-110080, issued by the Chandra
X-ray Observatory Center, which is operated by the Smithsonian
Astrophysical Observatory for and on behalf of the National
Aeronautics Space Administration under contract NAS8-03060. NY also
acknowledges support from NASA grant NNX11AI49G, under sub-award
00001944. TT is supported by the Grant-in-Aid for Scientific Research
(Nos. 21244033, 21111006 and 22111507). This work is also supported in
part by the Grant-in-Aid for the Global COE Program ``The Next
Generation of Physics, Spun from Universality and Emergence'' from the
Ministry of Education, Culture, Sports, Science and Technology (MEXT)
of Japan.


\appendix

\section{The Balding of Neutron Stars in EDGB Gravity}
\label{sec:balding}

In this appendix, we consider the scalar field equation in EDGB gravity for 
isolated NSs. Integrating the evolution equation, we find
\be
\int \sqrt{-g}  \square \vartheta d^{4}x \propto
\int \sqrt{-g}  {\cal{R}}_{\rm GB}^{2} d^{4}x\,,
\label{first-eq}
\ee
where we have defined the Gauss-Bonnet invariant ${\cal{R}}_{\rm GB}^{2} 
\equiv R^{2} - 4 R_{\mu \nu} R^{\mu \nu} + R_{\mu \nu \delta \sigma} 
R^{\mu \nu \delta \sigma}$. Since the Gauss-Bonnet combination is a
topological invariant, the right-hand side identically vanishes 
for any simply-connected, asymptotically flat spacetime.
Moreover, since we are considering isolated NSs, these must be 
stationary, and so the time integration can be removed. 

With all of this and using Stokes' theorem, Eq.~\eqref{first-eq} becomes
\be
\int \sqrt{-g} (\pd_i \vartheta) n^i dS = \int \sqrt{-g} (\pd_r \vartheta) dS =  0\,,
\label{condition}
\ee
where $n^i$ is the radial unit vector and the integral is performed over 
the 2-sphere at spatial infinity.
Notice that $\sqrt{-g} \sim r^{2}$, while the 
scalar field must decay at infinity 
for it to have a finite energy.

Equation~\eqref{condition} does not vanish at spatial infinity for all scalar 
field solutions, i.e.~ if we model $\vartheta = \vartheta_{n}/r$ with $\vartheta_{n}$ a  constant, then Eq.~\eqref{condition} leads to the unique solution $\vartheta_{n} = 0$. This is a physicists's proof that the EDGB 
scalar field cannot have scalar monopole charge for a spherically symmetric NS.
Similarly, one can show that NSs cannot have scalar monopole charge in dynamical CS gravity; the proof laid out above carries through with the replacement ${\cal{R}}_{\rm GB}^{2} \to \pont$, since $\pont$ is also a topological invariant.

\section{Integration techniques}
\label{int-tech}

In this appendix, we provide some useful integration techniques. 
When computing near-zone integrals, we are faced many times with integrals of the form
\be
\int d^3x \frac{x_{\langle L\rangle}}{r_1 r_2}\,.
\ee
When the point-particle approximation is valid, such near-zone integrals can be Hadamard regularized by keeping only the finite part. Let us then define~\cite{2002PhRvD..65f4005B}
\be
Y_{\langle L\rangle}(\bm{x}_1, \bm{x}_2) = -\frac{1}{2\pi} {\mathop{\mathrm{FP}}_{B=0}} \int d^3x |\tilde{\bm{x}}|^B \frac{x_{\langle L\rangle}}{r_1 r_2}\,,
\ee
to be evaluated in the near-zone and where ${\mathop{\mathrm{FP}}_{B=0}}$ stands for the finite part operator (in the limit $B \to 0$) and $|\tilde{\bm{x}}|$ is an analytic continuation factor~\cite{2002PhRvD..65f4005B}. The solution to this integral is 
\be
Y_{\langle L\rangle}=\frac{b}{l+1} \sum^{l}_{q=0} x_1^{\langle L-Q} x_2^{Q\rangle}\,.
\label{Y}
\ee
The first few $Y_{\langle L\rangle}$ are simply
\begin{align}
\label{Y0-Y1}
Y_{0} =&\, Y = b\,,
\qquad
Y_{i} = \frac{b}{2} \left( x_{1}^{i} + x_{2}^{i}\right)\,,
\\
\label{Yij}
Y_{\langle  ij\rangle} =&\, \frac{b}{3} \left(x_{1}^{\langle  ij\rangle}+
x_{1}^{\langle  i} x_{2}^{j\rangle} + x_{2}^{\langle  ij\rangle} \right)\,,
\\
\label{Yijk}
Y_{\langle  ijk\rangle} =&\, \frac{b}{4} 
\left(x_{1}^{\langle  ijk\rangle} + x_{1}^{\langle  ij} x_{2}^{k\rangle} 
+ x_{1}^{\langle  i} x_{2}^{jk\rangle} + x_{2}^{\langle  ijk\rangle} \right)\,. \qquad
\end{align}

The solution to the $Y_{\langle L\rangle}$ integral can also be derived by using certain Poisson integral identities~\cite{2000PhRvD..62l4015P}:
\be
P\left(f_{,i} g_{,i}\right) = - \frac{1}{2} \left[ f g + P\left(f g_{,ii}\right) + P\left(g f_{,ii}\right) - \mathcal{B}_{p}(fg) \right]\,,
\label{poisson-relations}
\ee
where we have defined
\be
P(f) \equiv \frac{1}{4 \pi} \int_\mathcal{M} \frac{f(t,x')}{|\mb{x} - \mb{x}'|} d^{3}x'\,,
\ee
and the boundary term is
\be
\mathcal{B}_{p}(g) \equiv \frac{1}{4 \pi} \oint_{\partial \mathcal{R}} \left[ \frac{g(t,x')}{|\mb{x} - \mb{x}'|} \partial_{r}' \ln\left[g(t,x') |\mb{x} - \mb{x}'|\right] \right]_{r' = \mathcal{R}} \hspace{-0.5cm} \mathcal{R}^{2} d \Omega'\,.
\ee
As usual, we retain only those terms that are independent of the boundary $\mathcal{R}$.

Finally, there is yet another type of integral that commonly appears in near-zone integration:
\be
\int_{\mathcal{M}} \frac{d^3x'}{|\bm{x}' - \bm{x}_{1}| |\bm{x}' - \bm{x}_{2}| |\bm{x}'-\bm{x}|}\,.
\ee
Let us then define the so-called {\emph{triangle}} potential~\cite{Pati:2002ux}
\be
\mathcal{G}(\bm{x}_1, \bm{x}_2, \bm{x}_3) \equiv \frac{1}{4 \pi} \int_{\mathcal{M}} \frac{d^3x'}{|\bm{x}' - \bm{x}_{1}| |\bm{x}' - \bm{x}_{2}| |\bm{x}'-\bm{x}_3|}\,.
\label{triangle}
\ee
It is a bit of a miracle that the above integral has the closed-form
solution $\mathcal{G}(\bm{x}_A, \bm{x}_B, \bm{x}_C) = 1 - \ln
\Delta(ABC)$, with $\Delta (ABC) \equiv |\bm{x}_{A}-\bm{x}_B|+|\bm{x}_{B}-\bm{x}_C|+|\bm{x}_{C}-\bm{x}_A|$. 

One can show that the triangle potential satisfies a set of relations, including~\cite{Pati:2002ux}
\ba
\partial^{(1)}_i \partial^{(2)}_i \mathcal{G}(\bm{x}_1, \bm{x}_2, \bm{x}) &\!\! =&\!\!  \frac{1}{2} \left[ \frac{1}{b} \left( \frac{1}{r_{1}} + \frac{1}{r_{2}} \right) - \frac{1}{r_{1} r_{2}} \right]\,, \nn
\\ 
\partial^{(1)}_{il} \partial^{(2)}_{jl} \mathcal{G}(\bm{x}_1, \bm{x}_2, \bm{x}) &\!\! =&\!\!  -\frac{1}{2} \left[ \frac{n_{1}^{i} n_{2}^{j}}{r_{1}^{2} r_{2}^{2}}
+ \frac{n_{12}^{i} n_{2}^{j}}{b^{2} r_{2}^{2}} - \frac{n_{12}^{j} n_{1}^{i}}{b^{2} r_{1}^{2}} 
\right.
\nonumber \\
&&\!\! + \left.
3 \frac{n_{12}^{\langle  ij\rangle}}{b^{3}} \left(\frac{1}{r_{1}} + \frac{1}{r_{2}} \right) \right]\,,
\ea
and more generally
\begin{align}
\partial^{(B)}_{i} \partial^{(C)}_{j} \mathcal{G}(ABC) =&\, \frac{1}{\Delta (ABC)^2} (n_{AB}^i-n_{BC}^i) (n_{AC}^j+n_{BC}^j) \nonumber \\
&+ \frac{1}{r_{BC} \Delta (ABC)} (\delta_{ij}-n_{BC}^i n_{BC}^j)\,,
\label{triangle-gen-rel}
\end{align}
where $\mathcal{G}(ABC) \equiv \mathcal{G}(\bm{x}_A, \bm{x}_B, \bm{x}_C)$.


\section{Odd-Parity, Non-Spinning, 
Regularized Contribution in the Metric Correction}
\label{non-spin-reg-cont}

We consider here the odd-parity sector for non-spinning binaries, where, for the scalar field, the magnetic-type dipole moment vanishes, $\mu_{A}^{i} = 0$, since $\chi_{A} = 0$.
For the regularized contribution, we only need to consider the cross-interaction terms since the isolated non-spinning BH solution in the odd-parity case is simply the Schwarzschild metric.
The $\tilde{K}^{(1)}_{ij}$ source term gives the largest contribution and one is then left only with
the pseudo-scalar generated by interaction terms, as given in Eq.~\eqref{final-PP-odd}. 

The metric deformation is given by Eq.~\eqref{NZ-FZ-metric-sol}, the $m=0$ piece of which can be split as in Eqs.~\eqref{hij-cs-1}-\eqref{hij-cs-4}. Before tackling each of these terms separately, let us point out that many of them identically vanish. For example, one of the contribution in Eq.~\eqref{hij-cs-1} is proportional to
\begin{align}
I_{ijqn} \equiv&\,  m_1 \epsilon_{jkl} \int_{\mathcal{M}}  \partial^{(1)}_{qnk} \left( \frac{1}{r_1} \right) \partial^{(1)}_{il} \left( \frac{1}{r_1} \right) d^3x \nonumber \\
& + m_2 \epsilon_{jkl} \int_{\mathcal{M}} \partial^{(1)}_{qnk} \left( \frac{1}{r_1} \right) \partial^{(2)}_{il} \left( \frac{1}{r_2} \right) d^3x \nonumber \\
=&\, -2\pi m_1\epsilon_{jkl} \lim_{2 \rightarrow 1}  \partial^{(1)}_{qnk} \partial^{(2)}_{il}  Y(\bm{x}_1, \bm{x}_2) \nonumber \\
& -2\pi m_2 \epsilon_{jkl}  \partial^{(1)}_{qnk} \partial^{(2)}_{il} Y(\bm{x}_1, \bm{x}_2) = 0\,. 
\end{align}
It is critical in this calculation and in the calculations that follow to replace the $x^{i}$ derivatives by particles derivatives, i.e.~derivatives with respect to $x_{1}^{i}$ and $x_{2}^{i}$. 

Let us then tackle the first contribution to the dissipative metric deformation. Equations~\eqref{hij-cs-1}-\eqref{hij-cs-4} can then be rewritten as 
\begin{align}
\hDef_{ij}^{(1)} =&\,   2048 \pi \frac{\alpha_4^2}{\beta} \frac{m_1^2 m_2}{r} \Big[ b\omega^{2} \left(I_{1ij} + I_{2ij} \right) 
\nonumber \\
&- v_{1n} (I_{3ijn}+I_{4ijn}) 
- v_{2n} (I_{5ijn}+I_{6ijn}) 
\nonumber \\
&\plusitoj \Big] \plusonetotwo\,,
\label{hij-cs-1-rest} \\
\hDef_{ij}^{(2)} =&\, -4096 \pi \frac{\alpha_4^2}{\beta} \frac{m_1^2 m_2}{r} v_{1[n} \Big[  I_{3i]jn}+I_{4i]jn} 
\nonumber \\
&+ I_{5i]jn}+I_{6i]jn} \plusitoj \Big] \plusonetotwo\,,
\label{hij-cs-2-in-I} \\
\hDef_{ij}^{(3)} =&\, 4096 \pi \frac{\alpha_4^2}{\beta} \frac{m_1^2 m_2}{r} v_{1[n} 
\nonumber \\
&\times \Big[  I_{7i]jn}+I_{8i]jn} \plusitoj \Big] \plusonetotwo\,, 
\label{hij-cs-3-in-I} \\
\hDef_{ij}^{(4)} =&\, -2048 \pi \frac{\alpha_4^2}{\beta} \frac{m_1^2 m_2}{r} v_{1n} 
\nonumber \\
&\times
\Big[  I_{7ijn}+I_{8ijn} \plusitoj \Big] \plusonetotwo\,,
\label{hij-cs-4-in-I} 
\end{align}
where we have defined
\begin{equation}
\begin{split}
I_{1ij} & \equiv \epsilon_{jkl} \epsilon_{pqs} n_{12s} 
J^{(1)}_{pk,q,il}\,, \\
I_{2ij} & \equiv \epsilon_{jkl} \epsilon_{pqs} n_{12s} 
J^{(1)}_{p,qk,il}\,, \\
I_{3ijn} & \equiv  \epsilon_{jkl} \epsilon_{pqs} v_{12s} 
J^{(1)}_{pkn,q,il}\,,\\
I_{4ijn} & \equiv  \epsilon_{jkl} \epsilon_{pqs} v_{12s} 
J^{(1)}_{pn,qk,il}\,, \\
I_{5ijn} & \equiv  \epsilon_{jkl} \epsilon_{pqs} v_{12s} 
J^{(1)}_{pk,qn,il}\,, \\
I_{6ijn} & \equiv  \epsilon_{jkl} \epsilon_{pqs} v_{12s} 
J^{(1)}_{p,qkn,il}\,, \\
I_{7ijn} & \equiv  \epsilon_{jkl} \epsilon_{pqs} v_{12s} 
J^{(1)}_{pk,q,iln}\,, \\
I_{8ijn} & \equiv \epsilon_{jkl} \epsilon_{pqs} v_{12s} 
J^{(1)}_{p,qk,iln}\,,
\end{split}
\end{equation}
and 
\be
\JJ{p}{A}{B}{C}
=
\Jdef{p}{A}{B}{C}\,,
\ee
with $A,B,C$ denoting the multi-index lists. We provide a more detailed discussion of 
$J$ tensors in Appendix~\ref{sec:Jtensors}. One can then show through explicit 
computation that the two terms combine to give $I_{1ij} + I_{2ij} = 0$, $I_{3ijn}
+I_{4ijn}=0$, $I_{5ijn} + I_{6ijn}=0$, and
$I_{7ijn}+I_{8ijn}=0$. Therefore
$\hDef_{ij}^{(1 \cdots 4)}=0$ at
leading order.

Let us now look at contributions that are smaller by $\mathcal{O}(v)$. 
Such a correction can arise from two different terms: (i) the
$\mathcal{O}(v)$ correction to the source term with $m=0$ in the sum
of Eq.~\eqref{NZ-FZ-metric-sol}, or (ii) the
$\mathcal{O}(v^{0})$ correction to the source term with $m=1$
in the sum of Eq.~\eqref{NZ-FZ-metric-sol}. For case (i),  the
next-order terms consist of two time derivatives and one factor of
$\hGR_{0i}$ (or three time derivatives and one factor of $\hGR_{ij}$),
which when combined are $\mathcal{O}(v^{2})$ smaller than the
$\mathcal{O}(v^{0})$ contribution shown to vanish previously. 
Also, the next-order terms in the PN metric appears at
$\mathcal{O}(v^{2})$ higher relative to the leading-order
terms. Finally, $\vartheta^{\NZ}$ in Eq.~\eqref{odd-cross} expanded
as in Eq.~$(2.27)$ of~\cite{Will:1996zj} with $m=1$ in the sum, 
gives an $\mathcal{O}(v)$ relative contribution to
$\partial_{k}{\vartheta}$, but explicit calculation shows that
\ba
\vartheta^{\NZ} &\!\! =&\!\!  \frac{8}{\pi} \frac{\alpha}{\beta} m_1 m_2
\epsilon_{ijk} \frac{\partial }{\partial t} \bigg[ v_{12k}
\int_\mathcal{M} \left( \frac{1}{r_1} \right)_{\!\! ,il} \left(
\frac{1}{r_2} \right)_{\!\! ,jl} d^3x \bigg] \nonumber \\
&\!\! =&\!\!  \frac{8}{\pi} \frac{\alpha}{\beta} m_1 m_2 \epsilon_{ijk} \frac{\partial }{\partial t} \bigg[ v_{12k} \partial^{(1)}_{il} \partial^{(2)}_{jl} \int_\mathcal{M} \frac{1}{r_1} \frac{1}{r_2}  d^3x \bigg] \nonumber \\
&\!\! = &\!\!  16 \frac{\alpha}{\beta} m_1 m_2 \epsilon_{ijk} \frac{\partial }{\partial t} \bigg[ v_{12k} \partial^{(1)}_{l} \partial^{(2)}_{ijl}b \bigg] = 0\,.
\ea
For case (ii), the resulting $\dot{\hDef}_{ij}$ contains one $n^i$ vector. The correction to the energy flux consists of $
\dot{\hDef}_{ij}$ multiplied by $h^{\TT}_{ij}$ and averaged over a 2-sphere.  However, since the leading contribution 
in $h^{\TT}_{ij}$ contains even numbers of $n^i$ vectors, the correction only contains angular integrals of odd 
numbers of $n^i$'s which vanish exactly upon integration. 

Since there is no $\mathcal{O}(G^{3},v)$ relative contribution to
$\partial_{t}{\hDef}_{ij}$, the first, non-vanishing contribution must
be at least $\mathcal{O}(v^{2})$ smaller than what we computed in Eqs.~\eqref{hij-cs-1-rest}-\eqref{hij-cs-4-in-I}, which amounts to a 7PN correction to the energy flux 
carried by the metric deformation, in the odd-parity, non-spinning case.

\section{Evaluating $J$ tensors}
\label{sec:Jtensors}

Recall that the definition of the $J$ tensors is
\be
\JJ{p}{A}{B}{C}
=
\Jdef{p}{A}{B}{C}\,.
\ee
The limit $3\to p$ which appears must be taken with
care. There may be terms proportional to
\be
\lim_{3\to p} \frac{1}{r_{p3}}\,,
\ee
which have no finite part. In the evaluation of the $J$ tensors, only
the finite part of the limit is kept. That is, a function can be
expanded as a Laurent series about these points, and the finite
part scales as $(r_{p3})^0$ in the limit as $3\to p$.

Another type of problematic limit is
\be
\lim_{3\to p} n_{p3}^i\quad\textrm{or}\quad \lim_{3\to p} n_{p3}^{ij}\,,
\ee
which does not formally exist, since it depends on the path
taken as we describe below. Parameterize the path that particle 3 takes to the location of
particle $p$ by the continuously differentiable path $\gamma(\lambda)$,
with $\lambda$ a parameter of path length and $\lambda=0$ the
location of particle $p$. There are an infinite number of paths one
could choose, and each can be parameterized in two senses.
Taking the limit along this path ``from below'' (i.e. from smaller
values of $\lambda$ to larger values) yields
\be
\lim_{3\to p,~\gamma^{-}} n_{p3}^i \to -\hat{v}^i_\gamma(0)\,,
\ee
where $\hat{v}_\gamma$ is the tangent vector to the curve
$\gamma$. Taking the limit from above, we find
\be
\lim_{3\to p,~\gamma^{+}} n_{p3}^i \to +\hat{v}^i_\gamma(0)\,.
\ee
The limit depends on the path's tangent at the point of
particle $p$, and the direction in which the limit is taken. Clearly, the
final answer must be unique, which implies the limit must vanish.

A unique prescription to this problem is formalized as Hadamard
regularization~\cite{2005PhRvD..71b4004B}. This can be summarized as
follows. All possible paths are considered, with tangent vectors
$\hat{v}_\gamma$. The average is then taken by integrating, e.g.
\be
\lim_{3\to p} \cdots n_{p3}^{ij} \cdots
=
\int \frac{d\Omega(\hat{v}_\gamma)}{4\pi} \cdots \hat{v}_\gamma^{ij}
\cdots \,.
\ee
The first few such limits, for example, are
\begin{align}
\lim_{3\to p} n_{p3}^i =&\, 0\,,\\
\lim_{3\to p} n_{p3}^{ij} =&\, \frac{1}{3}\delta^{ij}\,.
\end{align}

\bibliography{master}

\begin{thebibliography}{8}%
\makeatletter
\providecommand \@ifxundefined [1]{%
 \@ifx{#1\undefined}
}%
\providecommand \@ifnum [1]{%
 \ifnum #1\expandafter \@firstoftwo
 \else \expandafter \@secondoftwo
 \fi
}%
\providecommand \@ifx [1]{%
 \ifx #1\expandafter \@firstoftwo
 \else \expandafter \@secondoftwo
 \fi
}%
\providecommand \natexlab [1]{#1}%
\providecommand \enquote  [1]{``#1''}%
\providecommand \bibnamefont  [1]{#1}%
\providecommand \bibfnamefont [1]{#1}%
\providecommand \citenamefont [1]{#1}%
\providecommand \href@noop [0]{\@secondoftwo}%
\providecommand \href [0]{\begingroup \@sanitize@url \@href}%
\providecommand \@href[1]{\@@startlink{#1}\@@href}%
\providecommand \@@href[1]{\endgroup#1\@@endlink}%
\providecommand \@sanitize@url [0]{\catcode `\\12\catcode `\$12\catcode
  `\&12\catcode `\#12\catcode `\^12\catcode `\_12\catcode `\%12\relax}%
\providecommand \@@startlink[1]{}%
\providecommand \@@endlink[0]{}%
\providecommand \url  [0]{\begingroup\@sanitize@url \@url }%
\providecommand \@url [1]{\endgroup\@href {#1}{\urlprefix }}%
\providecommand \urlprefix  [0]{URL }%
\providecommand \Eprint [0]{\href }%
\providecommand \doibase [0]{http://dx.doi.org/}%
\providecommand \selectlanguage [0]{\@gobble}%
\providecommand \bibinfo  [0]{\@secondoftwo}%
\providecommand \bibfield  [0]{\@secondoftwo}%
\providecommand \translation [1]{[#1]}%
\providecommand \BibitemOpen [0]{}%
\providecommand \bibitemStop [0]{}%
\providecommand \bibitemNoStop [0]{.\EOS\space}%
\providecommand \EOS [0]{\spacefactor3000\relax}%
\providecommand \BibitemShut  [1]{\csname bibitem#1\endcsname}%
\let\auto@bib@innerbib\@empty
\bibitem [{\citenamefont {Yagi}\ \emph
  {et~al.}(2012{\natexlab{a}})\citenamefont {Yagi}, \citenamefont {Stein},
  \citenamefont {Yunes},\ and\ \citenamefont {Tanaka}}]{quadratic}%
  \BibitemOpen
  \bibfield  {author} {\bibinfo {author} {\bibfnamefont {K.}~\bibnamefont
  {Yagi}}, \bibinfo {author} {\bibfnamefont {L.~C.}\ \bibnamefont {Stein}},
  \bibinfo {author} {\bibfnamefont {N.}~\bibnamefont {Yunes}}, \ and\ \bibinfo
  {author} {\bibfnamefont {T.}~\bibnamefont {Tanaka}},\ }\href {\doibase
  10.1103/PhysRevD.85.064022} {\bibfield  {journal} {\bibinfo  {journal} {Phys.
  Rev.}\ }\textbf {\bibinfo {volume} {D85}},\ \bibinfo {pages} {064022}
  (\bibinfo {year} {2012}{\natexlab{a}})},\ \Eprint
  {http://arxiv.org/abs/1110.5950} {arXiv:1110.5950 [gr-qc]} \BibitemShut
  {NoStop}%
\bibitem [{\citenamefont {{Pati}}\ and\ \citenamefont
  {{Will}}(2000)}]{2000PhRvD..62l4015P}%
  \BibitemOpen
  \bibfield  {author} {\bibinfo {author} {\bibfnamefont {M.~E.}\ \bibnamefont
  {{Pati}}}\ and\ \bibinfo {author} {\bibfnamefont {C.~M.}\ \bibnamefont
  {{Will}}},\ }\href {\doibase 10.1103/PhysRevD.62.124015} {\bibfield
  {journal} {\bibinfo  {journal} {\prd}\ }\textbf {\bibinfo {volume} {62}},\
  \bibinfo {pages} {124015} (\bibinfo {year} {2000})},\ \Eprint
  {http://arxiv.org/abs/gr-qc/0007087} {gr-qc/0007087} \BibitemShut {NoStop}%
\bibitem [{\citenamefont {Yagi}\ \emph
  {et~al.}(2012{\natexlab{b}})\citenamefont {Yagi}, \citenamefont {Yunes},\
  and\ \citenamefont {Tanaka}}]{Yagi:2012ya}%
  \BibitemOpen
  \bibfield  {author} {\bibinfo {author} {\bibfnamefont {K.}~\bibnamefont
  {Yagi}}, \bibinfo {author} {\bibfnamefont {N.}~\bibnamefont {Yunes}}, \ and\
  \bibinfo {author} {\bibfnamefont {T.}~\bibnamefont {Tanaka}},\ }\href
  {\doibase 10.1103/PhysRevD.89.049902, 10.1103/PhysRevD.86.044037} {\bibfield
  {journal} {\bibinfo  {journal} {Phys. Rev.}\ }\textbf {\bibinfo {volume}
  {D86}},\ \bibinfo {pages} {044037} (\bibinfo {year} {2012}{\natexlab{b}})},\
  \bibinfo {note} {[Erratum: Phys. Rev.D89,049902(2014)]},\ \Eprint
  {http://arxiv.org/abs/1206.6130} {arXiv:1206.6130 [gr-qc]} \BibitemShut
  {NoStop}%
\bibitem [{\citenamefont {Yagi}\ \emph {et~al.}(2013)\citenamefont {Yagi},
  \citenamefont {Stein}, \citenamefont {Yunes},\ and\ \citenamefont
  {Tanaka}}]{Yagi:2013mbt}%
  \BibitemOpen
  \bibfield  {author} {\bibinfo {author} {\bibfnamefont {K.}~\bibnamefont
  {Yagi}}, \bibinfo {author} {\bibfnamefont {L.~C.}\ \bibnamefont {Stein}},
  \bibinfo {author} {\bibfnamefont {N.}~\bibnamefont {Yunes}}, \ and\ \bibinfo
  {author} {\bibfnamefont {T.}~\bibnamefont {Tanaka}},\ }\href {\doibase
  10.1103/PhysRevD.87.084058} {\bibfield  {journal} {\bibinfo  {journal}
  {Phys.Rev.}\ }\textbf {\bibinfo {volume} {D87}},\ \bibinfo {pages} {084058}
  (\bibinfo {year} {2013})},\ \Eprint {http://arxiv.org/abs/1302.1918}
  {arXiv:1302.1918 [gr-qc]} \BibitemShut {NoStop}%
\bibitem [{\citenamefont {Maggiore}(2007)}]{Maggiore:1900zz}%
  \BibitemOpen
  \bibfield  {author} {\bibinfo {author} {\bibfnamefont {M.}~\bibnamefont
  {Maggiore}},\ }\href {http://www.oup.com/uk/catalogue/?ci=9780198570745}
  {\emph {\bibinfo {title} {{Gravitational Waves. Vol. 1: Theory and
  Experiments}}}},\ Oxford Master Series in Physics\ (\bibinfo  {publisher}
  {Oxford University Press},\ \bibinfo {year} {2007})\BibitemShut {NoStop}%
\bibitem [{\citenamefont {Yagi}\ \emph
  {et~al.}(2012{\natexlab{c}})\citenamefont {Yagi}, \citenamefont {Yunes},\
  and\ \citenamefont {Tanaka}}]{Yagi:2012vf}%
  \BibitemOpen
  \bibfield  {author} {\bibinfo {author} {\bibfnamefont {K.}~\bibnamefont
  {Yagi}}, \bibinfo {author} {\bibfnamefont {N.}~\bibnamefont {Yunes}}, \ and\
  \bibinfo {author} {\bibfnamefont {T.}~\bibnamefont {Tanaka}},\ }\href
  {\doibase 10.1103/PhysRevLett.109.251105} {\bibfield  {journal} {\bibinfo
  {journal} {Phys. Rev. Lett.}\ }\textbf {\bibinfo {volume} {109}},\ \bibinfo
  {pages} {251105} (\bibinfo {year} {2012}{\natexlab{c}})},\ \Eprint
  {http://arxiv.org/abs/1208.5102} {arXiv:1208.5102 [gr-qc]} \BibitemShut
  {NoStop}%
\bibitem [{\citenamefont {Blanchet}\ \emph {et~al.}(2011)\citenamefont
  {Blanchet}, \citenamefont {Buonanno},\ and\ \citenamefont
  {Faye}}]{Blanchet:2011zv}%
  \BibitemOpen
  \bibfield  {author} {\bibinfo {author} {\bibfnamefont {L.}~\bibnamefont
  {Blanchet}}, \bibinfo {author} {\bibfnamefont {A.}~\bibnamefont {Buonanno}},
  \ and\ \bibinfo {author} {\bibfnamefont {G.}~\bibnamefont {Faye}},\ }\href
  {\doibase 10.1103/PhysRevD.84.064041} {\bibfield  {journal} {\bibinfo
  {journal} {Phys. Rev.}\ }\textbf {\bibinfo {volume} {D84}},\ \bibinfo {pages}
  {064041} (\bibinfo {year} {2011})},\ \Eprint {http://arxiv.org/abs/1104.5659}
  {arXiv:1104.5659 [gr-qc]} \BibitemShut {NoStop}%
\bibitem [{\citenamefont {{Blanchet}}\ \emph {et~al.}(2002)\citenamefont
  {{Blanchet}}, \citenamefont {{Iyer}},\ and\ \citenamefont
  {{Joguet}}}]{2002PhRvD..65f4005B}%
  \BibitemOpen
  \bibfield  {author} {\bibinfo {author} {\bibfnamefont {L.}~\bibnamefont
  {{Blanchet}}}, \bibinfo {author} {\bibfnamefont {B.~R.}\ \bibnamefont
  {{Iyer}}}, \ and\ \bibinfo {author} {\bibfnamefont {B.}~\bibnamefont
  {{Joguet}}},\ }\href {\doibase 10.1103/PhysRevD.65.064005} {\bibfield
  {journal} {\bibinfo  {journal} {\prd}\ }\textbf {\bibinfo {volume} {65}},\
  \bibinfo {pages} {064005} (\bibinfo {year} {2002})},\ \Eprint
  {http://arxiv.org/abs/gr-qc/0105098} {gr-qc/0105098} \BibitemShut {NoStop}%
\end{thebibliography}%


\begin{thebibliography}{70}
\expandafter\ifx\csname natexlab\endcsname\relax\def\natexlab#1{#1}\fi
\expandafter\ifx\csname bibnamefont\endcsname\relax
  \def\bibnamefont#1{#1}\fi
\expandafter\ifx\csname bibfnamefont\endcsname\relax
  \def\bibfnamefont#1{#1}\fi
\expandafter\ifx\csname citenamefont\endcsname\relax
  \def\citenamefont#1{#1}\fi
\expandafter\ifx\csname url\endcsname\relax
  \def\url#1{\texttt{#1}}\fi
\expandafter\ifx\csname urlprefix\endcsname\relax\def\urlprefix{URL }\fi
\providecommand{\bibinfo}[2]{#2}
\providecommand{\eprint}[2][]{\url{#2}}

\bibitem[{\citenamefont{Will}(2006)}]{lrr-2006-3}
\bibinfo{author}{\bibfnamefont{C.~M.} \bibnamefont{Will}},
  \bibinfo{journal}{Living Reviews in Relativity} \textbf{\bibinfo{volume}{9}}
  (\bibinfo{year}{2006}), \eprint{gr-qc/0510072},
  \urlprefix\url{http://www.livingreviews.org/lrr-2006-3}.

\bibitem[{\citenamefont{Schutz et~al.}(2009)\citenamefont{Schutz, Centrella,
  Cutler, and Hughes}}]{Schutz:2009tz}
\bibinfo{author}{\bibfnamefont{B.~F.} \bibnamefont{Schutz}},
  \bibinfo{author}{\bibfnamefont{J.}~\bibnamefont{Centrella}},
  \bibinfo{author}{\bibfnamefont{C.}~\bibnamefont{Cutler}}, \bibnamefont{and}
  \bibinfo{author}{\bibfnamefont{S.~A.} \bibnamefont{Hughes}}
  (\bibinfo{year}{2009}), \eprint{0903.0100}.

\bibitem[{\citenamefont{{Sopuerta}}(2010)}]{2010GWN.....4....3S}
\bibinfo{author}{\bibfnamefont{C.~F.} \bibnamefont{{Sopuerta}}},
  \bibinfo{journal}{GW Notes, Vol.~4, p.~3-47} \textbf{\bibinfo{volume}{4}},
  \bibinfo{pages}{3} (\bibinfo{year}{2010}).

\bibitem[{\citenamefont{Piotr~Jaranowski}(2005)}]{lrr-2005-3}
\bibinfo{author}{\bibfnamefont{A.~K.} \bibnamefont{Piotr~Jaranowski}},
  \bibinfo{journal}{Living Reviews in Relativity} \textbf{\bibinfo{volume}{8}}
  (\bibinfo{year}{2005}),
  \urlprefix\url{http://www.livingreviews.org/lrr-2005-3}.

\bibitem[{\citenamefont{Misner et~al.}(1973)\citenamefont{Misner, Thorne, and
  Wheeler}}]{Misner:1973cw}
\bibinfo{author}{\bibfnamefont{C.~W.} \bibnamefont{Misner}},
  \bibinfo{author}{\bibfnamefont{K.}~\bibnamefont{Thorne}}, \bibnamefont{and}
  \bibinfo{author}{\bibfnamefont{J.~A.} \bibnamefont{Wheeler}},
  \emph{\bibinfo{title}{Gravitation}} (\bibinfo{publisher}{W. H. Freeman \&
  Co.}, \bibinfo{address}{San Francisco}, \bibinfo{year}{1973}).

\bibitem[{\citenamefont{Yunes and
  Pretorius}(2009{\natexlab{a}})}]{Yunes:2009hc}
\bibinfo{author}{\bibfnamefont{N.}~\bibnamefont{Yunes}} \bibnamefont{and}
  \bibinfo{author}{\bibfnamefont{F.}~\bibnamefont{Pretorius}},
  \bibinfo{journal}{Phys. Rev.} \textbf{\bibinfo{volume}{D79}},
  \bibinfo{pages}{084043} (\bibinfo{year}{2009}{\natexlab{a}}),
  \eprint{0902.4669}.

\bibitem[{\citenamefont{{Campanelli} et~al.}(1994)\citenamefont{{Campanelli},
  {Lousto}, and {Audretsch}}}]{1994PhRvD..49.5188C}
\bibinfo{author}{\bibfnamefont{M.}~\bibnamefont{{Campanelli}}},
  \bibinfo{author}{\bibfnamefont{C.~O.} \bibnamefont{{Lousto}}},
  \bibnamefont{and}
  \bibinfo{author}{\bibfnamefont{J.}~\bibnamefont{{Audretsch}}},
  \bibinfo{journal}{\prd} \textbf{\bibinfo{volume}{49}}, \bibinfo{pages}{5188}
  (\bibinfo{year}{1994}), \eprint{gr-qc/9401013}.

\bibitem[{\citenamefont{Woodard}(2007)}]{Woodard:2006nt}
\bibinfo{author}{\bibfnamefont{R.~P.} \bibnamefont{Woodard}},
  \bibinfo{journal}{Lect. Notes Phys.} \textbf{\bibinfo{volume}{720}},
  \bibinfo{pages}{403} (\bibinfo{year}{2007}), \eprint{astro-ph/0601672}.

\bibitem[{\citenamefont{{Cooney} et~al.}(2010)\citenamefont{{Cooney}, {Dedeo},
  and {Psaltis}}}]{2010PhRvD..82f4033C}
\bibinfo{author}{\bibfnamefont{A.}~\bibnamefont{{Cooney}}},
  \bibinfo{author}{\bibfnamefont{S.}~\bibnamefont{{Dedeo}}}, \bibnamefont{and}
  \bibinfo{author}{\bibfnamefont{D.}~\bibnamefont{{Psaltis}}},
  \bibinfo{journal}{\prd} \textbf{\bibinfo{volume}{82}},
  \bibinfo{pages}{064033} (\bibinfo{year}{2010}), \eprint{0910.5480}.

\bibitem[{lig()}]{ligo}
\emph{\bibinfo{title}{{LIGO}}}, \bibinfo{note}{{\tt www.ligo.caltech.edu}}.

\bibitem[{vir()}]{virgo}
\emph{\bibinfo{title}{{VIRGO}}}, \bibinfo{note}{{\tt www.virgo.infn.it}}.

\bibitem[{lcg()}]{lcgt}
\emph{\bibinfo{title}{Lcgt}}, \bibinfo{note}{{\tt
  gw.icrr.u-tokyo.ac.jp/lcgt/}}.

\bibitem[{lis()}]{lisa}
\emph{\bibinfo{title}{{LISA}}}, \bibinfo{note}{{\tt www.esa.int/science/lisa},
  {\tt lisa.jpl.nasa.gov}}.

\bibitem[{\citenamefont{Ashtekar and Lewandowski}(2004)}]{Ashtekar:2004eh}
\bibinfo{author}{\bibfnamefont{A.}~\bibnamefont{Ashtekar}} \bibnamefont{and}
  \bibinfo{author}{\bibfnamefont{J.}~\bibnamefont{Lewandowski}},
  \bibinfo{journal}{Class.Quant.Grav.} \textbf{\bibinfo{volume}{21}},
  \bibinfo{pages}{R53} (\bibinfo{year}{2004}), \eprint{gr-qc/0404018}.

\bibitem[{\citenamefont{Rovelli}(2004)}]{Rovelli:2004tv}
\bibinfo{author}{\bibfnamefont{C.}~\bibnamefont{Rovelli}}
  (\bibinfo{year}{2004}), \bibinfo{note}{published in Cambridge Monographs on
  Mathematical Physics, pages 1-480, year 2004}.

\bibitem[{\citenamefont{{Polchinski}}(1998)}]{1998stth.book.....P}
\bibinfo{author}{\bibfnamefont{J.}~\bibnamefont{{Polchinski}}},
  \emph{\bibinfo{title}{{String Theory}}} (\bibinfo{year}{1998}).

\bibitem[{\citenamefont{Moura and Schiappa}(2007)}]{Moura:2006pz}
\bibinfo{author}{\bibfnamefont{F.}~\bibnamefont{Moura}} \bibnamefont{and}
  \bibinfo{author}{\bibfnamefont{R.}~\bibnamefont{Schiappa}},
  \bibinfo{journal}{Class.Quant.Grav.} \textbf{\bibinfo{volume}{24}},
  \bibinfo{pages}{361} (\bibinfo{year}{2007}), \eprint{hep-th/0605001}.

\bibitem[{\citenamefont{Pani and Cardoso}(2009)}]{Pani:2009wy}
\bibinfo{author}{\bibfnamefont{P.}~\bibnamefont{Pani}} \bibnamefont{and}
  \bibinfo{author}{\bibfnamefont{V.}~\bibnamefont{Cardoso}},
  \bibinfo{journal}{Phys.Rev.} \textbf{\bibinfo{volume}{D79}},
  \bibinfo{pages}{084031} (\bibinfo{year}{2009}), \eprint{0902.1569}.

\bibitem[{\citenamefont{Jackiw and Pi}(2003)}]{jackiw:2003:cmo}
\bibinfo{author}{\bibfnamefont{R.}~\bibnamefont{Jackiw}} \bibnamefont{and}
  \bibinfo{author}{\bibfnamefont{S.~Y.} \bibnamefont{Pi}},
  \bibinfo{journal}{Phys. Rev.} \textbf{\bibinfo{volume}{D68}},
  \bibinfo{pages}{104012} (\bibinfo{year}{2003}), \eprint{gr-qc/0308071}.

\bibitem[{\citenamefont{Alexander and Yunes}(2008)}]{Alexander:2008wi}
\bibinfo{author}{\bibfnamefont{S.}~\bibnamefont{Alexander}} \bibnamefont{and}
  \bibinfo{author}{\bibfnamefont{N.}~\bibnamefont{Yunes}},
  \bibinfo{journal}{Phys. Rev.} \textbf{\bibinfo{volume}{D77}},
  \bibinfo{pages}{124040} (\bibinfo{year}{2008}), \eprint{0804.1797}.

\bibitem[{\citenamefont{{Damour}}(1983)}]{Damour:1983}
\bibinfo{author}{\bibfnamefont{T.}~\bibnamefont{{Damour}}}, in
  \emph{\bibinfo{booktitle}{Gravitational Radiation}}, edited by
  \bibinfo{editor}{\bibnamefont{{N.~Deruelle \& T.~Piran}}}
  (\bibinfo{year}{1983}), pp. \bibinfo{pages}{58--+}.

\bibitem[{\citenamefont{Tagoshi and Sasaki}(1994)}]{Tagoshi:1994sm}
\bibinfo{author}{\bibfnamefont{H.}~\bibnamefont{Tagoshi}} \bibnamefont{and}
  \bibinfo{author}{\bibfnamefont{M.}~\bibnamefont{Sasaki}},
  \bibinfo{journal}{Prog.Theor.Phys.} \textbf{\bibinfo{volume}{92}},
  \bibinfo{pages}{745} (\bibinfo{year}{1994}), \eprint{gr-qc/9405062}.

\bibitem[{\citenamefont{Tanaka et~al.}(1996)\citenamefont{Tanaka, Tagoshi, and
  Sasaki}}]{Tanaka:1997dj}
\bibinfo{author}{\bibfnamefont{T.}~\bibnamefont{Tanaka}},
  \bibinfo{author}{\bibfnamefont{H.}~\bibnamefont{Tagoshi}}, \bibnamefont{and}
  \bibinfo{author}{\bibfnamefont{M.}~\bibnamefont{Sasaki}},
  \bibinfo{journal}{Prog. Theor. Phys.} \textbf{\bibinfo{volume}{96}},
  \bibinfo{pages}{1087} (\bibinfo{year}{1996}).

\bibitem[{\citenamefont{Blanchet and Damour}(1992)}]{Blanchet:1992br}
\bibinfo{author}{\bibfnamefont{L.}~\bibnamefont{Blanchet}} \bibnamefont{and}
  \bibinfo{author}{\bibfnamefont{T.}~\bibnamefont{Damour}},
  \bibinfo{journal}{Phys.Rev.} \textbf{\bibinfo{volume}{D46}},
  \bibinfo{pages}{4304} (\bibinfo{year}{1992}).

\bibitem[{\citenamefont{Blanchet}(1995)}]{Blanchet:1995fr}
\bibinfo{author}{\bibfnamefont{L.}~\bibnamefont{Blanchet}},
  \bibinfo{journal}{Phys. Rev.} \textbf{\bibinfo{volume}{D51}},
  \bibinfo{pages}{2559} (\bibinfo{year}{1995}), \eprint{gr-qc/9501030}.

\bibitem[{\citenamefont{Blanchet et~al.}(1995)\citenamefont{Blanchet, Damour,
  and Iyer}}]{Blanchet:1995fg}
\bibinfo{author}{\bibfnamefont{L.}~\bibnamefont{Blanchet}},
  \bibinfo{author}{\bibfnamefont{T.}~\bibnamefont{Damour}}, \bibnamefont{and}
  \bibinfo{author}{\bibfnamefont{B.~R.} \bibnamefont{Iyer}},
  \bibinfo{journal}{Phys.Rev.} \textbf{\bibinfo{volume}{D51}},
  \bibinfo{pages}{5360} (\bibinfo{year}{1995}), \eprint{gr-qc/9501029}.

\bibitem[{\citenamefont{Blanchet}(2006)}]{Blanchet:2002av}
\bibinfo{author}{\bibfnamefont{L.}~\bibnamefont{Blanchet}},
  \bibinfo{journal}{Living Rev. Relativity} \textbf{\bibinfo{volume}{9}},
  \bibinfo{pages}{4} (\bibinfo{year}{2006}), \eprint{gr-qc/0202016}.

\bibitem[{\citenamefont{{Yunes} and {Stein}}(2011)}]{Yunes:2011we}
\bibinfo{author}{\bibfnamefont{N.}~\bibnamefont{{Yunes}}} \bibnamefont{and}
  \bibinfo{author}{\bibfnamefont{L.~C.} \bibnamefont{{Stein}}},
  \bibinfo{journal}{\prd} \textbf{\bibinfo{volume}{83}},
  \bibinfo{pages}{104002} (\bibinfo{year}{2011}), \eprint{1101.2921}.

\bibitem[{\citenamefont{Ali-Haimoud}(2011)}]{AliHaimoud:2011bk}
\bibinfo{author}{\bibfnamefont{Y.}~\bibnamefont{Ali-Haimoud}},
  \bibinfo{journal}{Phys. Rev.} \textbf{\bibinfo{volume}{D83}},
  \bibinfo{pages}{124050} (\bibinfo{year}{2011}), \eprint{1105.0009}.

\bibitem[{\citenamefont{Ali-Haimoud and Chen}(2011)}]{AliHaimoud:2011fw}
\bibinfo{author}{\bibfnamefont{Y.}~\bibnamefont{Ali-Haimoud}} \bibnamefont{and}
  \bibinfo{author}{\bibfnamefont{Y.}~\bibnamefont{Chen}}
  (\bibinfo{year}{2011}), \eprint{1110.5329}.

\bibitem[{\citenamefont{Yunes and
  Pretorius}(2009{\natexlab{b}})}]{Yunes:2009ke}
\bibinfo{author}{\bibfnamefont{N.}~\bibnamefont{Yunes}} \bibnamefont{and}
  \bibinfo{author}{\bibfnamefont{F.}~\bibnamefont{Pretorius}},
  \bibinfo{journal}{Phys.Rev.} \textbf{\bibinfo{volume}{D80}},
  \bibinfo{pages}{122003} (\bibinfo{year}{2009}{\natexlab{b}}),
  \eprint{0909.3328}.

\bibitem[{\citenamefont{Cornish et~al.}(2011)\citenamefont{Cornish, Sampson,
  Yunes, and Pretorius}}]{Cornish:2011ys}
\bibinfo{author}{\bibfnamefont{N.}~\bibnamefont{Cornish}},
  \bibinfo{author}{\bibfnamefont{L.}~\bibnamefont{Sampson}},
  \bibinfo{author}{\bibfnamefont{N.}~\bibnamefont{Yunes}}, \bibnamefont{and}
  \bibinfo{author}{\bibfnamefont{F.}~\bibnamefont{Pretorius}}
  (\bibinfo{year}{2011}), \bibinfo{note}{accepted to Phys. Rev. D},
  \eprint{1105.2088}.

\bibitem[{\citenamefont{Alexander and Yunes}(2007)}]{Alexander:2007vt}
\bibinfo{author}{\bibfnamefont{S.}~\bibnamefont{Alexander}} \bibnamefont{and}
  \bibinfo{author}{\bibfnamefont{N.}~\bibnamefont{Yunes}},
  \bibinfo{journal}{Phys. Rev.} \textbf{\bibinfo{volume}{D75}},
  \bibinfo{pages}{124022} (\bibinfo{year}{2007}), \eprint{0704.0299}.

\bibitem[{\citenamefont{Smith et~al.}(2008)\citenamefont{Smith, Erickcek,
  Caldwell, and Kamionkowski}}]{Smith:2007jm}
\bibinfo{author}{\bibfnamefont{T.~L.} \bibnamefont{Smith}},
  \bibinfo{author}{\bibfnamefont{A.~L.} \bibnamefont{Erickcek}},
  \bibinfo{author}{\bibfnamefont{R.~R.} \bibnamefont{Caldwell}},
  \bibnamefont{and}
  \bibinfo{author}{\bibfnamefont{M.}~\bibnamefont{Kamionkowski}},
  \bibinfo{journal}{Phys. Rev.} \textbf{\bibinfo{volume}{D77}},
  \bibinfo{pages}{024015} (\bibinfo{year}{2008}), \eprint{0708.0001}.

\bibitem[{\citenamefont{Amendola et~al.}(2007)\citenamefont{Amendola,
  Charmousis, and Davis}}]{Amendola:2007ni}
\bibinfo{author}{\bibfnamefont{L.}~\bibnamefont{Amendola}},
  \bibinfo{author}{\bibfnamefont{C.}~\bibnamefont{Charmousis}},
  \bibnamefont{and} \bibinfo{author}{\bibfnamefont{S.~C.} \bibnamefont{Davis}},
  \bibinfo{journal}{JCAP} \textbf{\bibinfo{volume}{0710}}, \bibinfo{pages}{004}
  (\bibinfo{year}{2007}), \eprint{0704.0175}.

\bibitem[{\citenamefont{{Mart{\'{\i}}n-Garc{\'{\i}}a}}(2008)}]{2008CoPhC.179..597M}
\bibinfo{author}{\bibfnamefont{J.~M.}
  \bibnamefont{{Mart{\'{\i}}n-Garc{\'{\i}}a}}}, \bibinfo{journal}{Computer
  Physics Communications} \textbf{\bibinfo{volume}{179}}, \bibinfo{pages}{597}
  (\bibinfo{year}{2008}), \eprint{0803.0862}.

\bibitem[{\citenamefont{{Brizuela} et~al.}(2009)\citenamefont{{Brizuela},
  {Mart{\'{\i}}n-Garc{\'{\i}}a}, and {Mena Marug{\'a}n}}}]{2009GReGr..41.2415B}
\bibinfo{author}{\bibfnamefont{D.}~\bibnamefont{{Brizuela}}},
  \bibinfo{author}{\bibfnamefont{J.~M.}
  \bibnamefont{{Mart{\'{\i}}n-Garc{\'{\i}}a}}}, \bibnamefont{and}
  \bibinfo{author}{\bibfnamefont{G.~A.} \bibnamefont{{Mena Marug{\'a}n}}},
  \bibinfo{journal}{General Relativity and Gravitation}
  \textbf{\bibinfo{volume}{41}}, \bibinfo{pages}{2415} (\bibinfo{year}{2009}),
  \eprint{0807.0824}.

\bibitem[{\citenamefont{{Alexander} and {Yunes}}(2009)}]{Alexander:2009tp}
\bibinfo{author}{\bibfnamefont{S.}~\bibnamefont{{Alexander}}} \bibnamefont{and}
  \bibinfo{author}{\bibfnamefont{N.}~\bibnamefont{{Yunes}}},
  \bibinfo{journal}{{Phys. Rep.}} \textbf{\bibinfo{volume}{480}},
  \bibinfo{pages}{1} (\bibinfo{year}{2009}), \eprint{0907.2562}.

\bibitem[{\citenamefont{Sopuerta and Yunes}(2009)}]{Sopuerta:2009iy}
\bibinfo{author}{\bibfnamefont{C.~F.} \bibnamefont{Sopuerta}} \bibnamefont{and}
  \bibinfo{author}{\bibfnamefont{N.}~\bibnamefont{Yunes}},
  \bibinfo{journal}{Phys.Rev.} \textbf{\bibinfo{volume}{D80}},
  \bibinfo{pages}{064006} (\bibinfo{year}{2009}), \eprint{0904.4501}.

\bibitem[{\citenamefont{{Stein} and {Yunes}}(2011)}]{Stein:2010pn}
\bibinfo{author}{\bibfnamefont{L.~C.} \bibnamefont{{Stein}}} \bibnamefont{and}
  \bibinfo{author}{\bibfnamefont{N.}~\bibnamefont{{Yunes}}},
  \bibinfo{journal}{\prd} \textbf{\bibinfo{volume}{83}},
  \bibinfo{pages}{064038} (\bibinfo{year}{2011}), \eprint{1012.3144}.

\bibitem[{\citenamefont{Green et~al.}(1987{\natexlab{a}})\citenamefont{Green,
  Schwarz, and Witten}}]{Green:1987sp}
\bibinfo{author}{\bibfnamefont{M.~B.} \bibnamefont{Green}},
  \bibinfo{author}{\bibfnamefont{J.~H.} \bibnamefont{Schwarz}},
  \bibnamefont{and} \bibinfo{author}{\bibfnamefont{E.}~\bibnamefont{Witten}},
  \emph{\bibinfo{title}{Superstring Theory. Vol. 1: Introduction}}
  (\bibinfo{publisher}{Cambridge University Press},
  \bibinfo{address}{Cambridge, UK}, \bibinfo{year}{1987}{\natexlab{a}}).

\bibitem[{\citenamefont{Green et~al.}(1987{\natexlab{b}})\citenamefont{Green,
  Schwarz, and Witten}}]{Green:1987mn}
\bibinfo{author}{\bibfnamefont{M.~B.} \bibnamefont{Green}},
  \bibinfo{author}{\bibfnamefont{J.~H.} \bibnamefont{Schwarz}},
  \bibnamefont{and} \bibinfo{author}{\bibfnamefont{E.}~\bibnamefont{Witten}},
  \emph{\bibinfo{title}{Superstring Theory. Vol. 2: Loop Amplitides, Anomalies
  and Phenomenology}} (\bibinfo{publisher}{Cambridge University Press},
  \bibinfo{address}{Cambridge, UK}, \bibinfo{year}{1987}{\natexlab{b}}).

\bibitem[{\citenamefont{Alexander and Gates}(2006)}]{Alexander:2004xd}
\bibinfo{author}{\bibfnamefont{S.~H.~S.} \bibnamefont{Alexander}}
  \bibnamefont{and} \bibinfo{author}{\bibfnamefont{J.}~\bibnamefont{Gates},
  \bibfnamefont{S.~James}}, \bibinfo{journal}{JCAP}
  \textbf{\bibinfo{volume}{0606}}, \bibinfo{pages}{018} (\bibinfo{year}{2006}),
  \eprint{hep-th/0409014}.

\bibitem[{\citenamefont{Burgess}(2004)}]{lrr-2004-5}
\bibinfo{author}{\bibfnamefont{C.~P.} \bibnamefont{Burgess}},
  \bibinfo{journal}{Living Reviews in Relativity} \textbf{\bibinfo{volume}{7}}
  (\bibinfo{year}{2004}),
  \urlprefix\url{http://www.livingreviews.org/lrr-2004-5}.

\bibitem[{\citenamefont{Taveras and Yunes}(2008)}]{Taveras:2008yf}
\bibinfo{author}{\bibfnamefont{V.}~\bibnamefont{Taveras}} \bibnamefont{and}
  \bibinfo{author}{\bibfnamefont{N.}~\bibnamefont{Yunes}},
  \bibinfo{journal}{Phys. Rev.} \textbf{\bibinfo{volume}{D78}},
  \bibinfo{pages}{064070} (\bibinfo{year}{2008}), \eprint{0807.2652}.

\bibitem[{\citenamefont{Mercuri and Taveras}(2009)}]{Mercuri:2009zt}
\bibinfo{author}{\bibfnamefont{S.}~\bibnamefont{Mercuri}} \bibnamefont{and}
  \bibinfo{author}{\bibfnamefont{V.}~\bibnamefont{Taveras}},
  \bibinfo{journal}{Phys. Rev.} \textbf{\bibinfo{volume}{D80}},
  \bibinfo{pages}{104007} (\bibinfo{year}{2009}), \eprint{0903.4407}.

\bibitem[{\citenamefont{Gates et~al.}(2009)\citenamefont{Gates, Ketov, and
  Yunes}}]{Gates:2009pt}
\bibinfo{author}{\bibfnamefont{J.}~\bibnamefont{Gates},
  \bibfnamefont{S.James}}, \bibinfo{author}{\bibfnamefont{S.~V.}
  \bibnamefont{Ketov}}, \bibnamefont{and}
  \bibinfo{author}{\bibfnamefont{N.}~\bibnamefont{Yunes}},
  \bibinfo{journal}{Phys.Rev.} \textbf{\bibinfo{volume}{D80}},
  \bibinfo{pages}{065003} (\bibinfo{year}{2009}), \eprint{0906.4978}.

\bibitem[{\citenamefont{Yunes and Sopuerta}(2008)}]{Yunes:2007ss}
\bibinfo{author}{\bibfnamefont{N.}~\bibnamefont{Yunes}} \bibnamefont{and}
  \bibinfo{author}{\bibfnamefont{C.~F.} \bibnamefont{Sopuerta}},
  \bibinfo{journal}{Phys. Rev.} \textbf{\bibinfo{volume}{D77}},
  \bibinfo{pages}{064007} (\bibinfo{year}{2008}), \eprint{0712.1028}.

\bibitem[{\citenamefont{Grumiller and Yunes}(2008)}]{Grumiller:2007rv}
\bibinfo{author}{\bibfnamefont{D.}~\bibnamefont{Grumiller}} \bibnamefont{and}
  \bibinfo{author}{\bibfnamefont{N.}~\bibnamefont{Yunes}},
  \bibinfo{journal}{Phys. Rev.} \textbf{\bibinfo{volume}{D77}},
  \bibinfo{pages}{044015} (\bibinfo{year}{2008}), \eprint{0711.1868}.

\bibitem[{\citenamefont{Boulware and Deser}(1985)}]{Boulware:1985wk}
\bibinfo{author}{\bibfnamefont{D.~G.} \bibnamefont{Boulware}} \bibnamefont{and}
  \bibinfo{author}{\bibfnamefont{S.}~\bibnamefont{Deser}},
  \bibinfo{journal}{Phys.Rev.Lett.} \textbf{\bibinfo{volume}{55}},
  \bibinfo{pages}{2656} (\bibinfo{year}{1985}).

\bibitem[{\citenamefont{{Campbell} et~al.}(1992)\citenamefont{{Campbell},
  {Kaloper}, and {Olive}}}]{1992PhLB..285..199C}
\bibinfo{author}{\bibfnamefont{B.~A.} \bibnamefont{{Campbell}}},
  \bibinfo{author}{\bibfnamefont{N.}~\bibnamefont{{Kaloper}}},
  \bibnamefont{and} \bibinfo{author}{\bibfnamefont{K.~A.}
  \bibnamefont{{Olive}}}, \bibinfo{journal}{Physics Letters B}
  \textbf{\bibinfo{volume}{285}}, \bibinfo{pages}{199} (\bibinfo{year}{1992}).

\bibitem[{\citenamefont{Blanchet et~al.}(1998)\citenamefont{Blanchet, Faye, and
  Ponsot}}]{Blanchet:1998vx}
\bibinfo{author}{\bibfnamefont{L.}~\bibnamefont{Blanchet}},
  \bibinfo{author}{\bibfnamefont{G.}~\bibnamefont{Faye}}, \bibnamefont{and}
  \bibinfo{author}{\bibfnamefont{B.}~\bibnamefont{Ponsot}},
  \bibinfo{journal}{Phys. Rev.} \textbf{\bibinfo{volume}{D58}},
  \bibinfo{pages}{124002} (\bibinfo{year}{1998}), \eprint{gr-qc/9804079}.

\bibitem[{\citenamefont{Tagoshi et~al.}(2001)\citenamefont{Tagoshi, Ohashi, and
  Owen}}]{Tagoshi:2000zg}
\bibinfo{author}{\bibfnamefont{H.}~\bibnamefont{Tagoshi}},
  \bibinfo{author}{\bibfnamefont{A.}~\bibnamefont{Ohashi}}, \bibnamefont{and}
  \bibinfo{author}{\bibfnamefont{B.~J.} \bibnamefont{Owen}},
  \bibinfo{journal}{Phys.Rev.} \textbf{\bibinfo{volume}{D63}},
  \bibinfo{pages}{044006} (\bibinfo{year}{2001}), \eprint{gr-qc/0010014}.

\bibitem[{\citenamefont{Alvi}(2000)}]{Alvi:1999cw}
\bibinfo{author}{\bibfnamefont{K.}~\bibnamefont{Alvi}}, \bibinfo{journal}{Phys.
  Rev.} \textbf{\bibinfo{volume}{D61}}, \bibinfo{pages}{124013}
  (\bibinfo{year}{2000}), \eprint{gr-qc/9912113}.

\bibitem[{\citenamefont{Yunes et~al.}(2005)\citenamefont{Yunes, Tichy, Owen,
  and Bruegmann}}]{Yunes:2005nn}
\bibinfo{author}{\bibfnamefont{N.}~\bibnamefont{Yunes}},
  \bibinfo{author}{\bibfnamefont{W.}~\bibnamefont{Tichy}},
  \bibinfo{author}{\bibfnamefont{B.~J.} \bibnamefont{Owen}}, \bibnamefont{and}
  \bibinfo{author}{\bibfnamefont{B.}~\bibnamefont{Bruegmann}}
  (\bibinfo{year}{2005}), \eprint{gr-qc/0503011}.

\bibitem[{\citenamefont{Johnson-McDaniel
  et~al.}(2009)\citenamefont{Johnson-McDaniel, Yunes, Tichy, and
  Owen}}]{JohnsonMcDaniel:2009dq}
\bibinfo{author}{\bibfnamefont{N.~K.} \bibnamefont{Johnson-McDaniel}},
  \bibinfo{author}{\bibfnamefont{N.}~\bibnamefont{Yunes}},
  \bibinfo{author}{\bibfnamefont{W.}~\bibnamefont{Tichy}}, \bibnamefont{and}
  \bibinfo{author}{\bibfnamefont{B.~J.} \bibnamefont{Owen}},
  \bibinfo{journal}{Phys.Rev.} \textbf{\bibinfo{volume}{D80}},
  \bibinfo{pages}{124039} (\bibinfo{year}{2009}), \eprint{0907.0891}.

\bibitem[{\citenamefont{D'Eath}(1975)}]{D'Eath:1975qs}
\bibinfo{author}{\bibfnamefont{P.~D.} \bibnamefont{D'Eath}},
  \bibinfo{journal}{Phys. Rev.} \textbf{\bibinfo{volume}{D11}},
  \bibinfo{pages}{1387} (\bibinfo{year}{1975}).

\bibitem[{\citenamefont{Thorne and Hartle}(1984)}]{Thorne:1984mz}
\bibinfo{author}{\bibfnamefont{K.~S.} \bibnamefont{Thorne}} \bibnamefont{and}
  \bibinfo{author}{\bibfnamefont{J.~B.} \bibnamefont{Hartle}},
  \bibinfo{journal}{Phys. Rev.} \textbf{\bibinfo{volume}{D31}},
  \bibinfo{pages}{1815} (\bibinfo{year}{1984}).

\bibitem[{\citenamefont{Martel and Poisson}(2005)}]{Martel:2005ir}
\bibinfo{author}{\bibfnamefont{K.}~\bibnamefont{Martel}} \bibnamefont{and}
  \bibinfo{author}{\bibfnamefont{E.}~\bibnamefont{Poisson}},
  \bibinfo{journal}{Phys. Rev.} \textbf{\bibinfo{volume}{D71}},
  \bibinfo{pages}{104003} (\bibinfo{year}{2005}), \eprint{gr-qc/0502028}.

\bibitem[{\citenamefont{Poisson}(2005)}]{Poisson:2005pi}
\bibinfo{author}{\bibfnamefont{E.}~\bibnamefont{Poisson}},
  \bibinfo{journal}{Phys.Rev.Lett.} \textbf{\bibinfo{volume}{94}},
  \bibinfo{pages}{161103} (\bibinfo{year}{2005}), \eprint{gr-qc/0501032}.

\bibitem[{\citenamefont{Yunes and Tichy}(2006)}]{Yunes:2006iw}
\bibinfo{author}{\bibfnamefont{N.}~\bibnamefont{Yunes}} \bibnamefont{and}
  \bibinfo{author}{\bibfnamefont{W.}~\bibnamefont{Tichy}},
  \bibinfo{journal}{Phys. Rev.} \textbf{\bibinfo{volume}{D74}},
  \bibinfo{pages}{064013} (\bibinfo{year}{2006}), \eprint{gr-qc/0601046}.

\bibitem[{\citenamefont{Will and Wiseman}(1996)}]{Will:1996zj}
\bibinfo{author}{\bibfnamefont{C.~M.} \bibnamefont{Will}} \bibnamefont{and}
  \bibinfo{author}{\bibfnamefont{A.~G.} \bibnamefont{Wiseman}},
  \bibinfo{journal}{Phys.Rev.} \textbf{\bibinfo{volume}{D54}},
  \bibinfo{pages}{4813} (\bibinfo{year}{1996}), \eprint{gr-qc/9608012}.

\bibitem[{\citenamefont{{Pati} and {Will}}(2000)}]{2000PhRvD..62l4015P}
\bibinfo{author}{\bibfnamefont{M.~E.} \bibnamefont{{Pati}}} \bibnamefont{and}
  \bibinfo{author}{\bibfnamefont{C.~M.} \bibnamefont{{Will}}},
  \bibinfo{journal}{\prd} \textbf{\bibinfo{volume}{62}},
  \bibinfo{pages}{124015} (\bibinfo{year}{2000}), \eprint{gr-qc/0007087}.

\bibitem[{\citenamefont{Blanchet and Faye}(2000)}]{Blanchet:2000nu}
\bibinfo{author}{\bibfnamefont{L.}~\bibnamefont{Blanchet}} \bibnamefont{and}
  \bibinfo{author}{\bibfnamefont{G.}~\bibnamefont{Faye}}, \bibinfo{journal}{J.
  Math. Phys.} \textbf{\bibinfo{volume}{41}}, \bibinfo{pages}{7675}
  (\bibinfo{year}{2000}), \eprint{gr-qc/0004008}.

\bibitem[{\citenamefont{Pani et~al.}(2011{\natexlab{a}})\citenamefont{Pani,
  Cardoso, and Gualtieri}}]{Panietal}
\bibinfo{author}{\bibfnamefont{P.}~\bibnamefont{Pani}},
  \bibinfo{author}{\bibfnamefont{V.}~\bibnamefont{Cardoso}}, \bibnamefont{and}
  \bibinfo{author}{\bibfnamefont{L.}~\bibnamefont{Gualtieri}},
  \bibinfo{journal}{Phys. Rev.} \textbf{\bibinfo{volume}{D83}},
  \bibinfo{pages}{104048} (\bibinfo{year}{2011}{\natexlab{a}}),
  \eprint{1104.1183}.

\bibitem[{\citenamefont{Droz et~al.}(1999)\citenamefont{Droz, Knapp, Poisson,
  and Owen}}]{Droz:1999qx}
\bibinfo{author}{\bibfnamefont{S.}~\bibnamefont{Droz}},
  \bibinfo{author}{\bibfnamefont{D.~J.} \bibnamefont{Knapp}},
  \bibinfo{author}{\bibfnamefont{E.}~\bibnamefont{Poisson}}, \bibnamefont{and}
  \bibinfo{author}{\bibfnamefont{B.~J.} \bibnamefont{Owen}},
  \bibinfo{journal}{Phys.Rev.} \textbf{\bibinfo{volume}{D59}},
  \bibinfo{pages}{124016} (\bibinfo{year}{1999}), \eprint{gr-qc/9901076}.

\bibitem[{\citenamefont{Pani et~al.}(2011{\natexlab{b}})\citenamefont{Pani,
  Berti, Cardoso, and Read}}]{Pani:2011xm}
\bibinfo{author}{\bibfnamefont{P.}~\bibnamefont{Pani}},
  \bibinfo{author}{\bibfnamefont{E.}~\bibnamefont{Berti}},
  \bibinfo{author}{\bibfnamefont{V.}~\bibnamefont{Cardoso}}, \bibnamefont{and}
  \bibinfo{author}{\bibfnamefont{J.}~\bibnamefont{Read}}
  (\bibinfo{year}{2011}{\natexlab{b}}), \bibinfo{note}{* Temporary entry *},
  \eprint{1109.0928}.

\bibitem[{\citenamefont{{Blanchet} et~al.}(2002)\citenamefont{{Blanchet},
  {Iyer}, and {Joguet}}}]{2002PhRvD..65f4005B}
\bibinfo{author}{\bibfnamefont{L.}~\bibnamefont{{Blanchet}}},
  \bibinfo{author}{\bibfnamefont{B.~R.} \bibnamefont{{Iyer}}},
  \bibnamefont{and} \bibinfo{author}{\bibfnamefont{B.}~\bibnamefont{{Joguet}}},
  \bibinfo{journal}{\prd} \textbf{\bibinfo{volume}{65}},
  \bibinfo{pages}{064005} (\bibinfo{year}{2002}), \eprint{gr-qc/0105098}.

\bibitem[{\citenamefont{Pati and Will}(2002)}]{Pati:2002ux}
\bibinfo{author}{\bibfnamefont{M.~E.} \bibnamefont{Pati}} \bibnamefont{and}
  \bibinfo{author}{\bibfnamefont{C.~M.} \bibnamefont{Will}},
  \bibinfo{journal}{Phys.Rev.} \textbf{\bibinfo{volume}{D65}},
  \bibinfo{pages}{104008} (\bibinfo{year}{2002}), \eprint{gr-qc/0201001}.

\bibitem[{\citenamefont{{Blanchet} and {Iyer}}(2005)}]{2005PhRvD..71b4004B}
\bibinfo{author}{\bibfnamefont{L.}~\bibnamefont{{Blanchet}}} \bibnamefont{and}
  \bibinfo{author}{\bibfnamefont{B.~R.} \bibnamefont{{Iyer}}},
  \bibinfo{journal}{\prd} \textbf{\bibinfo{volume}{71}},
  \bibinfo{pages}{024004} (\bibinfo{year}{2005}), \eprint{gr-qc/0409094}.

\end{thebibliography}
\end{document}


\title{Erratum: Post-Newtonian, Quasi-Circular Binary Inspirals \\ in Quadratic Modified Gravity}

\author{Kent Yagi}
\affiliation{Department of Physics, Montana State University, Bozeman, MT 59717, USA.}
\affiliation{Department of Physics, Princeton University, Princeton, NJ 08544, USA.}

\author{Leo C. Stein}
\affiliation{Cornell Center for Astrophysics and
Planetary Science (CCAPS), Cornell University, Ithaca, NY 14853, USA}
\affiliation{Theoretical Astrophysics,
Walter Burke Institute for Theoretical Physics,
California Institute of Technology, Pasadena, CA
91125, USA}

\author{Nicol\'as Yunes}
\affiliation{Department of Physics, Montana State University, Bozeman, MT 59717, USA.}

\author{Takahiro Tanaka}
\affiliation{Yukawa Institute for Theoretical Physics, Kyoto University, Kyoto, 606-8502, Japan.}
\affiliation{Department of Physics, Kyoto University, Kyoto, 606-8502, Japan.}

\date{\today}



\begin{abstract}
  In this erratum, we point out that the correction to the
  gravitational energy flux from compact binary inspirals in dynamical
  Chern-Simons gravity enters at third post-Newtonian order instead of
  second post-Newtonian order. We also correct the scalar energy flux
  at second post-Newtonian order by considering terms that we missed
  in the original paper.
\end{abstract}

\maketitle

\section{Gravitational Radiation}

In~\cite{quadratic}, we claimed that the correction to the
gravitational energy flux, i.e.~the energy flux from binary inspirals
induced by the metric perturbation in dynamical Chern-Simons (dCS)
gravity, enters at second post-Newtonian (PN) order. Here, we point
out that there are terms that we overlooked, and if we take these
terms into account properly, the 2PN terms cancel and the correction
enters at 3PN order.\footnote{Recall that a term in an expression is
  said to be of $N$th PN order if it is proportional to $v^{2N}$
  relative to the leading-order term in that same expression.}

First, let us remind ourselves what we did in~\cite{quadratic}. Our
starting point was perturbing the metric around a flat background as
\be
g_{\mu\nu} = g_{\mu\nu}^\GR + \hDef_{\mu\nu}\,, \qquad g_{\mu\nu}^\GR = \eta_{\mu\nu} + h_{\mu\nu}\,,
\ee
where $h_{\mu\nu}$ and $\hDef_{\mu\nu}$ are the General Relativity
(GR) and dCS metric perturbations respectively. We worked in harmonic
gauge,
$\bar{\hGR}^{\mu\nu}{}_{,\nu} = 0 = \bar{\hDef}^{\mu\nu}{}_{,\nu}$,
where we defined the metric perturbations via
\ba
\bar{\hGR}^{\mu\nu} &\equiv & \eta^{\mu\nu}-\sqrt{-g_{\GR}}g^{\mu\nu}_{\GR}\,, \\
\bar{\hDef}^{\mu\nu} & \equiv & \left( \eta^{\mu\nu}-\sqrt{-g}g^{\mu\nu}\right) - \bar{\hGR}^{\mu\nu}\,.
\ea
We then expanded the modified Einstein equations, focusing on the dCS
correction and keeping terms up to linear order in $h_{\mu\nu}$ and
$\bar \hDef_{\mu\nu}$:
\be
\kappa \square_{\eta} \bar \hDef_{\mu\nu} = 2 \alpha_4 \tilde{\mathcal{K}}^{(1)}_{\mu \nu} - T_{\mu\nu}^{(\vartheta)} \,,
\label{wave-eq-h-pert}
\ee
where $\square_{\eta}$ is the d'Alembertian operator of flat spacetime
and
\ba
\tilde{\mathcal{K}}^{(1)}_{\mu\nu} &=& \vartheta^{,\delta}{}_{,\sigma} \eta_{\nu\alpha}\epsilon^{\alpha\sigma\beta\gamma}
\left(\hGR{}_{\mu [\gamma,\beta] \delta}+\hGR{}_{\delta [\beta,\gamma] \mu}\right) \nn \\
& & - 2 \vartheta{}^{,\delta} \epsilon_{\delta\sigma\chi\mu} \hGR{}^{\sigma}{}_{[\alpha}{}^{,\alpha\chi}{}_{\nu]} + \mathcal{O}(h^2)  +(\mu\leftrightarrow \nu)\,, \\
T_{\mu\nu}^{(\vartheta)} &=& \beta\left( \vartheta_{,\mu} \vartheta_{,\nu} - \frac{1}{2} \eta_{\mu\nu} \vartheta_{,\delta} \vartheta{}^{,\delta} \right) + \mathcal{O}(h)\,.
\ea

We then solved this wave equation in the far zone (FZ), for field
points much larger than the gravitational wave (GW) wavelength of the
system, using a Green's function. The Green's function solution
requires the integration of $\tilde{\mathcal{K}}^{(1)}_{\mu\nu} $ and
$T_{\mu\nu}^{(\vartheta)}$, and thus of the fields $h_{\mu \nu}$ and
$\vartheta$ evaluated in the near zone (NZ), i.e.~at field point
distances smaller than the GW wavelength of the system, but far from
the location of the objects.  To leading PN order, the NZ fields are
given by
\ba
\label{NZmetric1}
\hGR_{00}^\NZ & = &   2 \frac{m_{1}}{r_{1}} + \mathcal{O}\left( \frac{m_1^2}{r_1^2} \right) \plusonetotwo \,,
\\
\hGR_{ij}^\NZ & = & 2 \frac{m_{1}}{r_{1}}  \delta_{ij}  + \mathcal{O}\left( \frac{m_1^2}{r_1^2} \right) \plusonetotwo\,,
\\
\vartheta^\NZ & = & - \mu_1^i \partial_i \frac{1}{r_{1}} + \mathcal{O}\left( \frac{m_1^3}{r_1^3} \right) \plusonetotwo  \nn \\
 & = &  \frac{\mu_1^i n_i^i}{r_{1}^2} + \mathcal{O}\left( \frac{m_1^3}{r_1^3} \right) \plusonetotwo\,,
\label{NZmetric3}
\ea
where $\mu_A^i$ is the NZ scalar dipole charge, with
\be
\label{eq:dipole-charge-BH}
\mu_A^i \equiv \frac{5}{2} \frac{\alpha_4}{\beta} \chi_A \hat{S}_A^i
\ee
for a black hole, where $\chi_A$ and $\hat{S}_A^i$ represent
respectively the dimensionless spin parameter and the unit spin
angular momentum vector of the $A$th binary constituent.  The dominant
contribution to the FZ solution comes from the integration of the
$T_{\mu\nu}^{(\vartheta)}$ term in Eq.~\eqref{wave-eq-h-pert}, which
is roughly of the following PN order:
\ba
\bar \hDef_{ij}^\FZ &\sim & \delta_{ij} \int \vartheta_{,k}^\NZ \vartheta_{,k}^\NZ \frac{d^3 x'}{|\bm{x} - \bm{x}'|} \nn \\
&  \propto & \frac{1}{r} \frac{1}{b^3}\frac{1}{b^3} b^3 \propto \frac{v^6}{r}\,,
\ea
where $b$ is the binary's separation, $v$ is the binary's (relative)
velocity and $r$ is the FZ field point distance. Comparing this with
the GR, leading PN contribution, $\bar h_{ij}^\FZ \sim v^2/r$\,, we
concluded that the dCS correction first enters at 2PN order, i.e.~at
${\cal{O}}(v^{4}/c^{4})$ relative to the leading PN order GR
contribution.

Now, let us look at terms that we neglected in
Eq.~\eqref{wave-eq-h-pert}. Let us begin by expanding out this
equation to find
\ba
\label{wave-eq-h-pert2}
\kappa \square_{\eta} \bar \hDef_{\mu\nu} & = & S_{\mu\nu}\,, \\
\label{eq:S}
S_{\mu\nu} & \equiv & 2 \alpha_4 \tilde{\mathcal{K}}^{(1)}_{\mu \nu} - T_{\mu\nu}^{(\vartheta)} - \hDef T_{\mu\nu}^{\MAT} - t_{\mu\nu}^{(1)} + \mathcal{O}(\hDef^2) \,. \nn \\
\ea
One of the terms we neglected in Eq.~\eqref{wave-eq-h-pert} is the
third one in Eq.~\eqref{eq:S}, which comes from expanding the metric
determinant in the term $(g T_{\mu\nu}^{\MAT})$, where
$T_{\mu\nu}^{\MAT}$ is the matter stress-energy tensor (see the
relaxed Einstein equations in GR given by Eqs.~(2.4) and (2.5)
in~\cite{2000PhRvD..62l4015P}). Another term we neglected is the
fourth term in Eq.~\eqref{eq:S}, which is the linear order in
$\bar \hDef$ piece of $t_{\mu\nu}$, where the latter nonlinear tensor
is defined by
\be
t_{\mu\nu} \equiv (-g) t_{\mu\nu}^{\LL} + \kappa \left( \tilde h_{\mu \alpha}{}^{,\beta} \tilde h_{\nu \beta}{}^{, \alpha} - \tilde h^{\alpha \beta} \tilde h_{\mu\nu}{}_{,\alpha \beta} \right)\,,
\ee
with $t_{\mu\nu}^{\LL}$ the Landau-Lifshitz pseudotensor, which
depends on
$\tilde h_{\mu\nu} \equiv \bar h_{\mu\nu} + \bar \hDef_{\mu\nu}$.

Clearly then, in order to solve Eq.~\eqref{wave-eq-h-pert2} in terms
of Green's functions, we need to integrate $(\hDef T_{\mu\nu}^{\MAT})$
and $t_{\mu\nu}^{(1)}$ in the FZ using the NZ solution for
$\hDef$. The latter is given by~\cite{Yagi:2012ya,Yagi:2013mbt}
\be
\hDef^\NZ \sim \zeta_4 \frac{m_1^3}{r_1^3} \chi_1^2 \plusonetotwo\,. 
\ee
Denoting the contribution of the third and fourth terms in
Eq.~\eqref{eq:S} to $\bar \hDef_{ij}^\FZ$ as
$\bar \hDef_{ij}^{\FZ,T_{ij}^\MAT}$ and $\bar \hDef_{ij}^{\FZ,t_{ij}}$
respectively, one then finds
\ba
\bar \hDef_{ij}^{\FZ,T_{ij}^\MAT} &\sim & \int \hDef^\NZ T_{ij}^\MAT \frac{d^3 x'}{|\bm{x} - \bm{x}'|} \nn \\ 
&\propto & \frac{1}{r} \frac{1}{b^3} v^2 \propto \frac{v^8}{r}\,, \\
\bar \hDef_{ij}^{\FZ,t_{ij}} &\sim & \int \hDef_{kl}^{\NZ} h_{ij,kl}^\NZ \frac{d^3 x'}{|\bm{x} - \bm{x}'|} \nn \\ 
& \propto & \frac{1}{r} \frac{1}{b^3} \frac{1}{b^3} b^3 \propto \frac{v^6}{r}\,,
\ea
where we used that $T_{ij}^\MAT \sim \rho v_i v_j$ with $\rho$
representing the energy density of one of the bodies. From the above
equations, we see that the $(\hDef T_{\mu\nu}^{\MAT})$ and
$t_{\mu\nu}^{(1)}$ terms give 3PN and 2PN corrections to the
gravitational radiation respectively. Notice that the latter is of the
same order as the contribution from $T_{\mu\nu}^{(\vartheta)}$ that we
calculated in~\cite{quadratic}.

Given this, what is the correct dCS corrections to the gravitational
energy flux? One can answer this question by using a very useful (and
standard) trick employed regularly in the PN expansion of GR. Due to
the gauge condition $\bar \hDef_{\mu\nu}{}^{,\nu}=0$, the tensor
$S_{\mu \nu}$ in Eq.~\eqref{eq:S} is conserved,
i.e.~$S_{\mu\nu}{}^{,\nu}=0$. Thanks to this conservation law, one can
write (see e.g.~\cite{Maggiore:1900zz})
\be
\bar \hDef_{ij}^\FZ \sim \frac{1}{r} \int S_{ij} d^3x'  \sim \frac{1}{r} \frac{\partial^2}{\partial t^2} \int S_{00} x'^i x'^j d^3x' \,,
\ee
after integrating by parts a few times. Then, denoting the
contribution of $T_{00}^{(\vartheta)}$, $T^{\MAT}_{00}$ and $t_{00}$
to $\bar \hDef_{ij}^\FZ$ as
$\bar \hDef_{ij}^{\FZ,T_{00}^{(\vartheta)}}$,
$\bar \hDef_{ij}^{\FZ,T_{00}^\MAT}$ and $\bar \hDef_{ij}^{\FZ,t_{00}}$
respectively, one finds
\ba
\bar \hDef_{ij}^{\FZ,T_{00}^{(\vartheta)}} &\sim &\frac{1}{r} \delta_{00} \frac{\partial^2}{\partial t^2} \int \vartheta_{,k}^{\NZ} \vartheta_{,k}^{\NZ} x'^i x'^j d^3 x' \nn \\ 
& \propto & \frac{1}{r} \omega^2 \frac{1}{b^3} \frac{1}{b^3} b^2 b^3 \propto \frac{v^8}{r}\,, \\
\bar \hDef_{ij}^{\FZ,T_{00}^\MAT} &\sim &\frac{1}{r} \frac{\partial^2}{\partial t^2} \int \hDef^\NZ T_{00}^\MAT x'^i x'^j d^3 x' \nn \\ 
& \propto & \frac{1}{r} \omega^2 \frac{1}{b^3} b^2 \propto \frac{v^8}{r}\,, \\
\bar \hDef_{ij}^{\FZ,t_{00}} &\sim &\frac{1}{r} \frac{\partial^2}{\partial t^2} \int \hDef_{kl}^{\NZ} h_{00,kl}^\NZ x'^i x'^j d^3 x' \nn \\ 
& \propto & \frac{1}{r} \omega^2  \frac{1}{b^3} \frac{1}{b^3} b^2 b^3 \propto \frac{v^8}{r}\,,
\ea
where we used $T_{00}^\MAT \sim \rho$.  Observe that all of these
contributions enter at 3PN order relative to the leading GR
expression, and not at 2PN order. This implies that certain terms we
neglected actually cancel the 2PN correction to the metric
perturbation, forcing the correction to enter at 3PN order. A similar
result can be obtained in GR when deriving an expression for the FZ
metric perturbation in terms of the source multipole moments. Using
the conservation law of $S_{\mu \nu}$ in GR, one can show that
$t_{\mu\nu}$ in GR sources higher PN contributions compared to the
leading quadrupolar term sourced by the matter stress-energy tensor.

A very important consequence of all of this is that the contribution
of the dCS correction to the metric perturbation to the energy flux is
now smaller (in a PN sense) than that induced by the scalar
field. Therefore, one does not need to include the dCS correction to
the metric perturbation when computing the energy flux to leading dCS
and PN order in compact binary inspirals~\cite{Yagi:2012vf} and in the
orbital decay rate of binary pulsars~\cite{Yagi:2013mbt}.

\section{Scalar Radiation}

We now look at the scalar energy flux. The scalar field evolution
equation is
\be
\label{eq:scalar-field-eq-true}
\Box \vartheta = -\alpha_{4} R \, {}^{*}\!R\,,
\ee
where the source is the Pontryagin density (see e.g.~Eq.~$(8)$
in~\cite{quadratic}). When one solves this equation through a Green's
function approach, one must employ finite part regularization to avoid
divergences introduced by the point-particle approximation. A matching
calculation gives rise to an effective source term to the scalar field
evolution equation that gives the correct scalar field solution.  To
leading PN order and in \emph{covariant form}, the evolution equation
then becomes
\be
\label{eq:scalar-field-eq}
\Box \vartheta = 4 \pi \sum_{A=1,2} \int d\tau \mu_A^\alpha \left( \frac{\delta^{(4)}[x^\mu - x_A^\mu(\tau)]}{\sqrt{-g}} \right)_{,\alpha} -\alpha_{4} R \, {}^{*}\!R\,,
\ee
where $x_A^\mu$ is the trajectory of body $A$.  We will neglect
$-\alpha_{4} R \, {}^{*}\!R$ on the right-hand side because it only
sources higher PN order contributions to the FZ scalar field, as shown
in~\cite{quadratic}.  In~\cite{quadratic}, we did not use the
covariant form of the effective source term (compare
Eq.~\eqref{eq:scalar-field-eq} to Eq.~$(74)$ in~\cite{quadratic}); as
we will see, there are terms that arise from the covariant form of the
source that enter at the same PN order as those computed
in~\cite{quadratic}.

As we are interested in $\vartheta^{\FZ}$, $\mu_A^\alpha$ in
Eq.~\eqref{eq:scalar-field-eq} corresponds to the FZ scalar dipole
moment.  In principle, this moment must be determined by matching it
to the NZ moment in the buffer zone, i.e.~where the FZ and the NZ
overlap. As in the case of the (source and radiative) multipole
moments of metric perturbations in GR, the NZ scalar dipole moment is
equal to the FZ scalar dipole moment up to non-linear terms.  These
correspond, for example, to the propagation of the scalar field on a
non-flat background, leading to scalar field scattering analogous to
the ``tail'' terms of the metric perturbation in GR. As such, these
non-linear terms enter at higher PN order, and thus, they can be
neglected to the PN order we consider here. The FZ scalar dipole
moment is then equal to the NZ scalar dipole moment, which was related
in~\cite{quadratic} to the inner zone (IZ) integral of the self
interaction contribution of both the Pontryagin density and the
$\Box \vartheta$ terms in the scalar field equation. Therefore, the
scalar dipole moment $\mu_{A}^{\alpha}$ can be obtained by promoting
the spin 3-vector $\hat S^i$ in Eq.~\eqref{eq:dipole-charge-BH} to a
spin 4-vector $\hat S^\alpha$, where the condition that determines
$\mu_A^0$ will be discussed later.

Let us now manipulate Eq.~\eqref{eq:scalar-field-eq} so that one can
solve it using the Green's function approach for a flat metric. We
decompose the left-hand side of Eq.~\eqref{eq:scalar-field-eq} as
$\Box = \Box_\eta + \Box_h$, where $\Box_\eta$ is the flat space
d'Alembertian operator. Then, Eq.~\eqref{eq:scalar-field-eq}
becomes
%
%
\begin{align}
\Box_\eta  \vartheta ={}& 4 \pi  \sum_{A=1,2}\int d\tau \left\{ \mu_A^i \left( \frac{\delta^{(4)}[x^\mu - x_A^\mu(\tau)]}{\sqrt{-g}} \right)_{,i}  \right. \nn \\
&  \left. +  \mu_A^0 \left( \frac{\delta^{(4)}[x^\mu - x_A^\mu(\tau)]}{\sqrt{-g}} \right)_{,0} \right\} - \Box_h \vartheta \nn \\
\approx{}&  4 \pi  \sum_{A=1,2} \left\{   \mu_A^i \delta^{(3)}(\bm{x} - \bm{x}_A)_{,i}  \right. \nn \\
& \left. + \int d\tau \mu_A^0 \delta^{(4)}[x^\mu - x_A^\mu(\tau)]_{,0} \right\} - \Box_h \vartheta\,,
\label{eq:scalar-eq-with-mu0}
\end{align}
where we have only kept leading PN order terms. In~\cite{quadratic},
we only included the first term on the right-hand side of
Eq.~\eqref{eq:scalar-eq-with-mu0}, which gives a quadrupolar scalar
radiation term from a cross interaction contribution:
\be
\label{eq:scalar-FZ-original}
\vartheta^{\FZ,\mrm{cross}}_{(1)} = \frac{1}{r} \ddot{\mu}_{ij} n^{ij} = - \frac{1}{r} \omega^2 \mu_{ij} n^{ij} \sim \frac{1}{r} v^6 b \sim \frac{1}{r} v^4\,,
\ee
where $\mu_{ij} \equiv x_1^{(i}\mu_1^{j)} \plusonetotwo$. Below, we
look at the contribution from the other terms that we did not consider
in~\cite{quadratic}.

We first look at the FZ scalar field $\vartheta^\FZ_{(2)}$ sourced by
the second term in Eq.~\eqref{eq:scalar-eq-with-mu0}:
\allowdisplaybreaks[4]
\bw
\ba
4 \pi \sum_{A=1,2}  \int d\tau \mu_A^0 \delta^{(4)}[x^\mu - x_A^\mu(\tau)]_{,0} &=& 4 \pi \sum_{A=1,2}  \int d\tau \mu_A^0 \delta^{(3)}[x^i - x_A^i (\tau)] \delta [t - t_A (\tau)]_{,0} \nn \\
& = & - 4 \pi \sum_{A=1,2}  \int d\tau \mu_A^0 \delta^{(3)}[x^i - x_A^i (\tau)] \delta [t - t_A (\tau)]_{,\tau} \left( \frac{dt_A}{d\tau} \right)^{-1} \nn \\
& \approx & - 4 \pi \sum_{A=1,2}  \int d\tau \mu_A^0 \delta^{(3)}[x^i - x_A^i (\tau)] \delta [t - \tau]_{,\tau} + \mathcal{O}(v^2) \nn \\
& \approx & 4 \pi \sum_{A=1,2}  \int d\tau \left\{ \mu_A^0 \delta^{(3)}[x^i - x_A^i (\tau)] \right\}_{,\tau} \delta [t - \tau] \nn \\
& \approx & 4 \pi \sum_{A=1,2}   \left\{ \mu_A^0 \delta^{(3)}[x^i - x_A^i (t)] \right\}_{,0} \nn \\
& \approx & 4 \pi \sum_{A=1,2}  \left\{ \frac{d  \mu_A^0}{dt} \delta^{(3)}[x^i - x_A^i (t)] - \mu_A^0 \delta^{(3)}[x^i - x_A^i (t)]_{,j} \, v_A^j(t) \right\}\,,
\label{eq:source-for-theta2}
\ea
\ew
where in the third line, we used $t_A \approx \tau$ to leading PN
order.
To obtain $\mu^0_A$, we follow~\cite{Blanchet:2011zv} and assume that
the spin vector is purely spatial in the stellar rest frame. Then, in
any frame, one has
\be
\label{eq:spin-condition}
\mu_{A}^{\mu} u^A_{\mu}=0\,, 
\ee
because the spin vector is assumed orthogonal to the four-velocity and
such an assumption gives $\mu_a^0 \propto S^0=0$ in the rest frame;
this leads to
\be
\mu^0_A = \mu_A^i v^A_i \left[1  + \mathcal{O} \left( v^{2} \right)\right]\,,
\ee
where we used $u_0^A=-1$ to leading PN order. Substituting this and
Eq.~\eqref{eq:source-for-theta2} into
Eq.~\eqref{eq:scalar-eq-with-mu0}, the field equation for
$\vartheta^\FZ_{(2)}$ becomes
\ba
\Box_\eta \vartheta^\FZ_{(2)} &=& 4 \pi \sum_{A=1,2} \left[ \frac{d}{dt} \left( \mu_A^i v^A_i \right) \delta^{(3)}(\bm{x} - \bm{x}_A) \right. \nn \\
& - & \left. \mu_A^i v^A_i v_A^j \delta^{(3)}(\bm{x} - \bm{x}_A)_{,j} \right]\,.
\ea
When solving this equation for the FZ solution, the first term gives a
monopolar contribution while the second term gives a dipolar one, and
hence, the former is dominant.  One then finds
\be
\label{eq:scalar-FZ-3}
 \vartheta^\FZ_{(2)} = - \frac{1}{r} \ddot{\mu}_{ii}.
\ee

We next look at the FZ scalar field $\vartheta^\FZ_{(3)}$ sourced by
the third term in Eq.~\eqref{eq:scalar-eq-with-mu0}.  Since
\ba
\Box \vartheta &=&\frac{1}{\sqrt{-g}} \partial_\mu \left( \sqrt{-g} \partial^{\mu} \vartheta \right) \nn \\
& = & \left( 1 - \frac{h}{2} \right) \partial_\mu \left[ \left( 1 + \frac{h}{2} \right) \left( \eta^{\mu\nu} - h^{\mu\nu}  \right)\partial_\nu \vartheta \right] + \mathcal{O}(h^2 \vartheta) \nn \\
& = & \partial_\mu \left[ \left( \eta^{\mu\nu} - h^{\mu\nu} + \frac{h}{2} \eta^{\mu\nu} \right)\partial_\nu \vartheta \right] - \frac{h}{2} \eta^{\mu\nu} \partial_{\mu\nu} \vartheta  \nn \\
& & + \mathcal{O}(h^2 \vartheta) \nn \\
& = & \eta^{\mu\nu} \partial_{\mu\nu} \vartheta - (\partial_\mu h^{\mu\nu}) (\partial_\nu \vartheta)  - h^{\mu\nu} \partial_{\mu\nu}  \vartheta \nn \\
& & + \frac{1}{2} \eta^{\mu\nu} (\partial_\mu h) (\partial_\nu \vartheta) + \mathcal{O}(h^2 \vartheta)\,,
\ea
one finds
\ba
\Box_\eta \vartheta^\FZ_{(3)} &=& - \Box_h \vartheta \nn \\
&=& (\partial_\mu h^{\mu\nu}) (\partial_\nu \vartheta)  + h^{\mu\nu} \partial_{\mu\nu}  \vartheta - \frac{1}{2} \eta^{\mu\nu} (\partial_\mu h) (\partial_\nu \vartheta)\,. \nn \\
\ea
The leading contribution comes from spatial derivatives (not time
derivatives). When solving the above wave equation by using the
Green's function method and performing the volume integral of the
source terms on the right-hand side, the contribution from the first
and second terms to the FZ solution vanish upon integration by
parts. Therefore, one is left with
\be
\label{eq:scalar-eq-boxh}
\Box_\eta \vartheta^\FZ_{(3)} = - \frac{1}{2} (\partial_i h) (\partial_i \vartheta)\,.
\ee

Let us first comment on the self-interaction contribution of the
source term in Eq.~\eqref{eq:scalar-eq-boxh} to the FZ scalar field
solution. To leading PN order, one finds
\ba
\vartheta^{\FZ,\mrm{self}}_{(3)} & = & -\frac{1}{4\pi} \frac{1}{r} \int \left( - \frac{1}{2} h^\NZ_{1,i} \vartheta^\NZ_{1,i} \right) d^3x \plusonetotwo \nn \\
& = & - \frac{1}{2\pi} \frac{1}{r} m_1 \mu_{1}^j \int \partial_i \left( \frac{1}{r_1} \right) \partial_{ij} \left( \frac{1}{r_1} \right) d^3x \plusonetotwo\,. \nn \\
\ea
Such an integral diverges at the body's position.
The effective source term in Eq.~\eqref{eq:scalar-field-eq} captures the regular part of this contribution.

Let us then focus on the cross-interaction contribution of
Eq.~\eqref{eq:scalar-eq-boxh}. The monopole radiation piece of
$\vartheta^\FZ_{(3)}$ is given by
\ba
\vartheta^\FZ_{(3)} & = & -\frac{1}{4\pi} \frac{1}{r} \int \left( - \frac{1}{2} h^\NZ_{1,i} \vartheta^\NZ_{2,i} \right) d^3x \plusonetotwo \nn \\
& = & - \frac{1}{2\pi} \frac{1}{r} m_1 \mu_{2}^j \int \partial_i \left( \frac{1}{r_1} \right) \partial_{ij} \left( \frac{1}{r_2} \right) d^3x \plusonetotwo \nn \\
& = & \frac{1}{2\pi} \frac{1}{r} m_1 \mu_{2}^j \partial_i^{(1)} \partial_{ij}^{(2)}  \int \frac{1}{r_1}\frac{1}{r_2} d^3x \plusonetotwo\,,  
\ea
where $\partial_i^{(A)}$ is the particle derivative with respect to
the $A$th body. Since the finite part of the remaining integral is
given by
$\int 1/(r_1 r_2) d^3x = - 2 \pi b$~\cite{2002PhRvD..65f4005B}, one
finds
\ba
\vartheta^\FZ_{(3)} & = & - \frac{1}{r} m_1 \mu_{2}^j \partial_i^{(1)} \partial_{ij}^{(2)} b \plusonetotwo \nn \\
&=& - \frac{1}{r} m_1 \mu_{2}^j \partial_{jii}^{(1)} b \plusonetotwo\,.
\ea
and since
\ba
\partial_{jii}^{(1)} b &=& \partial_{ji}^{(1)} n_{12i} = \partial_{j}^{(1)} \left( \frac{\delta_{ii}-n_{12ii}}{b} \right) \nn \\
&=& 2 \partial_{j}^{(1)} \left( \frac{1}{b} \right) = - \frac{2}{b^2} n_{12j}\,,
\ea
one finally finds
\ba
\vartheta^\FZ_{(3)} & = & \frac{2}{r} m_1 \mu_{2}^j \frac{n_{12}^j}{b^2} \plusonetotwo \nn \\
& = & \frac{2}{r} (m_1 \mu_2^j - m_2 \mu_1^j) \frac{n_{12}^j}{b^2} = \frac{2}{r} \ddot{\mu}_{ii}\,.
\label{eq:scalar-FZ-4}
\ea
Combining Eqs.~\eqref{eq:scalar-FZ-original}, \eqref{eq:scalar-FZ-3}
and~\eqref{eq:scalar-FZ-4}, the new FZ scalar field solution is given
by
\be
\label{eq:scalar-FZ-new}
\vartheta^\FZ = \sum_{k=1}^3 \vartheta^\FZ_{(k)} = \frac{1}{r} \ddot{\mu}_{ij} \left(n^{ij} + \delta^{ij} \right)\,.
\ee

We now derive the new scalar energy flux by substituting
Eq.~\eqref{eq:scalar-FZ-new} into the formula
\be
\delta \dot{E}^{(\vartheta)} = - \beta \lim_{r\to\infty} \int_{S^2_r} 
\left< \left( \dot{\vartheta}^{\FZ} \right)^2 \right>_{\omega} r^2 d\Omega\,.
\ee
Doing so, one finds
\ba
\delta  \dot{E}^{(\vartheta)} &=& -\beta
\left\langle \dddot{\mu}_{ij}\dddot{\mu}_{kl} \right\rangle_{\omega} \int_{S^2_\infty} \left( n^{ij} +  \delta^{ij} \right) \left( n^{kl} +  \delta^{kl} \right)  d\Omega\,. \nn\\\ea
Using the angle-averaging formulas
\ba
\int_{S^2_\infty} n^{ij} d\Omega &=& \frac{4\pi}{3} \delta^{ij}\,, \\
\int_{S^2_\infty} n^{ijkl} d\Omega &=& \frac{4\pi}{15} \left( \delta^{ij} \delta^{kl} + \delta^{ik} \delta^{jl} + \delta^{il} \delta^{jk} \right)\,, 
\ea
we find
\ba
\delta  \dot{E}^{(\vartheta)} &=& - \frac{4\pi}{15} \beta \left\langle \dddot{\mu}_{ij}\dddot{\mu}_{kl} \right\rangle_{\omega} \left( \delta^{ij} \delta^{kl} + \delta^{ik} \delta^{jl} + \delta^{il} \delta^{jk} \right. \nn \\
& & \left.  + 5  \delta^{ij} \delta^{kl}  + 5  \delta^{ij} \delta^{kl} + 15  \delta^{ij} \delta^{kl} \right) \nn \\
&=& - \frac{4\pi}{15} \beta \left\langle \dddot{\mu}_{ij}\dddot{\mu}_{kl} \right\rangle_{\omega} \left( \delta^{ik} \delta^{jl} + \delta^{il} \delta^{jk} + 26  \delta^{ij} \delta^{kl}  \right) \nn \\
&=& - \frac{4\pi}{15} \beta \left\langle \left[ \dddot{\mu}_{ij}\dddot{\mu}^{ij} + 27 \left(\dddot{\mu}^i{}_i\right)^2 \right] \right\rangle_{\omega}\,.
\label{eq:Edot-scalar}
\ea
Equation~(136) of~\cite{quadratic} gives
\be
\dddot{\mu}^{ij} =\frac{1}{b^3} 
  \left( m_1 v_{12}^{(i} \mu_2^{j)} - m_2
  v_{12}^{(i} \mu_1^{j)} \right)\,,
\ee
and substituting Eq.~\eqref{eq:dipole-charge-BH} into this equation,
one finds
\be
\dddot{\mu}^{ij} =- \frac{5}{2} \frac{\alpha_4}{\beta} \frac{m}{b^3} 
    v_{12}^{(i} \bar \Delta^{j)}\,,
\ee
where 
\be
\bar \Delta \equiv \frac{m_2}{m} \chi_1 \hat{S}_1^i - \frac{m_1}{m} \chi_2 \hat{S}_2^i\,.
\ee
Substituting this equation into Eq.~\eqref{eq:Edot-scalar}, we
finally find the new scalar energy flux:
\be
\delta  \dot{E}^{(\vartheta)} = - \frac{5}{48} \zeta_4 \left[ \bar{\Delta}^2+27
\left<  (\bar{\Delta}\cdot \hat{v}_{12})^2\right>_{\omega} \right] v^{14}\,.
\ee
This expression is very similar to Eq.~(137) in~\cite{quadratic}. The
only difference is that the factor of ``2'' in front of
$\left< (\bar{\Delta}\cdot \hat{v}_{12})^2\right>_{\omega}$ has now
become ``27''. We remind the reader that Eq.~\eqref{eq:spin-condition}
does not imply that
$\left< (\bar{\Delta}\cdot \hat{v}_{12})^2\right>_{\omega}$ vanishes
because the former is a condition on the 4-velocity. Observe, however,
that $\left<(\bar{\Delta}\cdot \hat{v}_{12})^2\right>_{\omega}$ does
vanish for spin-aligned binaries, and hence, this correction does not
affect the main results in~\cite{Yagi:2012vf}.


\bibliography{master}